\documentclass[3p,authoryear,twocolumn]{elsarticle}

\usepackage{ecrc}
\usepackage{newtxtext,newtxmath}

\volume{00}

\firstpage{1}

\journalname{New Astronomy}

\runauth{}

\jid{NewA}

\jnltitlelogo{New Astronomy}

\CopyrightLine{2011}{Published by Elsevier Ltd.}

\usepackage[figuresright]{rotating}

\newcommand{\Lf}{L1 }
\newcommand{\Ls}{L2 }
\newcommand{\Mtot}{M_\mathrm{b}}
\newcommand{\Msol}{\mathrm{M}_\odot}
\newcommand{\Rsol}{\mathrm{R}_\odot}
\newcommand{\Lsol}{\mathrm{L}_\odot}

\begin{document}

\begin{frontmatter}



\dochead{}

\title{Statistical analysis of eclipsing binaries with monotonic orbital-period variations: A-type W UMa contact systems}


\author{Shinjirou Kouzuma}
\address{Faculty of Liberal Arts and Sciences, Chukyo University, Nagoya, Aichi 466-8666, Japan}

\ead{skouzuma@lets.chukyo-u.ac.jp}

\begin{abstract}
On the basis of monotonic orbital-period variations, this study aims to identify genuine relationships between binary parameters and the rates of mass transfer (MT), mass loss (ML), and angular momentum loss (AML). 
Sample binaries with monotonic period variations are collected from the literature, together with well-determined binary parameters. 
Assuming the monotonic variations are responsible for any one of the MT, ML, and AML, their rates are calculated with the rates of change of period. 
After selecting crucial parameters using partial least-squares analysis, a parameter that exhibits the closest correlation with any one of the derived rates is further selected using partial regression plots. 
Moreover, power-law relationships are found for the discovered correlations. 
The properties of the sample binaries are also investigated by examining associations between binary parameters. 
In the systems with negative period variations, it is found that the rate of MT from more- to less-massive stars is a function of the primary radius; 
the AML rate is a function of the fill-out factor. 
In addition, the relationships between the mass ratio and stellar masses indicate that the ML rate relative to the MT rate decreases with increasing mass ratio below $\sim 0.46$. 
Meanwhile, in the systems with positive variations, it is found that the rate of MT from less- to more-massive stars is a function of the luminosity ratio and/or mass ratio; 
the ML rate is a function of the secondary temperature. 
The discussion also addresses possible processes occurring in the sample binaries. 
\end{abstract}

\begin{keyword}
binaries: eclipsing \sep binaries: close \sep stars: statistics \sep catalogues
\end{keyword}

\end{frontmatter}


\section{Introduction}\label{Sec_Intro}
Orbital-period variations often contain useful information on physical processes in binary systems. 
Several patterns of period variations exist: e.g., sudden, cyclic, and monotonic variations. 
Various mechanisms have been proposed for describing such variations. 
For instance, mass ejection causes a sudden change \citep{Wood1950-ApJ}. 
Light-travel time effect (LTTE) related to the motion around a third body \citep{Irwin1959-AJ} is a typical mechanism for cyclic variations. 
Magnetic activity \citep[i.e., Applegate mechanism;][]{Applegate1992-ApJ} can also describe a kind of cyclic variations. 
Monotonic variation, usually monotonically increases or decreases at a constant rate for a long time, is attributed to mass transfer and angular momentum loss (AML) \citep{Kwee1958-BAN,Morton1960-ApJ,Tout1991-MNRAS}. 
Analyzing period variation provides clues about physical processes, such as the amount of ejected mass, the orbital elements of a third body, and the mass-transfer rate. 

Close binaries are often observed as eclipsing binaries, which exhibit periodic minima in light curves. 
Time intervals between every other (or perhaps neighbor) minima are equivalent to orbital periods. 
Accordingly, orbital periods are readily derived from a light curve by measuring the epochs of minima. 
When a long-term light-curve, which is not necessarily continuous, is available, a long-term period-variation can be investigated. 
Thus, eclipsing binaries are suitable to study the trend of orbital-period variation over several or several tens of years. 

W Ursae Majoris (W UMa) binaries are contact systems composed of two stars with spectra generally later than F0; both stars overfill their inner Roche lobes\footnote{
Strictly speaking, such binaries are overcontact systems, and in contact systems both stars exactly fill their Roche lobes. 
However, this paper covers both possibilities. } \citep{Lucy1968-ApJ877,Lucy1968-ApJ1123,Mochnacki1981-ApJ}.  
In a binary system with at least one overfilled component, mass transfer is likely to occur. 
The mass transfer results in the system varying the stellar masses, chemical composition, and orbital separation. 
These changes strongly affect the evolution of binaries \citep{Paczynski1971-ARAA}. 
Moreover, AML, due to such as magnetic braking via stellar wind \citep{Schatzman1962-AnAp,Mestel1968-MNRAS}, is another crucial process that controls the binary evolution \citep{vantVeer1979-AA}. 
Assuming the mass transfer and AML, theoretical studies have proposed various evolutionary paths \citep[e.g., ][]{Paczynski1972-AcA,Rahunen1981-AA,Vilhu1982-AA,Stepien2006-AcA199}. 
Nevertheless, in close binaries, the properties of the mass transfer and AML are observationally and statistically still unclear. 
Therefore, it is important to unravel their observational properties such as how they evolve and which binary parameters are closely associated with them. 
These will lead to constructing precise models, constraining theoretical models, and unraveling the evolution of close binaries. 

\citet{Kouzuma2018-PASJ} statistically studied the properties of mass transfer in contact systems, using the light curves of eclipsing binaries observed with \textit{Kepler}. 
The cited author demonstrated that the mass-transfer rate is correlated with several binary parameters. 
However, in contact systems, several pairs of parameters are strongly correlated with each other, which results in the detection of spurious correlations. 
The parameters showing spurious correlations are not genuinely correlated. 
Therefore, it is important to identify parameters that have genuine correlations.  
In addition, their parameters are estimated via neural-network analysis \citep{Prsa2011-AJ}. 
Parameters determined without synthetic light curve analysis can potentially have large uncertainties. 
Particularly, the mass ratio is essential for calculating the mass-transfer rate, as well as determining other binary parameters. 
Hence, an analysis using such parameters might lead to erroneous conclusions. 
Accordingly, Kouzuma's results need to be verified with more reliable parameters. 

This paper attempts to identify the parameters genuinely associated with the mass transfer and AML, using sample binaries with well-determined parameters. 
The properties of the sample binaries are also investigated by examining associations between binary parameters. 
This paper organizes as follows. 
Section \ref{Sec_Sample} describes a method for collecting sample binaries. 
Section \ref{Sec_Assumption} introduces the basic assumptions in this paper. 
Section \ref{Sec_Correlation} presents a method for identifying genuine correlations and their results. 
In Section \ref{Sec_Property}, the properties of the sample binaries are examined. 
Section \ref{Sec_Discussion} discusses the results obtained from this analysis. 
The summary and conclusions are presented in Section \ref{Sec_Summary}.

\section{Sample}\label{Sec_Sample}
Sample eclipsing binaries with monotonic orbital-period variations were collected from literature. 
An $O-C$ diagram is a widely used tool for examining the trend of period variation. 
The $O-C$ values are computed by subtracting the calculated times of the minima from those observed. 
The author visually inspected the $O-C$ curves of the eclipsing binaries in the literature and selected the systems having parabolic curves. 

However, the LTTE can mimic the parabolic curves in the $O-C$ diagrams, because this effect leads to cyclic period oscillation. 
If cyclic period oscillation results in a sinusoidal-like curve and only one local minimum or maximum appears on an $O-C$ diagram, it can be difficult to distinguish between the parabolic and the sinusoidal-like curve using the $O-C$ diagram alone. 
Accordingly, the author selected only binaries whose $O-C$ diagrams display the superposition of the cyclic oscillation upon the parabolic curve. 
Such a superposition, which often appears, has been interpreted as both mass-transfer and the LTTE \citep[e.g., ][and references therein]{Yang2011-PASP}. 
Though four or higher-order systems might contaminate the sample, its fraction is likely to be typically 10--20 percent, or perhaps it might reach up to 30 percent in the worst case \citep{Tokovinin2006-AA,Raghavan2010-ApJS}. 
The binary parameters of such contaminants should be outliers in this analysis. 
Therefore, most of the parabolic $O-C$ curves of the collected binaries are expected to be caused by monotonic period variations. 
The rates of orbital-period changes are taken from relevant literature. 

The parameters of the sample binaries were also gathered from relevant literature, together with their errors. 
These parameters were determined by a synthetic light curve analysis based on methods such as those of \citet{Wilson1971-ApJ} and \citet{Djurasevic1992-ApSS}. 
For a system with a low inclination, the uncertainties of photometric mass-ratio and other parameters could be large. 
Accordingly, if systems had no spectroscopic mass-ratio and inclinations smaller than 75$^\circ$, they were ruled out. 
In the absence of data on mass in the relevant literature, they are estimated on the basis of their spectral types and \citet{Cox2000-book}. 
Because the fill-out factors of three systems (BX Dra, DK Cyg, and V921 Her) were unavailable, they were estimated in the same manner as in \citet{Yakut2005-ApJ1055}. 
The errors of parameters were unavailable in some cases. 
\citet{Liu2021-PASP} demonstrated that the errors in parameters derived from the light-curve analysis of contact binaries are typically within 1–20 percent, although they vary with conditions such as total or partial eclipses and the presence of third light. 
Considering this situation, $\sim10$ percent of the parameter values was adopted as their errors. 

This paper focuses on A-type systems, which is a subclass of W UMa binaries \citep{Binnendijk1970-VA}. 
Hereinafter, An and Ap systems refer to A-type binaries with negative and positive orbital-period variations, respectively. 
The sample comprises 14 (10) An and 19 (13) Ap systems, with the numbers in parentheses indicating the systems whose parameters were determined based on spectroscopic mass ratios. 

Table \ref{Tab_Sample} lists the collected An and Ap systems, together with their principal parameters. 
In this paper, the subscripts `1' and `2' represent the primary (more massive) and secondary (less massive) stars, respectively. 
The mass ratio is $q=M_2/M_1$, which is always smaller than 1. 
The fill-out factor is 
\begin{equation}
	f=\frac{\Omega-\Omega_\mathrm{in}}{\Omega_\mathrm{out}-\Omega_\mathrm{in}}, 
\end{equation}
where $\Omega$, $\Omega_\mathrm{in}$, and $\Omega_\mathrm{out}$ are the potentials of the stellar surface, inner and outer Roche lobes, respectively. 
The inner and outer Roche lobes are the inner and outer Lagrangian zero-velocity equipotential surfaces crossing the Lagrange points \Lf and \Ls, respectively \citep{Mochnacki1981-ApJ, Stepien2015-AA}. 

\begin{sidewaystable*}
\centering
  \begin{minipage}{\textwidth}
 \caption[]{\label{Tab_Sample} Principal parameters of the collected A-type W UMa systems.}
 \scriptsize
\resizebox{\textwidth}{!}{%
\begin{tabular}{lccccccclccrcrccrrrrl}
 \hline \hline \\
\multicolumn{1}{c}{Name}  & \multicolumn{1}{c}{$P$} & \multicolumn{1}{c}{$a$} & \multicolumn{1}{c}{$R_1$}  & \multicolumn{1}{c}{$R_2$}  & \multicolumn{1}{c}{$M_1$}  & \multicolumn{1}{c}{$M_2$}  & \multicolumn{1}{c}{$M_\mathrm{b}$}  & \multicolumn{1}{c}{$q$}  & \multicolumn{1}{c}{$T_1$}  & \multicolumn{1}{c}{$T_2$}  & \multicolumn{1}{c}{$|\Delta T|$}  & \multicolumn{1}{c}{$f$}  & \multicolumn{1}{c}{$L_1$}  & \multicolumn{1}{c}{$L_2$}  & \multicolumn{1}{c}{$L_2/L_1$}  & \multicolumn{1}{c}{$J \times 10^{-51}$}  & \multicolumn{1}{c}{$\dot{P} \times 10^{7}$}  & \multicolumn{1}{c}{$\dot{M}_{12} \times 10^{7}$} & \multicolumn{1}{c}{$\dot{M}_\mathrm{b} \times 10^{7}$} & \multicolumn{1}{c}{Refs.}  \\
\multicolumn{1}{c}{} & \multicolumn{1}{c}{[d]} & \multicolumn{1}{c}{[$\Rsol$]} & \multicolumn{1}{c}{$\Rsol$} & \multicolumn{1}{c}{$\Rsol$} & \multicolumn{1}{c}{[$\Msol$]} & \multicolumn{1}{c}{[$\Msol$]} & \multicolumn{1}{c}{[$\Msol$]} & \multicolumn{1}{c}{} & \multicolumn{1}{c}{[K]} & \multicolumn{1}{c}{[K]} & \multicolumn{1}{c}{[K]} & \multicolumn{1}{c}{} & \multicolumn{1}{c}{[$\Lsol$]} & \multicolumn{1}{c}{[$\Lsol$]} & \multicolumn{1}{c}{} & \multicolumn{1}{c}{[g cm$^2$ s$^{-1}$]} & \multicolumn{1}{c}{[d yr$^{-1}$]} & \multicolumn{1}{c}{[$\Msol$ yr$^{-1}$]} & \multicolumn{1}{c}{[$\Msol$ yr$^{-1}$]} & \multicolumn{1}{c}{} \\ 
 \\ \hline
    AG Vir & $0.6427$ & $ 4.02$ & $ 2.07$ & $ 1.27$ & $ 1.61$ & $ 0.51$ & $ 2.12$ & $ 0.31^\mathrm{s}$ & $ 7400$ & $ 7000$ & $  400$ & $ 0.52$ & $11.48$ & $ 3.47$ & $ 0.30$ & $  7.35$ & $   0.40$ & $   0.16$ & $  0.67$ & 1, 2 \\
    AH Cnc & $0.3604$ & $ 2.32$ & $ 1.30$ & $ 0.62$ & $ 1.10$ & $ 0.19$ & $ 1.29$ & $ 0.17           $ & $ 6300$ & $ 6265$ & $   35$ & $ 0.58$ & $ 2.40$ & $ 0.53$ & $ 0.22$ & $  1.92$ & $   3.99$ & $   0.85$ & $  7.14$ & 3 \\
    AH Tau & $0.3327$ & $ 2.34$ & $ 1.05$ & $ 0.77$ & $ 1.04$ & $ 0.52$ & $ 1.56$ & $ 0.49           $ & $ 5900$ & $ 5887$ & $   13$ & $ 0.11$ & $ 1.19$ & $ 0.64$ & $ 0.54$ & $  4.19$ & $  -0.70$ & $  -0.73$ &     ---  & 4 \\
    AP Leo & $0.4304$ & $ 2.98$ & $ 1.49$ & $ 0.87$ & $ 1.47$ & $ 0.44$ & $ 1.91$ & $ 0.30^\mathrm{s}$ & $ 6150$ & $ 6201$ & $   51$ & $ 0.25$ & $ 2.85$ & $ 1.00$ & $ 0.35$ & $  5.23$ & $  -1.08$ & $  -0.53$ &     ---  & 5, 6 \\
    AU Ser & $0.3865$ & $ 2.57$ & $ 1.10$ & $ 0.94$ & $ 0.90$ & $ 0.64$ & $ 1.53$ & $ 0.71^\mathrm{s}$ & $ 5495$ & $ 5153$ & $  342$ & $ 0.20$ & $ 0.99$ & $ 0.56$ & $ 0.56$ & $  4.64$ & $  -1.38$ & $  -2.60$ &     ---  & 7, 8 \\
    AW UMa & $0.4387$ & $ 3.03$ & $ 1.87$ & $ 0.66$ & $ 1.79$ & $ 0.14$ & $ 1.93$ & $ 0.07^\mathrm{s}$ & $ 7175$ & $ 7022$ & $  153$ & $ 0.85$ & $ 7.27$ & $ 0.83$ & $ 0.11$ & $  2.54$ & $  -1.90$ & $  -0.22$ &     ---  & 9, 10 \\
    BX Dra & $0.5790$ & $ 4.06$ & $ 2.13$ & $ 1.28$ & $ 2.08$ & $ 0.60$ & $ 2.68$ & $ 0.29^\mathrm{s}$ & $ 6980$ & $ 6758$ & $  222$ & $ 0.57$ & $ 9.66$ & $ 3.05$ & $ 0.32$ & $ 10.05$ & $   5.65$ & $   2.74$ & $ 13.08$ & 11, 12 \\
    CK Boo & $0.3552$ & $ 2.44$ & $ 1.45$ & $ 0.59$ & $ 1.39$ & $ 0.15$ & $ 1.54$ & $ 0.11^\mathrm{s}$ & $ 6380$ & $ 6340$ & $   40$ & $ 0.72$ & $ 2.72$ & $ 0.43$ & $ 0.16$ & $  1.97$ & $   0.98$ & $   0.16$ & $  2.12$ & 13, 14 \\
    DK Cyg & $0.4707$ & $ 3.40$ & $ 1.74$ & $ 1.05$ & $ 1.82$ & $ 0.56$ & $ 2.38$ & $ 0.33^\mathrm{s}$ & $ 7500$ & $ 7011$ & $  489$ & $ 0.50$ & $ 8.50$ & $ 2.40$ & $ 0.28$ & $  7.91$ & $   1.00$ & $   0.57$ & $  2.53$ & 14, 15 \\
    DZ Psc & $0.3661$ & $ 2.50$ & $ 1.46$ & $ 0.67$ & $ 1.37$ & $ 0.19$ & $ 1.56$ & $ 0.14^\mathrm{s}$ & $ 6210$ & $ 6124$ & $   86$ & $ 0.90$ & $ 2.85$ & $ 0.57$ & $ 0.20$ & $  2.34$ & $   7.40$ & $   1.49$ & $ 15.77$ & 16, 17 \\
    EF Dra & $0.4240$ & $ 3.04$ & $ 1.70$ & $ 0.78$ & $ 1.81$ & $ 0.29$ & $ 2.10$ & $ 0.16^\mathrm{s}$ & $ 6250$ & $ 6186$ & $   64$ & $ 0.47$ & $ 3.96$ & $ 0.79$ & $ 0.20$ & $  4.37$ & $   5.44$ & $   1.48$ & $ 13.51$ & 6, 18 \\
    EM Psc & $0.3440$ & $ 2.22$ & $ 1.30$ & $ 0.64$ & $ 1.08$ & $ 0.16$ & $ 1.24$ & $ 0.15           $ & $ 5300$ & $ 4987$ & $  313$ & $ 0.95$ & $ 1.19$ & $ 0.23$ & $ 0.19$ & $  1.63$ & $  39.71$ & $   7.26$ & $ 71.62$ & 19, 20 \\
    GR Vir & $0.3470$ & $ 2.40$ & $ 1.42$ & $ 0.61$ & $ 1.37$ & $ 0.17$ & $ 1.54$ & $ 0.12^\mathrm{s}$ & $ 6300$ & $ 6163$ & $  137$ & $ 0.79$ & $ 2.87$ & $ 0.48$ & $ 0.17$ & $  2.10$ & $  -4.32$ & $  -0.80$ &     ---  & 14, 21 \\
    IK Per & $0.6760$ & $ 4.30$ & $ 2.40$ & $ 1.15$ & $ 1.99$ & $ 0.34$ & $ 2.33$ & $ 0.19           $ & $ 9070$ & $ 7470$ & $ 1600$ & $ 0.52$ & $35.04$ & $ 5.67$ & $ 0.16$ & $  6.30$ & $  -2.52$ & $  -0.51$ &     ---  & 22 \\
    LO And & $0.3804$ & $ 2.69$ & $ 1.30$ & $ 0.85$ & $ 1.31$ & $ 0.49$ & $ 1.80$ & $ 0.37           $ & $ 6500$ & $ 6465$ & $   35$ & $ 0.31$ & $ 2.70$ & $ 1.13$ & $ 0.42$ & $  5.03$ & $   2.46$ & $   1.69$ & $  5.81$ & 23 \\
    NO Cam & $0.4308$ & $ 3.03$ & $ 1.49$ & $ 1.07$ & $ 1.40$ & $ 0.61$ & $ 2.01$ & $ 0.44           $ & $ 6530$ & $ 6486$ & $   44$ & $ 0.56$ & $ 3.60$ & $ 1.80$ & $ 0.50$ & $  6.71$ & $   5.34$ & $   4.47$ & $ 12.46$ & 24 \\
    OO Aql & $0.5068$ & $ 3.34$ & $ 1.41$ & $ 1.31$ & $ 1.06$ & $ 0.90$ & $ 1.96$ & $ 0.85^\mathrm{s}$ & $ 6100$ & $ 5926$ & $  174$ & $ 0.37$ & $ 2.45$ & $ 1.89$ & $ 0.77$ & $  7.82$ & $  -0.36$ & $  -1.39$ &     ---  & 25, 26 \\
    RR Cen & $0.6057$ & $ 3.92$ & $ 2.10$ & $ 1.05$ & $ 1.82$ & $ 0.38$ & $ 2.20$ & $ 0.21^\mathrm{s}$ & $ 6912$ & $ 6891$ & $   21$ & $ 0.35$ & $ 8.89$ & $ 2.20$ & $ 0.25$ & $  6.17$ & $   1.21$ & $   0.32$ & $  2.19$ & 27, 28 \\
    RZ Tau & $0.4157$ & $ 3.11$ & $ 1.56$ & $ 1.04$ & $ 1.70$ & $ 0.64$ & $ 2.34$ & $ 0.38^\mathrm{s}$ & $ 7300$ & $ 7194$ & $  106$ & $ 0.55$ & $ 6.19$ & $ 2.60$ & $ 0.42$ & $  8.08$ & $   0.97$ & $   0.80$ & $  2.72$ & 29 \\
    TU Boo & $0.3243$ & $ 2.25$ & $ 1.05$ & $ 0.78$ & $ 0.97$ & $ 0.48$ & $ 1.45$ & $ 0.51           $ & $ 5800$ & $ 5737$ & $   63$ & $ 0.17$ & $ 1.10$ & $ 0.62$ & $ 0.56$ & $  3.68$ & $  -0.74$ & $  -0.72$ &     ---  & 30, 31 \\
    TV Mus & $0.4457$ & $ 2.85$ & $ 1.70$ & $ 0.83$ & $ 1.35$ & $ 0.22$ & $ 1.57$ & $ 0.12^\mathrm{s}$ & $ 5980$ & $ 5808$ & $  172$ & $ 0.74$ & $ 3.33$ & $ 0.71$ & $ 0.21$ & $  2.80$ & $  -2.16$ & $  -0.43$ &     ---  & 32, 33 \\
    TY Pup & $0.8192$ & $ 4.60$ & $ 2.64$ & $ 1.37$ & $ 1.65$ & $ 0.30$ & $ 1.95$ & $ 0.18           $ & $ 6900$ & $ 6915$ & $   15$ & $ 0.84$ & $14.11$ & $ 3.86$ & $ 0.27$ & $  5.26$ & $   0.56$ & $   0.08$ & $  0.66$ & 34 \\
    UZ Leo & $0.6180$ & $ 4.21$ & $ 2.23$ & $ 1.40$ & $ 2.01$ & $ 0.62$ & $ 2.63$ & $ 0.30^\mathrm{s}$ & $ 6980$ & $ 6772$ & $  208$ & $ 0.76$ & $10.60$ & $ 3.68$ & $ 0.35$ & $ 10.30$ & $   3.49$ & $   1.69$ & $  7.42$ & 14, 35 \\
 V1073 Cyg & $0.7859$ & $ 4.70$ & $ 2.33$ & $ 1.36$ & $ 1.73$ & $ 0.53$ & $ 2.26$ & $ 0.30^\mathrm{s}$ & $ 6700$ & $ 6520$ & $  180$ & $ 0.17$ & $ 9.77$ & $ 3.01$ & $ 0.31$ & $  8.55$ & $  -0.48$ & $  -0.16$ &     ---  & 36, 37, 38 \\
 V1918 Cyg & $0.4132$ & $ 2.90$ & $ 1.52$ & $ 0.87$ & $ 1.52$ & $ 0.40$ & $ 1.92$ & $ 0.26           $ & $ 7060$ & $ 6924$ & $  136$ & $ 0.49$ & $ 5.15$ & $ 1.56$ & $ 0.30$ & $  4.91$ & $  -4.31$ & $  -1.89$ &     ---  & 39 \\
  V366 Cas & $0.7293$ & $ 4.47$ & $ 1.90$ & $ 1.81$ & $ 1.19$ & $ 1.06$ & $ 2.25$ & $ 0.89           $ & $ 5860$ & $ 5907$ & $   47$ & $ 0.39$ & $ 3.82$ & $ 3.57$ & $ 0.93$ & $ 11.20$ & $   4.88$ & $  21.63$ & $  7.53$ & 40 \\
  V401 Cyg & $0.5827$ & $ 3.81$ & $ 1.95$ & $ 1.17$ & $ 1.68$ & $ 0.50$ & $ 2.18$ & $ 0.29^\mathrm{s}$ & $ 6700$ & $ 6650$ & $   50$ & $ 0.46$ & $ 6.87$ & $ 2.42$ & $ 0.35$ & $  7.22$ & $   1.50$ & $   0.61$ & $  2.81$ & 41, 42, 43 \\
  V508 Oph & $0.3448$ & $ 2.38$ & $ 1.06$ & $ 0.80$ & $ 1.01$ & $ 0.52$ & $ 1.53$ & $ 0.52^\mathrm{s}$ & $ 6000$ & $ 5810$ & $  190$ & $ 0.16$ & $ 1.22$ & $ 0.52$ & $ 0.42$ & $  4.14$ & $  -1.50$ & $  -1.56$ &     ---  & 44, 45, 46 \\
  V566 Oph & $0.4096$ & $ 2.86$ & $ 1.49$ & $ 0.81$ & $ 1.50$ & $ 0.38$ & $ 1.88$ & $ 0.25^\mathrm{s}$ & $ 6456$ & $ 6433$ & $   23$ & $ 0.34$ & $ 3.43$ & $ 1.01$ & $ 0.29$ & $  4.62$ & $   3.30$ & $   1.37$ & $  7.57$ & 47 \\
  V776 Cas & $0.4404$ & $ 2.97$ & $ 1.72$ & $ 0.80$ & $ 1.60$ & $ 0.21$ & $ 1.81$ & $ 0.13^\mathrm{s}$ & $ 7050$ & $ 6907$ & $  143$ & $ 0.79$ & $ 6.56$ & $ 1.31$ & $ 0.20$ & $  3.07$ & $  -5.16$ & $  -0.94$ &     ---  & 48, 49 \\
  V839 Oph & $0.4090$ & $ 2.94$ & $ 1.53$ & $ 0.87$ & $ 1.57$ & $ 0.46$ & $ 2.03$ & $ 0.30^\mathrm{s}$ & $ 6250$ & $ 6349$ & $   99$ & $ 0.53$ & $ 3.15$ & $ 1.10$ & $ 0.35$ & $  5.69$ & $   3.09$ & $   1.65$ & $  7.68$ & 14, 50, 51 \\
  V921 Her & $0.8774$ & $ 5.00$ & $ 2.56$ & $ 1.29$ & $ 1.78$ & $ 0.40$ & $ 2.19$ & $ 0.23^\mathrm{s}$ & $ 7700$ & $ 7003$ & $  697$ & $ 0.03$ & $12.20$ & $ 1.59$ & $ 0.13$ & $  7.17$ & $   2.79$ & $   0.55$ & $  3.48$ & 17, 52 \\
    YY CrB & $0.3766$ & $ 2.68$ & $ 1.45$ & $ 0.81$ & $ 1.47$ & $ 0.36$ & $ 1.82$ & $ 0.24^\mathrm{s}$ & $ 6077$ & $ 6198$ & $  121$ & $ 0.64$ & $ 2.56$ & $ 0.86$ & $ 0.34$ & $  4.21$ & $  -6.73$ & $  -2.81$ &     ---  & 53 \\ 
\hline
\end{tabular}
} 
  \small
  \parbox{\hsize}{\emph{Notes.} The values of mass ratio ($q$) with the superscript `s' are spectroscopic mass-ratios. 
  References: 
(1) \citet{Bell1990-MNRAS632}; (2) \citet{Qian2001-MNRAS914}; 
(3) \citet{Qian2006-AJ131}; 
(4) \citet{Yang2010-AJ}; 
(5) \citet{Qian2007-AJ357}; (6) \citet{Lu1999-AJ}; 
(7) \citet{Gurol2005-NewA}; (8) \citet{Pribulla2009-AJ3646}; 
(9) \citet{Pribulla1999-AA}; (10) \citet{McLean1981-MNRAS}; 
(11) \citet{Park2013-PASJ}; (12) \citet{Pych2004-AJ}; 
(13) \citet{Yang2012-AJ122}; (14) \citet{Rucinski1999-AJ}; 
(15) \citet{Lee2015-AJ}; 
(16) \citet{Yang2013-AJ}; (17) \citet{Rucinski2003-AJ}; 
(18) \citet{Yang2012-RAA}; 
(19) \citet{Qian2008-AJ1940}; (20) \citet{Yang2005-ApSS}; 
(21) \citet{Qian2004-AJ}; 
(22) \citet{Zhu2005-AJ}; 
(23) \citet{Gurol2005-AN}; 
(24) \citet{Zhou2017-PASJ}; 
(25) \citet{Li2016-RAA}; (26) \citet{Pribulla2007-AJ}; 
(27) \citet{Yang2005-PASJ}; (28) \citet{King1984-MNRAS}; 
(29) \citet{Yang2003-AJ}; 
(30) \citet{Lee2007-PASP}; (31) \citet{Niarchos1996-AAS}; 
(32) \citet{Qian2005-AJ224}; (33) \citet{Hilditch1989-MNRAS}; 
(34) \citet{Sarotsakulchai2018-AJ}; 
(35) \citet{Lee2018-PASP}; 
(36) \citet{Ekmekci2012-NewA}; (37) \citet{Tian2018-RAA}; (38) \citet{Pribulla2006-AJ}; 
(39) \citet{Yang2013-AJ60}; 
(40) \citet{Yang2013-NewA}; 
(41) \citet{Wolf2000-AAS}; (42) \citet{Rucinski2002-AJ}; (43) \citet{Zhu2013-AJ}; 
(44) \citet{Lapasset1990-AA}; (45) \citet{Xiang2015-AJ62}; (46) \citet{Lu1986-PASP}; 
(47) \citet{Selam2018-ApSS}; 
(48) \citet{Noori2017-NewA}; (49) \citet{Rucinski2001-AJ}; 
(50) \citet{Gazeas2006-AcA}; (51) \citet{Wolf1996-IBVS}; 
(52) \citet{Zhou2016-AdAst}; 
(53) \citet{Essam2010-NewA}. 
}
\end{minipage}
\end{sidewaystable*}

\section{Assumptions}\label{Sec_Assumption}
\subsection{Relation between physical processes and orbital-period variations}\label{Sec_Formulae}
Monotonic variations in orbital period are caused by physical processes in binary systems, such as mass transfer between components, mass loss (ML), and AML \citep{Kwee1958-BAN,Morton1960-ApJ,Tout1991-MNRAS}.  
A relation between period change and these processes is familiar and is formalized as follows. 
This section also introduces the notation used in this paper. 

The total angular momentum ($J$) of a binary system is the sum of the orbital and spin angular momenta: 
\begin{equation}\label{Eq_Jtot}
	J=J_\mathrm{orb} + J_\mathrm{spin}. 
\end{equation}
Assuming that a contact binary has a circular orbit, we obtain
\begin{equation}\label{Jorb}
	J_{\mathrm{orb}} = \frac{M_1 M_2}{M_1+M_2} a^2 \omega, 
\end{equation}
where $M_1$ and $M_2$ are the masses of the primary and secondary stars, $a$ is the orbital separation, and $\omega$ ($\omega=2\pi/P$, where $P$ is the orbital period) is the angular velocity. 
We assume that a contact binary has tidally-synchronized spins, then the spin angular momentum is 
\begin{eqnarray}
	J_{\mathrm{spin}} &=& J_{\mathrm{spin, 1}}  + J_{\mathrm{spin, 2}} \\
	&=& k_1^2 M_1 R_1^2 \omega + k_2^2 M_2 R_2^2 \omega, 
\end{eqnarray}
where $k_1^2$ and $k_2^2$ are the dimensionless gyration radii of the components. 
The dimensionless gyration radii are taken from \citet{Rasio1995-ApJ}, and we assume $k_1^2=k_2^2=0.06$. 
In close binaries, the spin angular momentum is much smaller than the orbital one. 
The sample has a mean of $J_{\mathrm{spin}}/J_{\mathrm{orb}}=0.10$ with a standard deviation of 0.06; 
all the sample systems have values smaller than 0.22 ($2\sigma$ level) except for AW UMa ($J_{\mathrm{spin}}/J_{\mathrm{orb}}=0.31$). 
Therefore, the spin angular momentum is sufficiently smaller than the orbital one, and it is possible to approximate $J \sim J_\mathrm{orb}$. 

In a contact binary, mass can be exchanged between component stars in either direction because both stars fill their Roche lobes. 
Moreover, mass can be also lost from each star. 
Hence, the mass-transfer rates of the two stars are 
\begin{eqnarray}
	\dot{M}_1&=&\dot{M}_{12}+\dot{m}_1, \\
	\dot{M}_2&=&-\dot{M}_{12}+\dot{m}_2, 
\end{eqnarray}
where $\dot{M}_{12}$ is the rate of mass transfer between the two stars, and $\dot{m}_1$ and $\dot{m}_2$ are the rates of ML from the primary and secondary stars, respectively. 
Assuming that the ML is due to spherical outflow such as isotropic stellar wind, the rate of AML is expressed as
\begin{eqnarray}
	\dot{J}=\dot{m}_1 \left(\frac{M_2}{\Mtot}a \right)^2 \omega + \dot{m}_2 \left(\frac{M_1}{\Mtot}a \right)^2 \omega + KJ, 
\end{eqnarray}
where $\Mtot$ is the total mass. 
The term $KJ$ represents the rate of extra AML, which should have a value ($K<0$) when extra AML due to other mechanisms occurs. 

The above equations and Kepler's law yield the following relation under the assumption $J\sim J_\mathrm{orb}$: 
\begin{equation}\label{Eq_Period_Change}
	\frac{\dot{P}}{P} = -2\frac{\dot{\Mtot}}{\Mtot} + \frac{3 (M_1-M_2)}{M_1 M_2} \dot{M}_{12} + 3 K, 
\end{equation}
where $\dot{\Mtot}=\dot{m}_1+\dot{m}_2$. 
When $\dot{M}_{12}$ is negative, mass transfer from more- to less-massive stars (MTML) occurs. 
On the contrary, when $\dot{M}_{12}$ is positive, mass transfer from less- to more-massive stars (MTLM) occurs. 

This paper assumes that the physical processes described in equation (\ref{Eq_Period_Change}) are responsible for the monotonic orbital-period variations of the sample binaries.

\subsection{Interpretation of orbital-period variations}\label{Sec_Assumption_Interpretation}
Equation (\ref{Eq_Period_Change}) indicates that a negative period variation arises from the MTML and/or extra AML whereas a positive one is due to the MTLM and/or ML. 
However, it is difficult to know precisely how each process contributes to an observed period variation because of a lack of previous studies. 
Accordingly, to simplify the situation, assuming that any one of these processes solely causes period variations, 
the rates of MTML, extra AML (for the An sample), MTLM, and ML (for the Ap sample) are calculated from the observed rate of period change.

\begin{figure*}
\centering
\begin{minipage}[b]{0.24\textwidth}
	\includegraphics[width=1.0\textwidth]{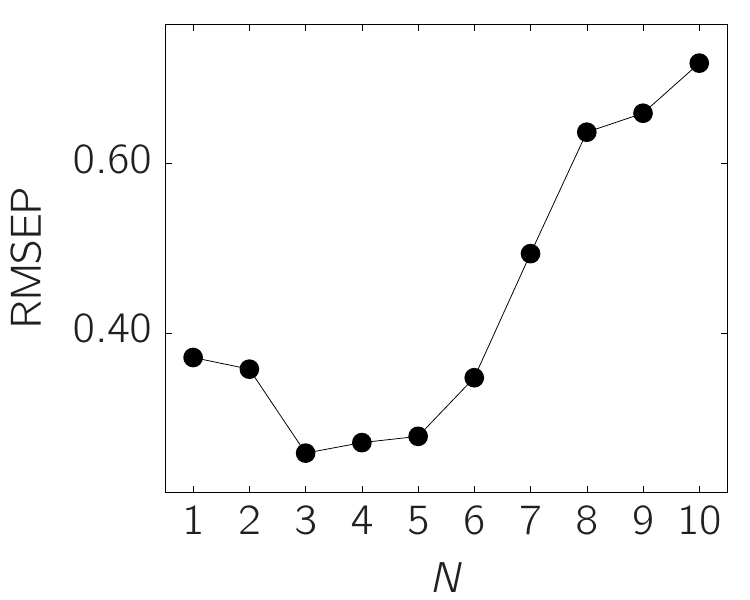}
	\centerline{(a)}
\end{minipage}
\begin{minipage}[b]{0.24\textwidth}
	\includegraphics[width=1.0\textwidth]{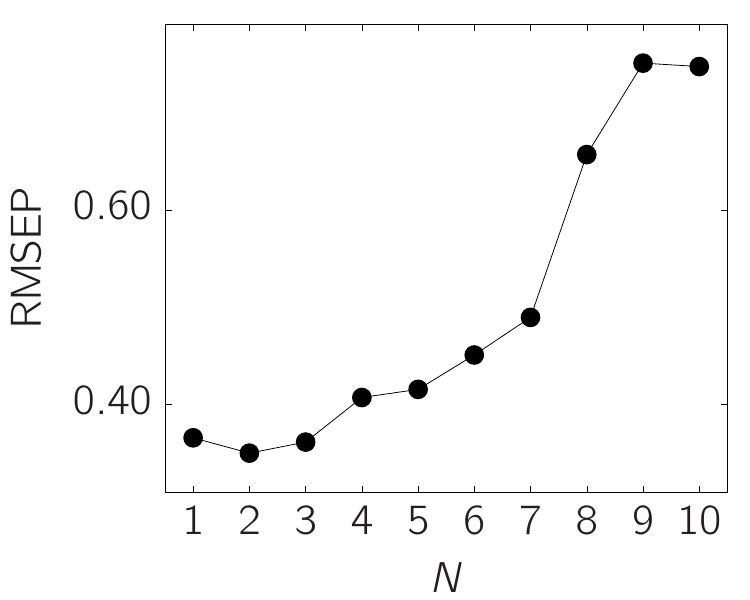}
	\centerline{(b)}
\end{minipage}
\begin{minipage}[b]{0.24\textwidth}
	\includegraphics[width=1.0\textwidth]{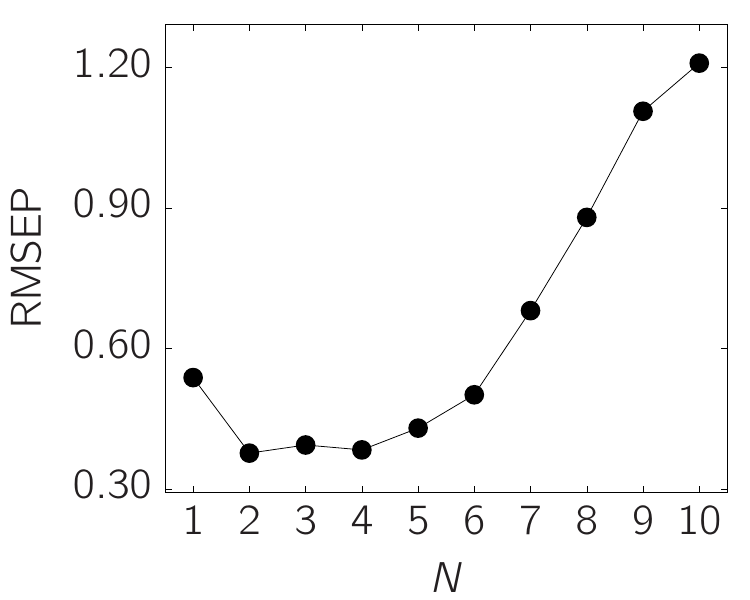}
	\centerline{(c)}
\end{minipage}
\begin{minipage}[b]{0.24\textwidth}
	\includegraphics[width=1.0\textwidth]{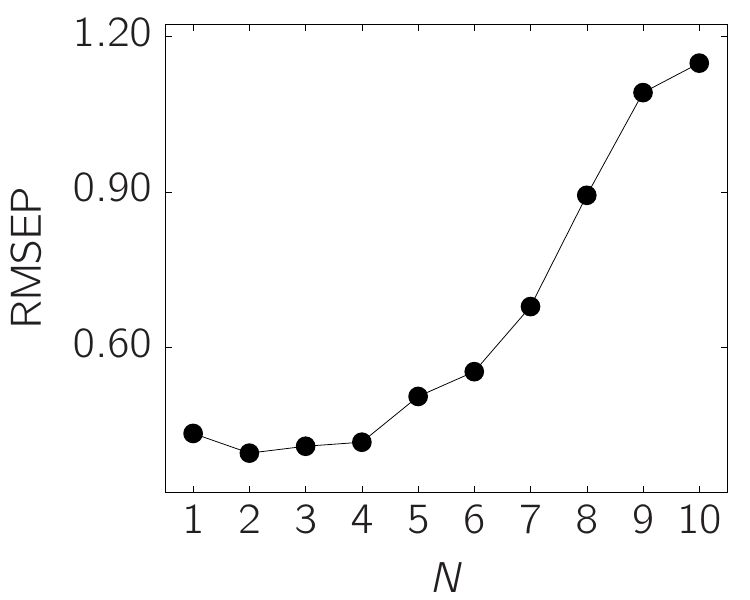}
	\centerline{(d)}
\end{minipage}
\caption{RMSEP value as a function of the number of PLS components ($N$) for MTML (a), extra AML (b), MTLM (c), and ML (d). 
\label{Fig_RMSEP}}
\end{figure*}

\begin{table*}
\centering
\caption[]{Variances of X- and Y-variables explained by latent variables. 
\label{Tab_EV}}
\begin{tabular}{rrrrrrrrrrrr}
\hline \hline
Sample $\backslash$ $N$ & & 1 & 2 & 3 & 4 & 5 & 6 & 7 & 8 & 9 & 10 \\
\hline
MTML 	& X  &  56.0  &  67.7  &  91.9  &  95.8  &  98.6  &  99.5  &  99.8  &  99.9  &  100 & 100  \\
				& Y  &  32.6  &  68.1  &  78.5  &  84.3  &  86.0  &  87.9  &  89.5  &  92.0  &  93.0  &  94.2  \\
K		    	& X  &   31.4  &  35.6  &  89.1  &  95.5  &  97.9  &  99.4  &  99.8  &  99.9  &  100  &  100  \\
				& Y  &   59.4  &  75.4  &  76.0  &  79.8  &  84.0  &  88.0  &  90.8  &  92.2  &  94.0  &  95.2  \\
MTLM 	& X  &   32.6  &  75.8  &  84.3  &  88.1  &  93.2  &  99.7  &  99.9  &  99.9  &  100  &  100  \\
				& Y  &  57.2  &  73.9  &  84.2  &  86.8  &  87.0  &  87.1  &  90.2  &  91.3  &  94.2  &  95.0  \\
ML 			& X  &   58.2  &  72.8  &  83.3  &  88.0  &  92.9  &  99.4  &  99.9  &  99.9  &  100  &  100  \\
				& Y  &  39.9  &  64.9  &  75.5  &  79.2  &  79.7  &  80.2  &  86.2  &  88.4  &  92.6  &  93.3  \\
\hline
\end{tabular}
  \parbox{\hsize}{\emph{Notes.} The values are the cumulative percentages of the explained variances for X and Y. }
\end{table*}

\begin{figure*}
\centering
\begin{minipage}[b]{0.4\textwidth}
	\includegraphics[width=1.0\textwidth]{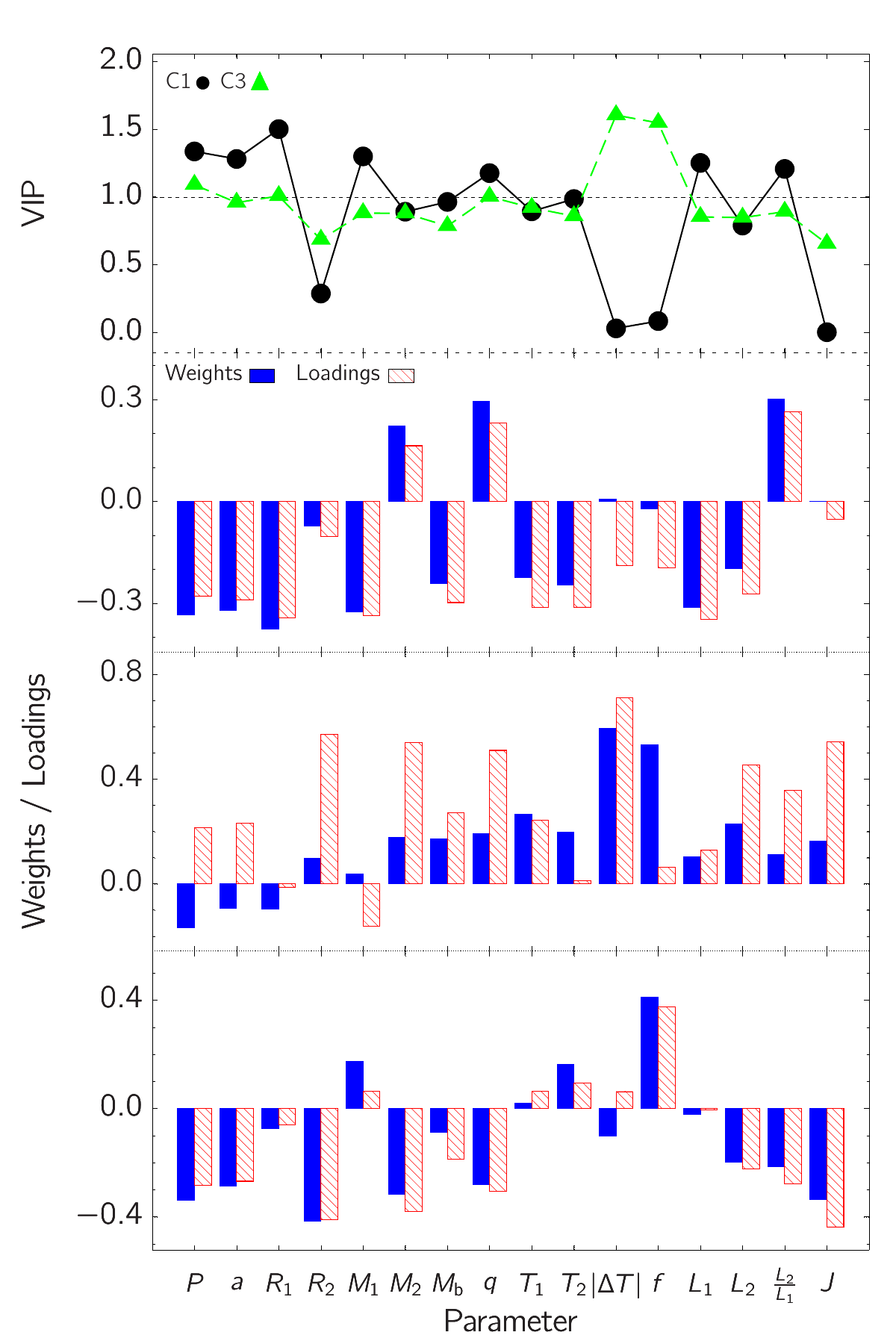}
	\centerline{(a)}
\end{minipage}
\begin{minipage}[b]{0.4\textwidth}
	\includegraphics[width=1.0\textwidth]{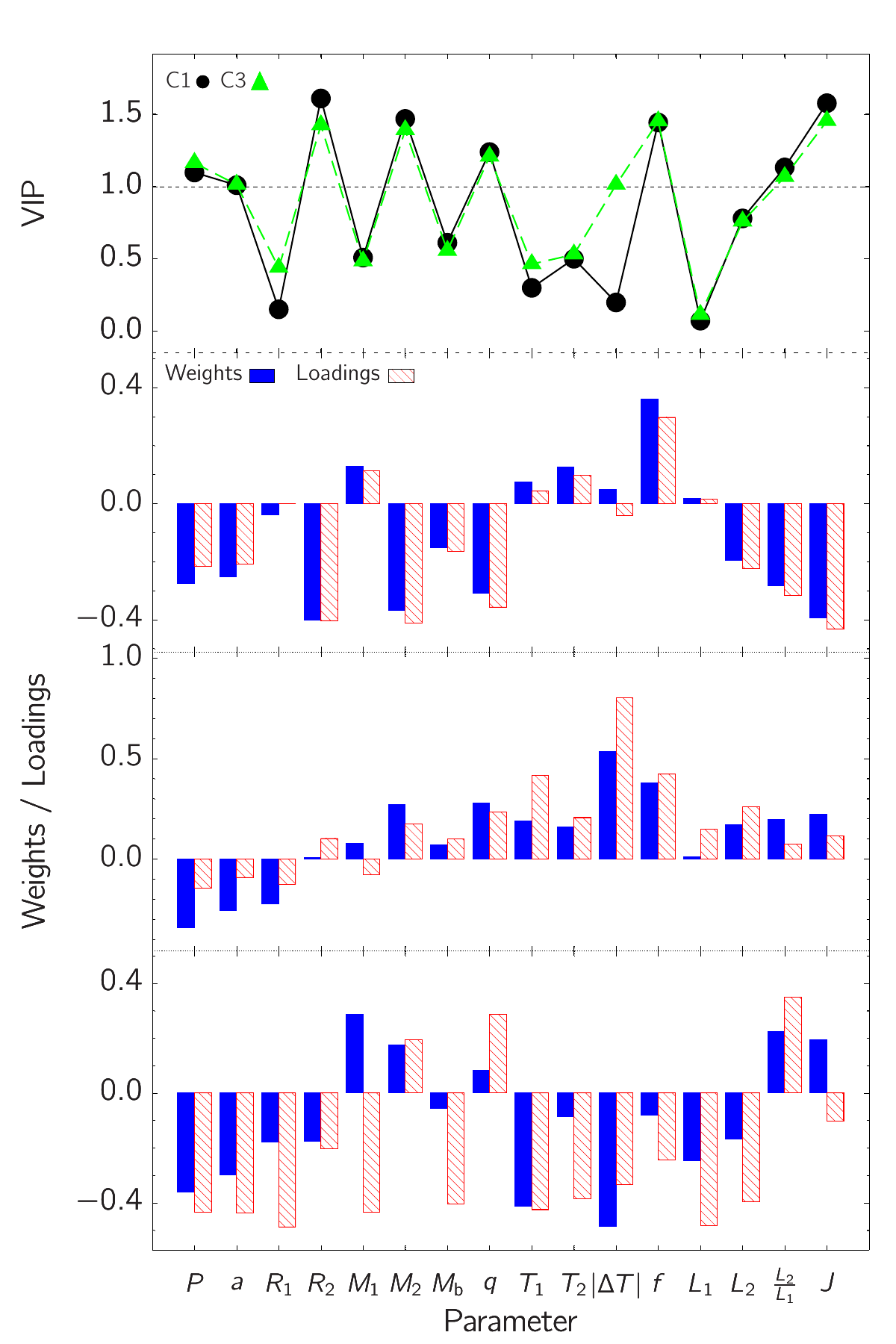}
	\centerline{(b)}
\end{minipage}
\begin{minipage}[b]{0.4\textwidth}
	\includegraphics[width=1.0\textwidth]{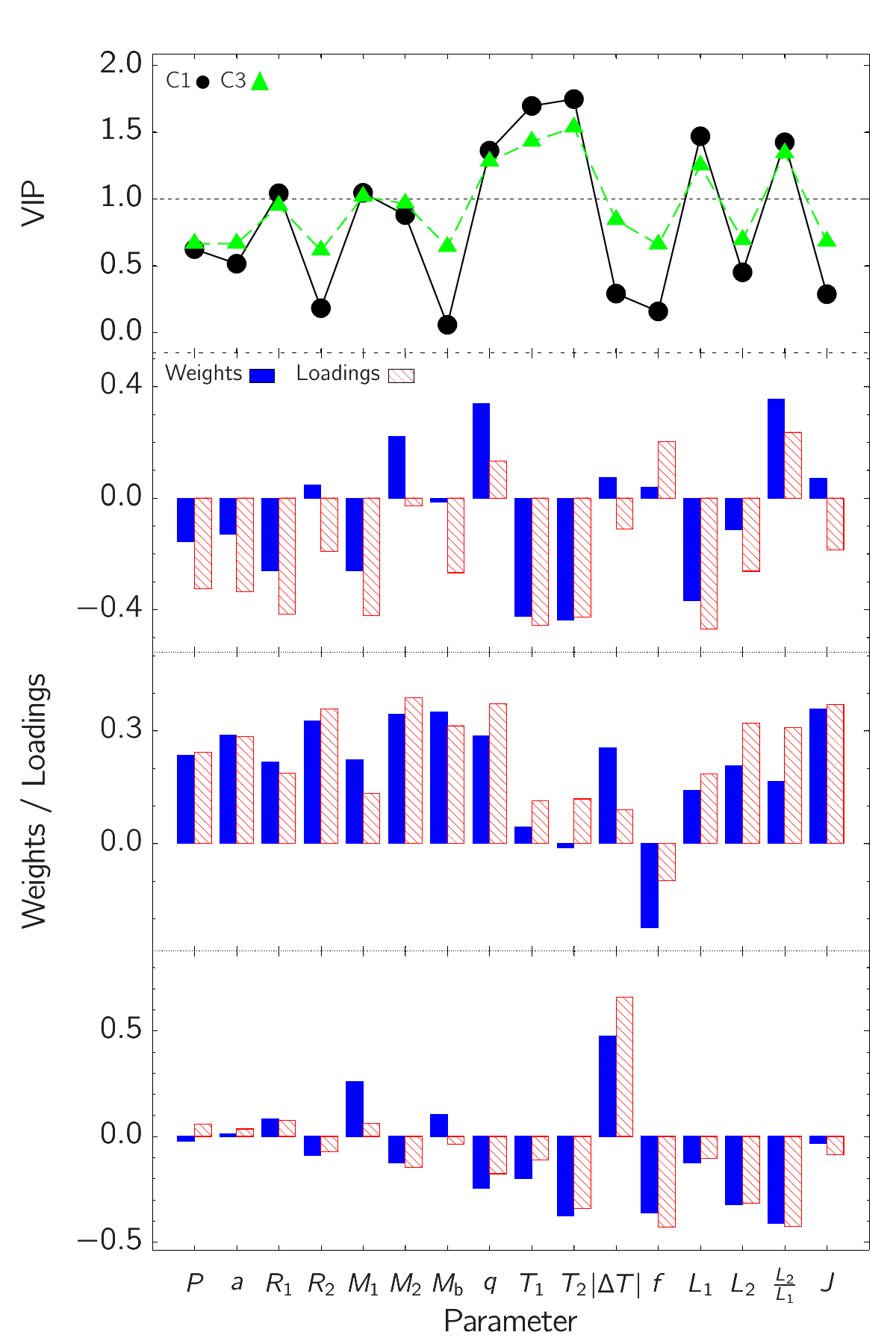}
	\centerline{(c)}
\end{minipage}
\begin{minipage}[b]{0.4\textwidth}
	\includegraphics[width=1.0\textwidth]{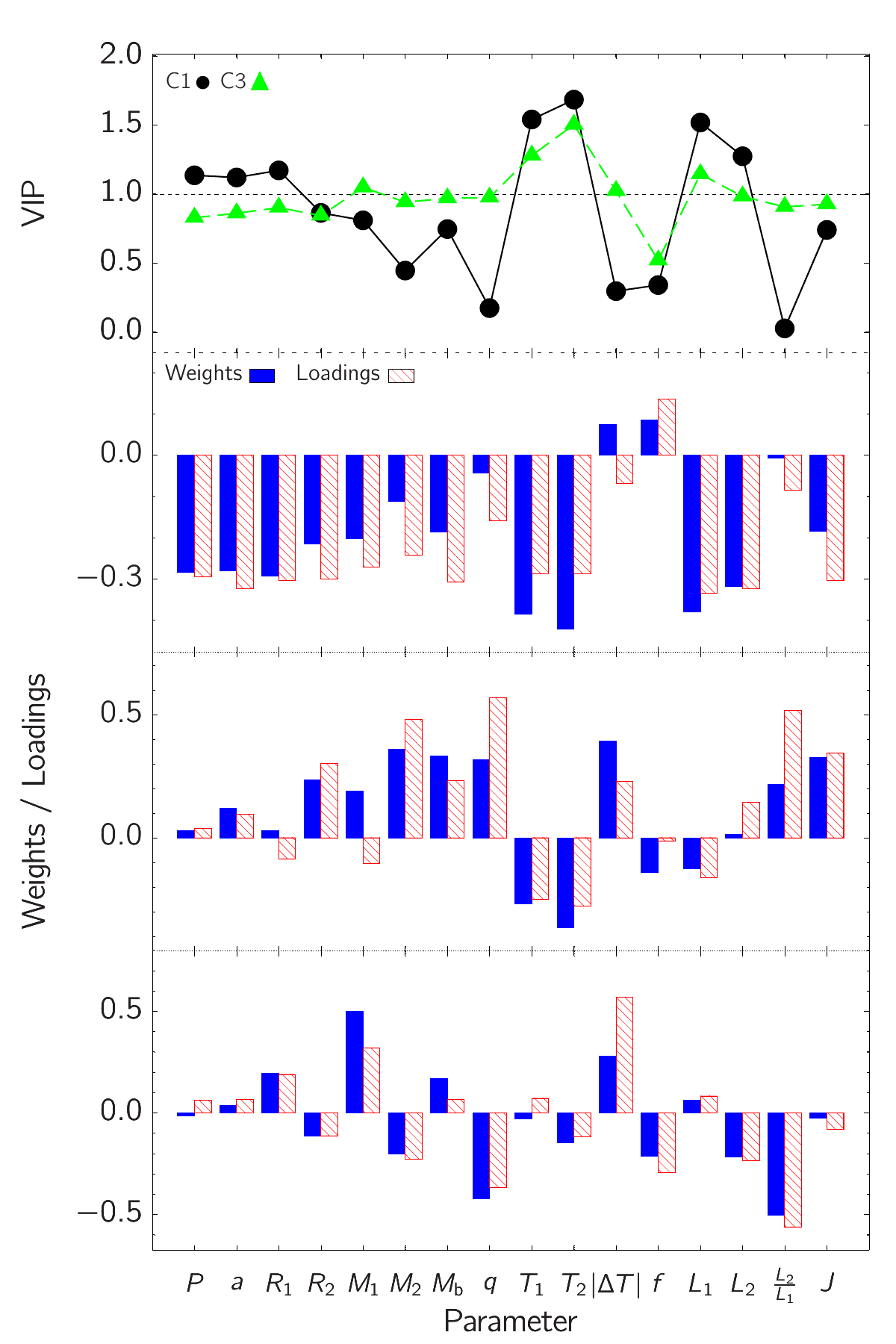}
	\centerline{(d)}
\end{minipage}
\caption{Plots of VIP scores (top), and histograms of weights (blue solid bars) and loadings (red shaded bars) for MTML (a), extra AML (b), MTLM (c), and ML (d). 
Black filled circles and green filled triangles represent the VIP scores for the first (C1) and the first to third (C3) components, respectively. 
The Nth histogram from the top displays the weights and loadings for the Nth component of the PLS model. 
\label{Fig_PWVIP}}
\end{figure*}
\section{Identifying key parameters}\label{Sec_Correlation}
\subsection{Method}\label{Sec_Correlation_Method}
\subsubsection{Overview}
This study examines the relations between the physical processes and 16 parameters: 
$P$, $a$, $R_1$, $R_2$, $M_1$, $M_2$, $\Mtot$, $q$, $T_1$, $T_2$, $\Delta T$, $f$, $L_1$, $L_2$, $L_2/L_1$, and $J$. 
In contact systems, several parameters are strongly correlated. 
In addition, the number of the parameters is similar to the sizes of the An and Ap samples. 
Accordingly, we first perform partial least-squares (PLS) analysis (see Section \ref{Sec_PLSR}) with the 16 parameters. 
The PLS method was adopted because it is particularly effective in managing multicollinearity. 
It is also effective in extracting the dominant components that explain the observed variations in the dependent variables. 
The number of latent variables that appropriately predict dependent variables is determined with cross-validation (Fig. \ref{Fig_RMSEP}) and the percentages of the variances explained by latent variables (Table \ref{Tab_EV}). 
On the basis of their variable importance in projection (VIP), weights, and loadings of the selected latent variables (Fig. \ref{Fig_PWVIP}), we select parameters importantly contributing to the latent variables. 

The selected parameters with a PLS analysis would be correlated with any one of the rates of MT, ML, and AML. 
However, strong correlations between parameters in contact systems can lead to detecting a spurious correlation between a binary parameter and one of the physical processes. 
To avoid spurious correlations, we further select parameters that have the closest correlation with any one of the physical processes by using partial regression plots (see Section \ref{Sec_PRP}).  
Finally, we obtain power laws for the closest correlations with ordinary least-squares (OLS) regressions. 
Note that the logarithms of the parameter values were used in these analyses.

\subsubsection{Partial least-squares regression}\label{Sec_PLSR}
PLS regression is a multivariate analysis that models dependent (response) variables Y by means of a set of independent (predictor) variables X. 
In PLS, dependent variables are predicted from latent variables, a set of orthogonal factors derived from independent variables. 
Latent variables are determined in a manner such that they have maximal covariance with the dependent variables. 
PLS is also a dimension reduction method, as well as principal component regression. 
This method can analyze data with strongly correlated, noisy, and numerous predictor variables \citep{Wold2001-CILS}. 

PLS regressions were performed with the NIPALS algorithm \citep{Wold1975-QS}. 
All X-variables (i.e., the 16 parameters) were centered and scaled with the standard deviations, whereas the Y-variables ($|\dot{M}_{12}|$, $|K|$, or $|\dot{\Mtot}|$) were only centered. 
To select important parameters that have close associations with any one of the physical processes, this work uses weights and loadings. 
The weights are used to calculate scores, and essential for the understanding of which X-variables are important \citep{Wold2004-UU}. 
The loadings indicate the strength of the correlation between parameters and scores. 

Cross-validation is used for selecting a predictive model. 
In general, a PLS model with too many latent variables (components) tends to be overfitting and thus be a poor predictive one. 
Accordingly, we select the number of components on the basis of cross-validation and obtain an appropriate model.  
This work applied leave-one-out method, and computed root mean square error of prediction (RMSEP) for the models with 1--10 components. 
The RMSEP estimates the predictive ability of a model. 

VIP is useful for selecting important variables. 
A VIP score gives a measure of the importance of an X-variable for both Y and X \citep{Wold2001-CILS}. 
This work uses VIP scores to select important binary parameters, together with weights and loadings. 
Note that a VIP score greater than one is often used as a criterion for variable selection \citep{Chong2005-CILS, Anderson2010-JC}.

\subsubsection{Partial regression plot}\label{Sec_PRP}
A partial regression plot is a plot of the residuals of the regression of $Y$ on $Z$ versus the residuals of the regression of $X$ on $Z$. 
This plot partials out the effect of $Z$ in the $X$--$Y$ relationship. 
Its partial correlation coefficient, which is calculated with the two variables used in the partial regression plot, measures the correct degree of the correlation between $X$ and $Y$. 

In this analysis, linear regression models are based on the OLS($Y|X$) and OLS bisector introduced by \citet{Isobe1990-ApJ}. 
The OLS bisector treats the variables symmetrically; it is suitable for deriving a regression line from bivariate data and for estimating the underlying functional relation between the variables. 
By comparing the results from both regressions for each individual plot, the most appropriate model was selected. 
Note that the author visually inspected the relevant plots in this work.

\subsection{Negative period variation}
\subsubsection{MTML}\label{Sec_Cor_MEML}
Figure \ref{Fig_RMSEP}a shows the result of cross-validation for a PLS analysis between the 16 parameters and MTML rate. 
The RMSEP values of the models with $N=3$--$5$ are relatively small, and the minimum is found at $N=3$. 
The models with $N\geq 3$ explain over 90 percent of the variance for X-variables (Table \ref{Tab_EV}). 
Moreover, in the models with $N\geq 4$, the explained variances for Y-variables increase by only a few percent. 
Therefore, the model with $N=3$ is optimal. 

Figure \ref{Fig_PWVIP}a shows the VIP scores, weights, and loadings of the 16 parameters for the model with $N=3$. 
For either of the first or the first to third components, nine have VIP scores greater than one: $P$, $a$, $R_1$, $M_1$, $q$, $|\Delta T|$, $f$, $L_1$, and $L_2/L_1$. 
In the histograms of weights and loadings, primary's parameters ($R_1$, $M_1$, and $L_1$) largely contribute to the first component. 
In addition, only $M_2$, $q$, and $L_2/L_1$ have weights and loadings of which signs opposite to those of the others. 
Therefore, it is deduced that the primary's parameters, particularly relative to the secondary's ones, affect the MTML. 
In the second component, $|\Delta T|$ largely contributes, and it may also affect the MTML.  
Note, however, that the $|\Delta T|$--$|\dot{M}_{12}|$ relation shows no strong correlation (see Table \ref{Tab_Pcor_MEML}). 
The third component is largely contributed by $R_2$, $M_2$, and $f$, which are the same as the parameters contributing to the first component of the PLS analysis for $K$ (Section \ref{Sec_Cor_aAML}). 
Accordingly, the third may be related to the thickness between the inner Roche lobe and stellar surface (see Section \ref{Sec_Disc_Processes}). 
However, the variance of Y-variables explained by the third is only $\sim 10$ percent. 

With the nine parameters, we next examine which parameter has the closest correlation with the MTML rate. 
Table \ref{Tab_Pcor_MEML} summarizes the partial correlation coefficients between the nine and MTML rate, controlling for any one of the nine. 
When a parameter is closely correlated with the MTML rate, its partial correlation coefficients should not dramatically differ from its Pearson correlation coefficient ($r_\mathrm{p}$). 
Accordingly, we rule out a parameter of which the partial correlation coefficient has the sign opposite to that of its $r_\mathrm{p}$. 
In addition, because $|\Delta T|$ and $f$ have extremely weak correlations with $|\dot{M}_{12}|$, both are ruled out. 
These criteria select four: $P$, $R_1$, $q$, and $L_2/L_1$. 
Of these four, $R_1$ is less affected by the other three while only $R_1$ affects the others. 
The parameters $q$ and $L_2/L_1$ affect each other. 
Thus, $R_1$ is the most plausible. 

\begin{figure*}
\centering
\begin{minipage}[b]{0.31\textwidth}
	\includegraphics[width=1.0\textwidth]{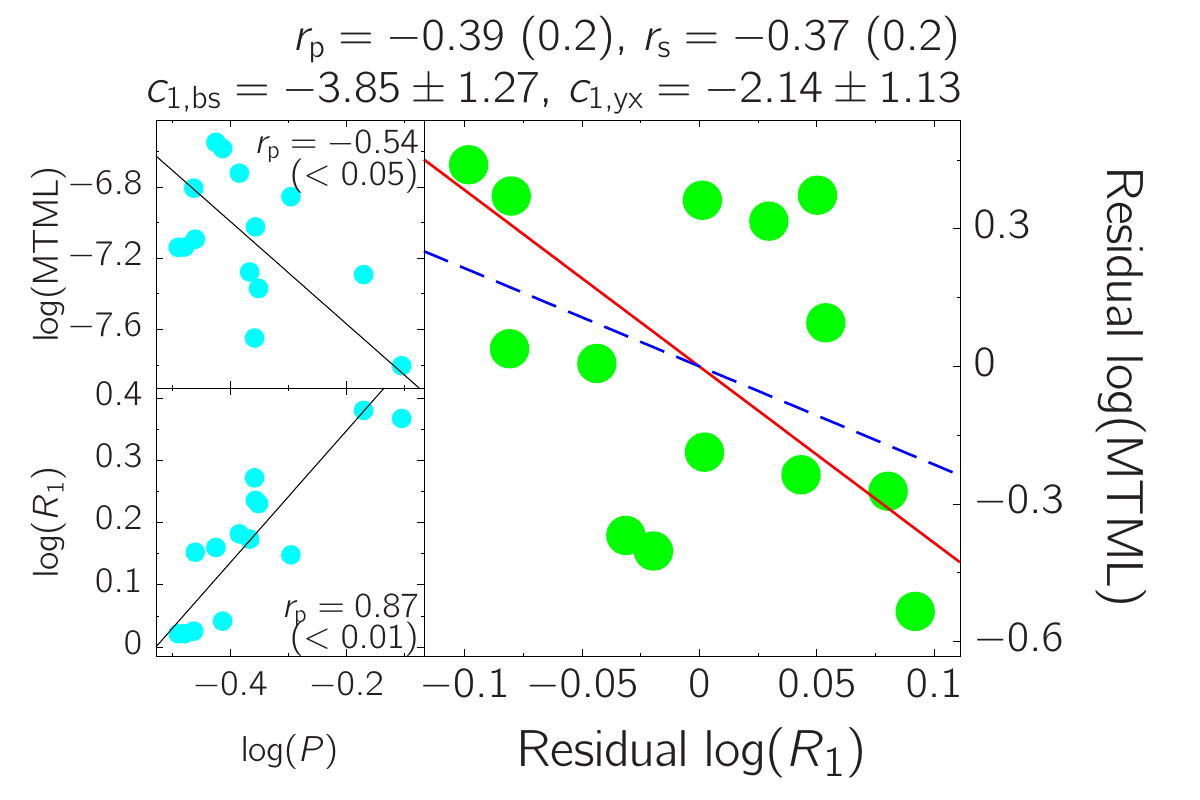}
	\centerline{(a)}
\end{minipage}
\begin{minipage}[b]{0.31\textwidth}
	\includegraphics[width=1.0\textwidth]{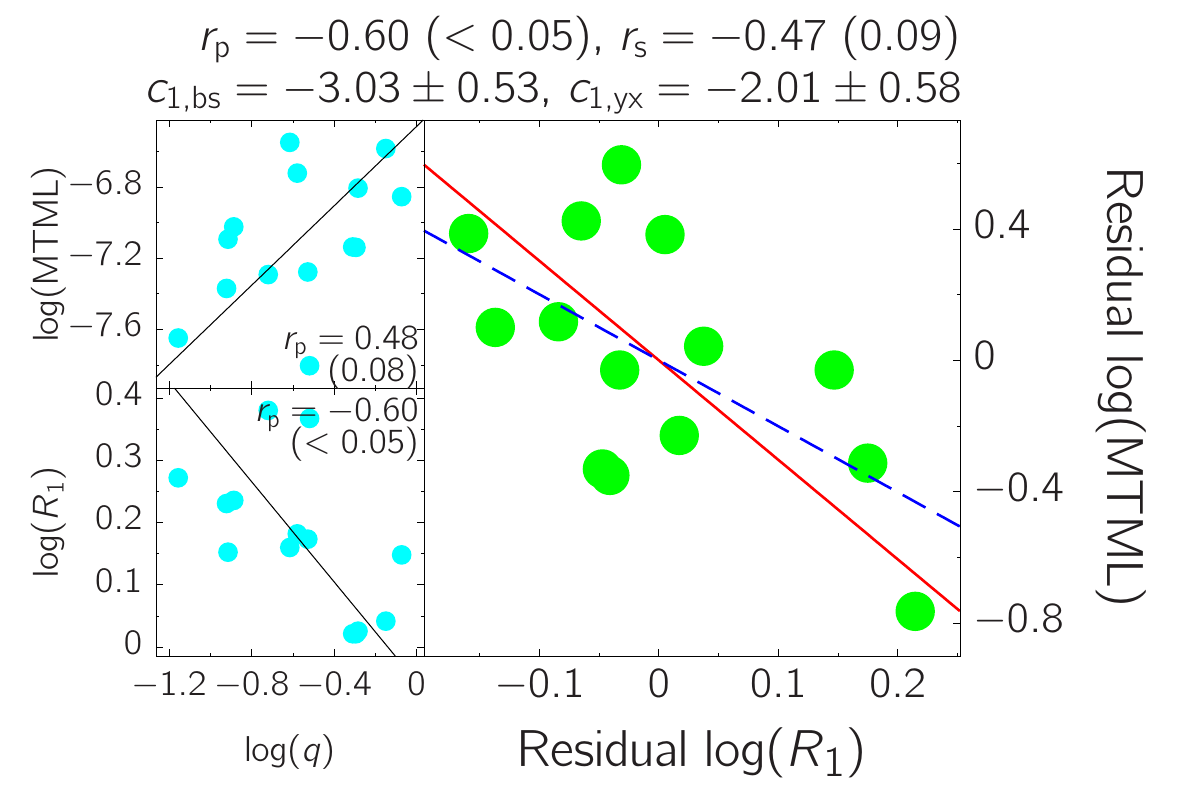}
	\centerline{(b)}
\end{minipage}
\begin{minipage}[b]{0.31\textwidth}
	\includegraphics[width=1.0\textwidth]{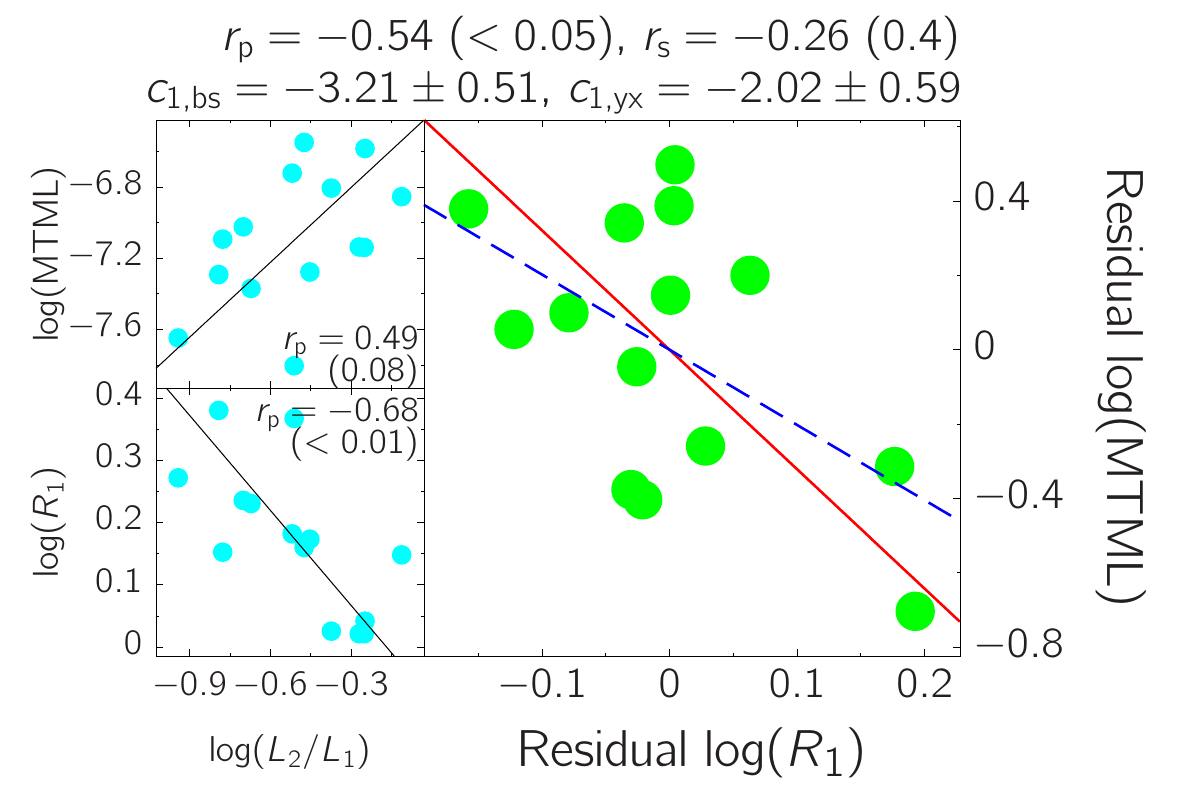}
	\centerline{(c)}
\end{minipage}
\caption{Partial regression plots (the right panel of each figure) of the primary radius vs. MTML rate, controlling for the orbital period (a), mass ratio (b), and luminosity ratio (c).
In each figure, the scatter plots of control variable vs. primary radius (bottom left) and vs. MTML rate (top left) are also shown. 
All residuals in the partial regression plots are based on the OLS bisector. 
The solid and dashed lines in the partial regression plots are the OLS bisector and OLS($Y|X$) lines of which slopes are $c_\mathrm{1, bs}$ and $c_\mathrm{1, yx}$, respectively. 
The values of $r_\mathrm{s}$ represent Spearman's rank correlation coefficients. 
Their corresponding $p$-values are shown in parentheses, which are computed with the null hypothesis that there is no linear (for $r_\mathrm{p}$) or monotonic (for $r_\mathrm{s}$) relationships between two variables.  
\label{Fig_PRP_An_ME}}
\end{figure*}

Figure \ref{Fig_PRP_An_ME} shows the partial regression plots of $R_1$ and $\dot{M}_{12}$, controlling for $P$, $q$, and $L_2/L_1$, together with the regression slopes. 
The parameter $R_1$ is hardly affected by the other three, and all three plots display negative correlations. 
Moreover, their regression slopes are similar. 
As a result, it is deduced that $R_1$ is genuinely correlated with the MTML rate. 
Power-law relations estimated from the slopes are 
\begin{equation}\label{Eq_MEML_depend}
	|\dot{M}_{12}| \propto R_1^{-3.2 \pm 0.4}  \hspace{3mm} \mathrm{and} \hspace{3mm} |\dot{M}_{12}| \propto R_1^{-2.0 \pm 0.4} \\
\end{equation}
with the OLS bisector and OLS($Y|X$), respectively. 

\begin{table*}
\centering
\caption[]{Partial correlation coefficients between the nine binary parameters ($X$) and MTML rate ($Y$) 
\label{Tab_Pcor_MEML}}
\begin{tabular}{crrrrrrrrrrr}
\hline \hline
\multicolumn{1}{c}{$X$ $\backslash$ $Z$} & \multicolumn{1}{c}{$r_\mathrm{p}$} & \multicolumn{1}{c}{$p$-value} & \multicolumn{1}{c}{$P$} &  \multicolumn{1}{c}{$a$} & \multicolumn{1}{c}{$R_1$} & \multicolumn{1}{c}{$M_1$}  & \multicolumn{1}{c}{$q$} & \multicolumn{1}{c}{$|\Delta T|$} & \multicolumn{1}{c}{$f$} & \multicolumn{1}{c}{$L_1$} & \multicolumn{1}{c}{$L_2/L_1$} \\
\hline
$P$           & $-$0.54 & $<0.05$ & ---   & $-$0.27 & $-$0.12  & $-$0.42  & $-$0.65  &  $-$0.67$^*$ & $-$0.54$^*$ & $-$0.39  & $-$0.62  \\ 
$a$           & $-$0.52 & $0.06$  &  0.17 & ---   & $-$0.04  & $-$0.36  & $-$0.51$^*$  & $-$0.63$^*$ & $-$0.52$^*$ & $-$0.33  & $-$0.60  \\ 
$R_1$         & $-$0.61 & $<0.05$ & $-$0.39 & $-$0.43 & ---    & $-$0.41  & $-$0.60  &  $-$0.74$^*$ & $-$0.70$^*$ & $-$0.48  & $-$0.54  \\ 
$M_1$         & $-$0.53 & $0.05$  & $-$0.39 & $-$0.37 &  0.01  & ---    & $-$0.42  &  $-$0.58$^*$ & $-$0.58$^*$ & $-$0.26  & $-$0.36  \\ 
$q$           &  0.48 & $0.08$	  &  0.54 &  0.52 &  0.28  &  0.24  & ---    & 0.49$^*$ &  0.74$^*$ &  0.45  &  0.09  \\ 
$|\Delta T|$   &  0.01 & $0.97$	  &  0.28 &  0.25 &  0.33  &  0.12  & 0.11$^*$  & ---   & 0.03$^*$ &  0.12  &  0.24$^*$  \\ 
$f$           & $-$0.03 & $0.91$  & $-$0.09 & $-$0.06 &  0.23  &  0.24  &  0.37  &  $-$0.04$^*$ & ---   & $-$0.05  &  0.32  \\ 
$L_1$         & $-$0.51 & $0.06$  & $-$0.24 & $-$0.24 &  0.21  & $-$0.13  & $-$0.46  & $-$0.64$^*$ & $-$0.57$^*$ & ---    & $-$0.38  \\ 
$L_2/L_1$     &  0.49 & 0.08   	  &  0.51 &  0.49 &  0.24  &  0.22  &  0.18  & 0.53$^*$ &  0.69$^*$ &  0.39  & ---    \\ 
\hline
\end{tabular}
  \parbox{\hsize}{\emph{Notes.} 
The parameter $Z$ is a control variable. 
The values $r_\mathrm{p}$ in column (2) are Pearson correlation coefficients between the nine parameters and MTML rate. 
Their $p$-values are shown in column (3). 
The other values are partial correlation coefficients. 
The coefficients with and without the superscript $^*$ are computed with the OLS($Y|X$) and OLS bisector, respectively. }
\end{table*}

\begin{table*}
\centering
\caption[]{Partial correlation coefficients between the nine binary parameters ($X$) and  extra AML rate ($Y$) 
\label{Tab_Pcor_K}}
\begin{tabular}{crrrrrrrrrrr}
\hline \hline
\multicolumn{1}{c}{$X$ $\backslash$ $Z$}  & \multicolumn{1}{c}{$r_\mathrm{p}$} & \multicolumn{1}{c}{$p$-value} & \multicolumn{1}{c}{$P$} & \multicolumn{1}{c}{$a$} & \multicolumn{1}{c}{$R_2$} & \multicolumn{1}{c}{$M_2$} & \multicolumn{1}{c}{$q$} & \multicolumn{1}{c}{$|\Delta T|$} &  \multicolumn{1}{c}{$f$} & \multicolumn{1}{c}{$L_2/L_1$} & \multicolumn{1}{c}{$J$}  \\
\hline
$P$            & $-$0.47 & 0.09 & ---   & $-$0.39  &  0.03 & $-$0.53$^*$ &  $-$0.65$^*$ &  $-$0.63$^*$ &  $-$0.70$^*$ & $-$0.71$^*$ & $-$0.23  \\ 
$a$            & $-$0.43 & 0.13 &  0.30 & ---    &  0.07 & $-$0.49$^*$ &  $-$0.62$^*$ &  $-$0.57$^*$ &  $-$0.67$^*$ & $-$0.68$^*$ & $-$0.17  \\ 
$R_2$          & $-$0.68 & $<0.05$ & $-$0.63 & $-$0.66  & ---   & $-$0.56 & $-$0.69 &  $-$0.79$^*$ & $-$0.67$^*$ & $-$0.70 & $-$0.28  \\ 
$M_2$          & $-$0.62 & $<0.05$ & $-$0.66$^*$ & $-$0.66$^*$  & $-$0.40 & ---   & $-$0.52 & $-$0.62$^*$ & $-$0.41 & $-$0.54 & $-$0.28  \\ 
$q$            & $-$0.53 & 0.05 & $-$0.68$^*$ & $-$0.67$^*$  & $-$0.40 &  0.28 & ---   & $-$0.52$^*$ & $-$0.20 & $-$0.30 & $-$0.30  \\ 
$|\Delta T|$    &  0.08 & 0.77 &  0.30 &  0.26  &  0.37 &  0.03$^*$ &  $-$0.01$^*$ & ---   &  0.01 &  $-$0.12$^*$ &  0.31$^*$  \\ 
$f$            &  0.61 & $<0.05$ &  0.77$^*$ &  0.76$^*$  &  0.65 &  0.36 &  0.47 &  0.64$^*$ & ---   &  0.52 &  0.53  \\ 
$L_2/L_1$      & $-$0.48 & 0.08 & $-$0.47 & $-$0.49  & $-$0.43 &  0.16 &  0.03 & $-$0.49$^*$ & $-$0.19 & ---   & $-$0.34  \\ 
$J$            & $-$0.67 & $<0.05$ & $-$0.61 & $-$0.62  & $-$0.18 & $-$0.44 & $-$0.62 &  $-$0.71$^*$ & $-$0.64 & $-$0.66 & ---    \\ 
\hline
\end{tabular}
\parbox{\hsize}{\emph{Notes.}
Correlation coefficients are summarized as in Table \ref{Tab_Pcor_MEML}. }
\end{table*}

\begin{figure*}
\centering
\begin{minipage}[b]{0.31\textwidth}
	\includegraphics[width=\textwidth]{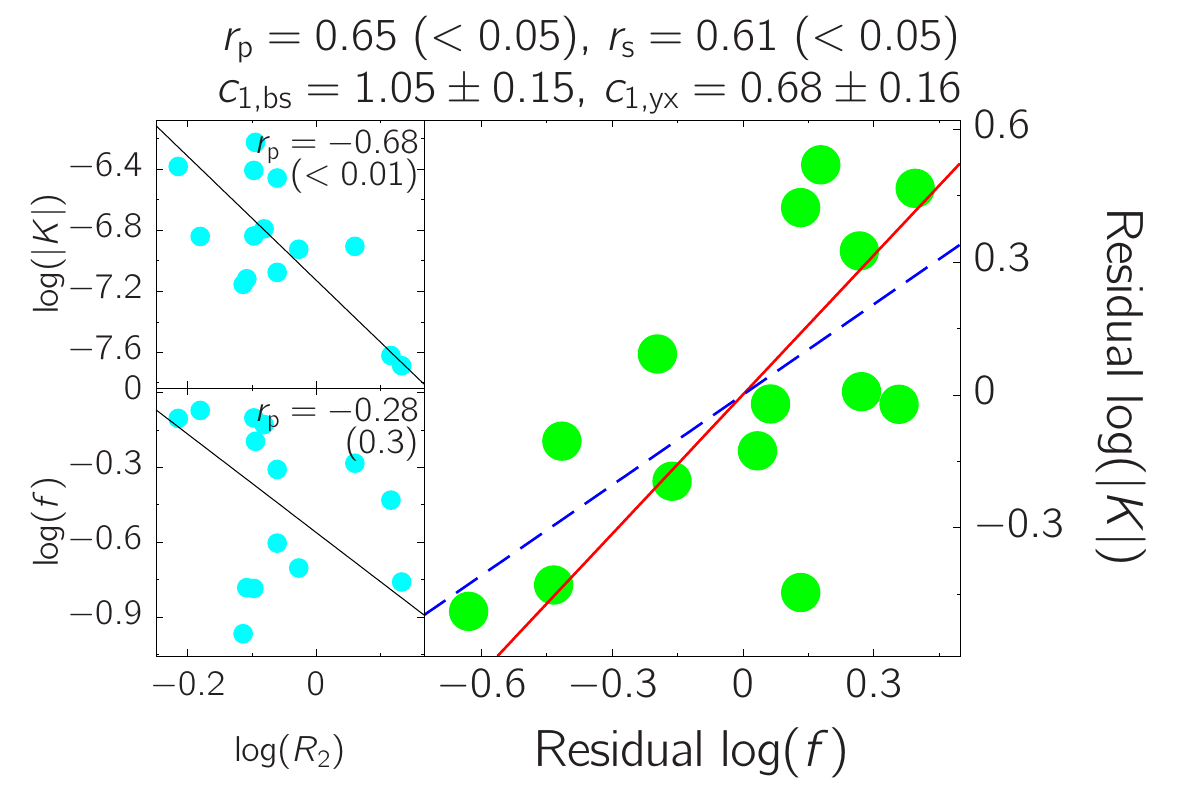}
	\centerline{(a)}
\end{minipage}
\begin{minipage}[b]{0.31\textwidth}
	\includegraphics[width=\textwidth]{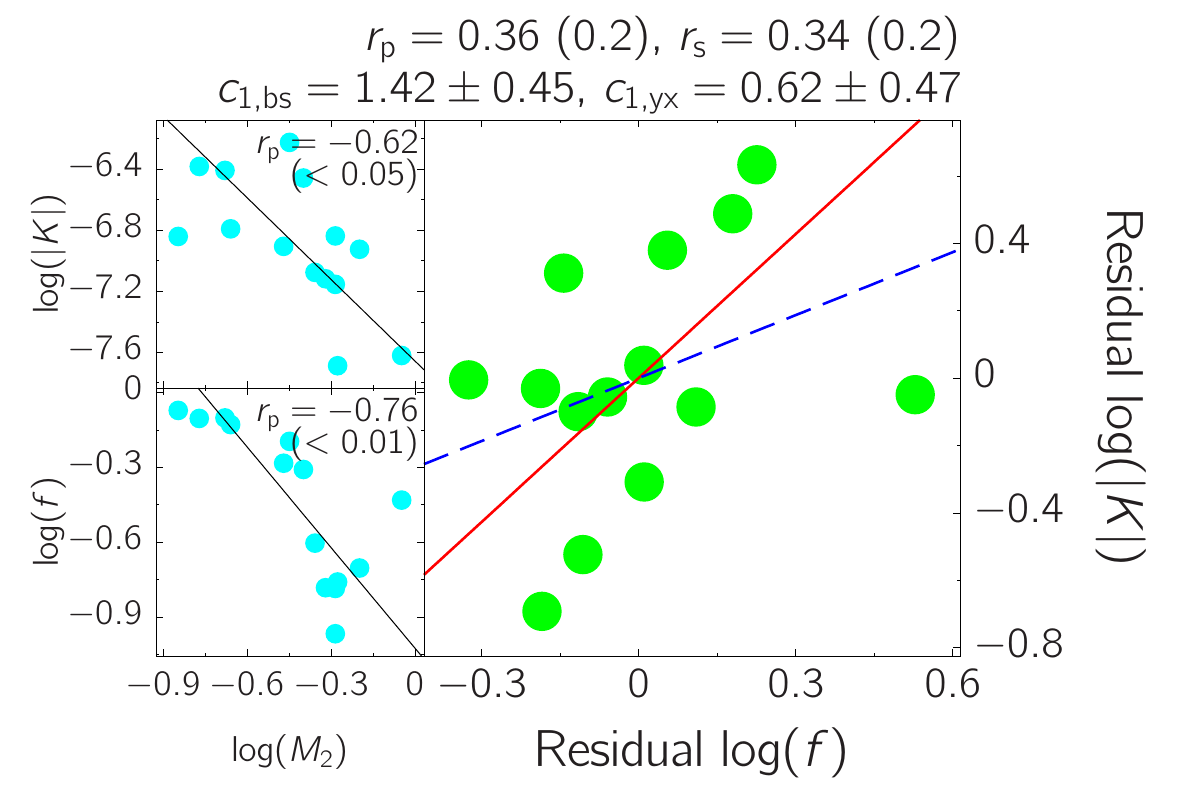}
	\centerline{(b)}
\end{minipage}
\begin{minipage}[b]{0.31\textwidth}
	\includegraphics[width=\textwidth]{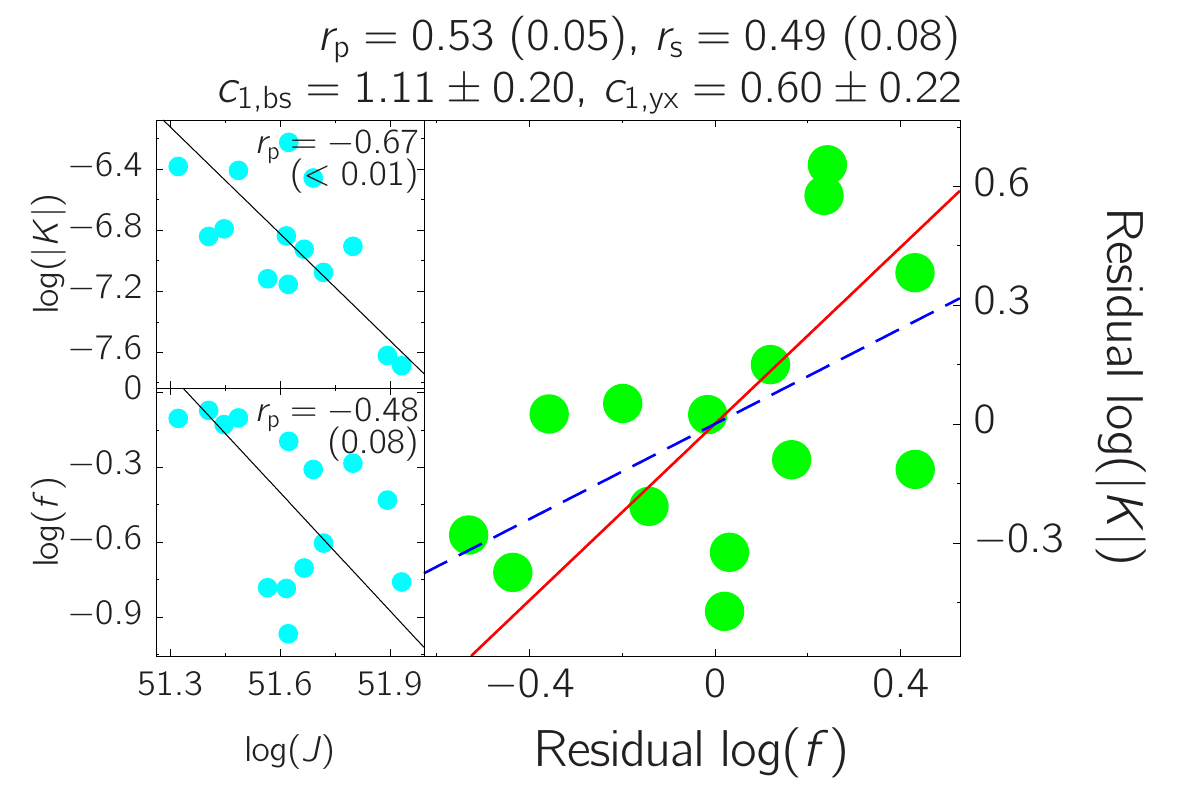}
	\centerline{(c)}
\end{minipage}
\caption{Partial regression plots of the fill-out factor vs. extra AML, controlling for the secondary radius (a), secondary mass (b), and angular momentum (c), together with scatter plots, as in Figure \ref{Fig_PRP_An_ME}. 
The residuals are based on the OLS bisector. 
\label{Fig_PRP_K_f}}
\end{figure*}
\subsubsection{Extra AML}\label{Sec_Cor_aAML}
In Fig. \ref{Fig_RMSEP}b, the RMSEP values at $N=1$--$3$ are approximately the same. 
The model with $N=3$ explains $\sim 90$ percent of the variance for X-variables, although the one with $N=2$ explains only $\sim 40$ percent (Table \ref{Tab_EV}). 
The explained variance for Y-variables increases by only a few percent when $N\geq 3$. 
Therefore, the model with $N=3$ is optimal. 

The VIP scores in Fig. \ref{Fig_PWVIP}b show that nine parameters have scores greater than one: 
$P$, $a$, $R_2$, $M_2$, $q$, $|\Delta T|$, $f$, $L_2/L_1$, and $J$. 
The first component is mainly contributed by the secondary's parameters ($R_2$ and $M_2$) and $J$. 
Therefore, the size of secondaries is likely to affect $K$. 
Furthermore, $f$ also has a large contribution, and its weights and loadings have the sign opposite to that for $R_2$ and $M_2$. 
As discussed in Section \ref{Sec_Disc_Processes}, the first component seems to be related to the thickness between the inner Roche lobe and stellar surface. 
In the second component, $|\Delta T|$ has the largest contribution, which can also be associated with $K$. 

Table \ref{Tab_Pcor_K} lists the partial correlation coefficients for the nine parameters. 
The sign of each coefficient for four ($R_2$, $M_2$, $f$, and $J$) is the same as that of the corresponding $r_\mathrm{p}$. 
Of these four, only $f$ is less affected by the other three. 
Though the coefficient between $M_2$ and $K$ controlling for $f$ is $-0.41$, its partial regression plot showed that one data point strongly affected the strength of the correlation. 
Furthermore, $M_2$ is slightly affected by $J$. 
For these reasons, $M_2$ is ruled out. 
Although $R_2$ and $J$ affect each other and their partial correlation coefficients are fairly small, their partial regression plots seem to display correlations. 
Their small coefficients should be due to dispersed distributions. 
Thus, these two cannot be completely ruled out. 

Figure \ref{Fig_PRP_K_f} shows the partial regression plots of $f$ and $K$, controlling for $R_2$, $M_2$, and $J$. 
Similar positive correlations are found, and thus $f$ is the most plausible. 
Power-law relations from the regression lines are 
\begin{equation}\label{Eq_K_depend}
	|K| \propto f^{1.1 \pm 0.1} \hspace{3mm} \mathrm{and} \hspace{3mm} |K| \propto f^{0.7 \pm 0.1}
\end{equation}
with the OLS bisector and OLS($Y|X$), respectively.

\begin{table*}
\centering
\caption{Partial correlation coefficients between the seven binary parameters ($X$) and  MTLM rate ($Y$) 
\label{Tab_Pcor_MELM}}
\begin{tabular}{crrrrrrrrr}
\hline \hline
\multicolumn{1}{c}{$X$ $\backslash$ $Z$} & \multicolumn{1}{c}{$r_\mathrm{p}$} & \multicolumn{1}{c}{$p$-value} & \multicolumn{1}{c}{$R_1$} & \multicolumn{1}{c}{$M_1$} & \multicolumn{1}{c}{$q$} & \multicolumn{1}{c}{$T_1$}  & \multicolumn{1}{c}{$T_2$}  & \multicolumn{1}{c}{$L_1$} & \multicolumn{1}{c}{$L_2/L_1$} \\
\hline
$R_1$       & $-$0.36 & 0.12 & ---   & $-$0.26 &  $-$0.50$^*$ & $-$0.13 & $-$0.15 &  0.17 & $-$0.41$^*$  \\ 
$M_1$       & $-$0.37 & 0.12 & $-$0.27 & ---   &  $-$0.45$^*$ &  0.01 &  0.02 & $-$0.05 & $-$0.37$^*$  \\ 
$q$         &  0.48 & $<0.05$ &  0.58$^*$ &  0.54$^*$ & ---   &  0.59 &  0.64 &  0.70$^*$ &  0.15  \\ 
$T_1$       & $-$0.59 & $<0.05$ & $-$0.57 & $-$0.57 &  $-$0.77$^*$ & ---   & $-$0.10 & $-$0.42 & $-$0.63$^*$  \\ 
$T_2$       & $-$0.61 & $<0.05$ & $-$0.59 & $-$0.59 & $-$0.84$^*$ & $-$0.23 & ---   & $-$0.46 &  $-$0.74$^*$  \\ 
$L_1$       & $-$0.51 & $<0.05$ & $-$0.50 & $-$0.45 & $-$0.37 & $-$0.09 & $-$0.10 & ---   & $-$0.61$^*$  \\ 
$L_2/L_1$   &  0.50 & $<0.05$ &  0.53$^*$ &  0.50$^*$ &  0.25 &  0.55$^*$ &  0.68$^*$ & 0.60$^*$ & ---    \\ 
\hline
\end{tabular}
\parbox{\hsize}{\emph{Notes.}
Correlation coefficients are summarized as in Table \ref{Tab_Pcor_MEML}. 
}
\end{table*}

\begin{figure*}
\centering
\begin{minipage}[b]{0.24\textwidth}
	\includegraphics[width=\textwidth]{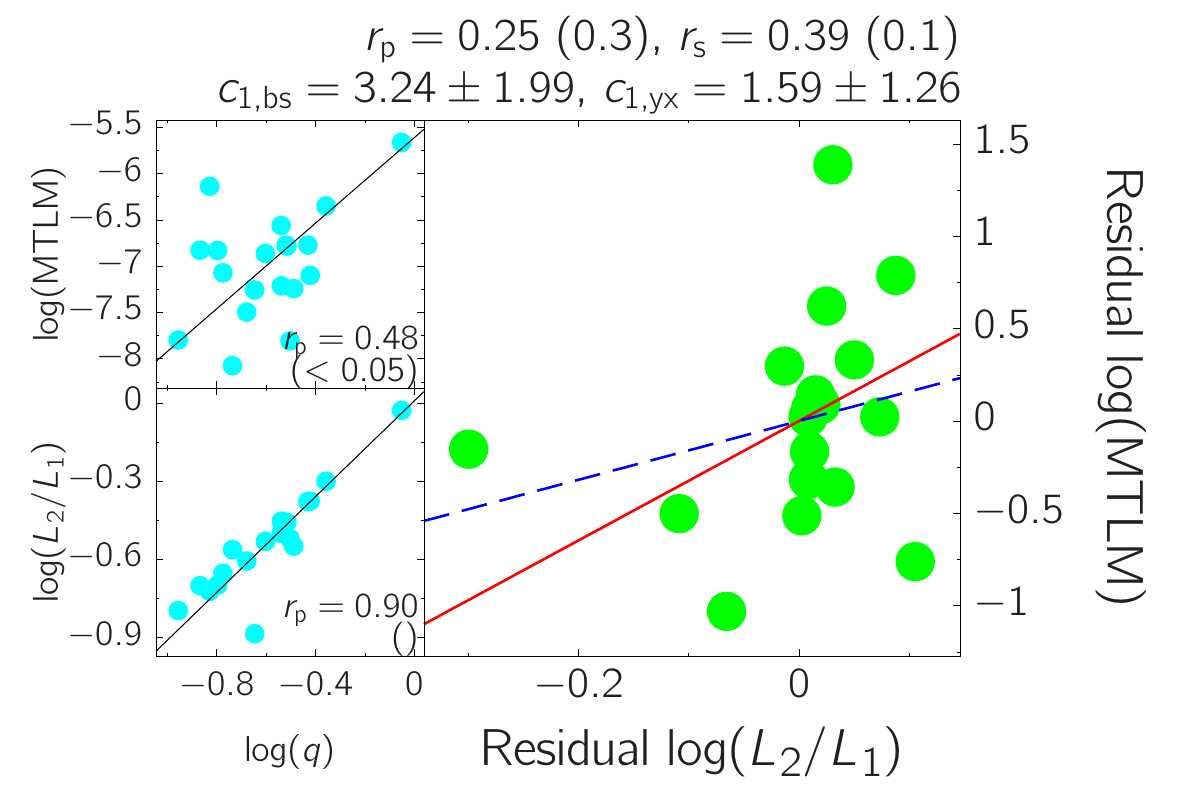}
	\centerline{(a)}
\end{minipage}
\begin{minipage}[b]{0.24\textwidth}
	\includegraphics[width=\textwidth]{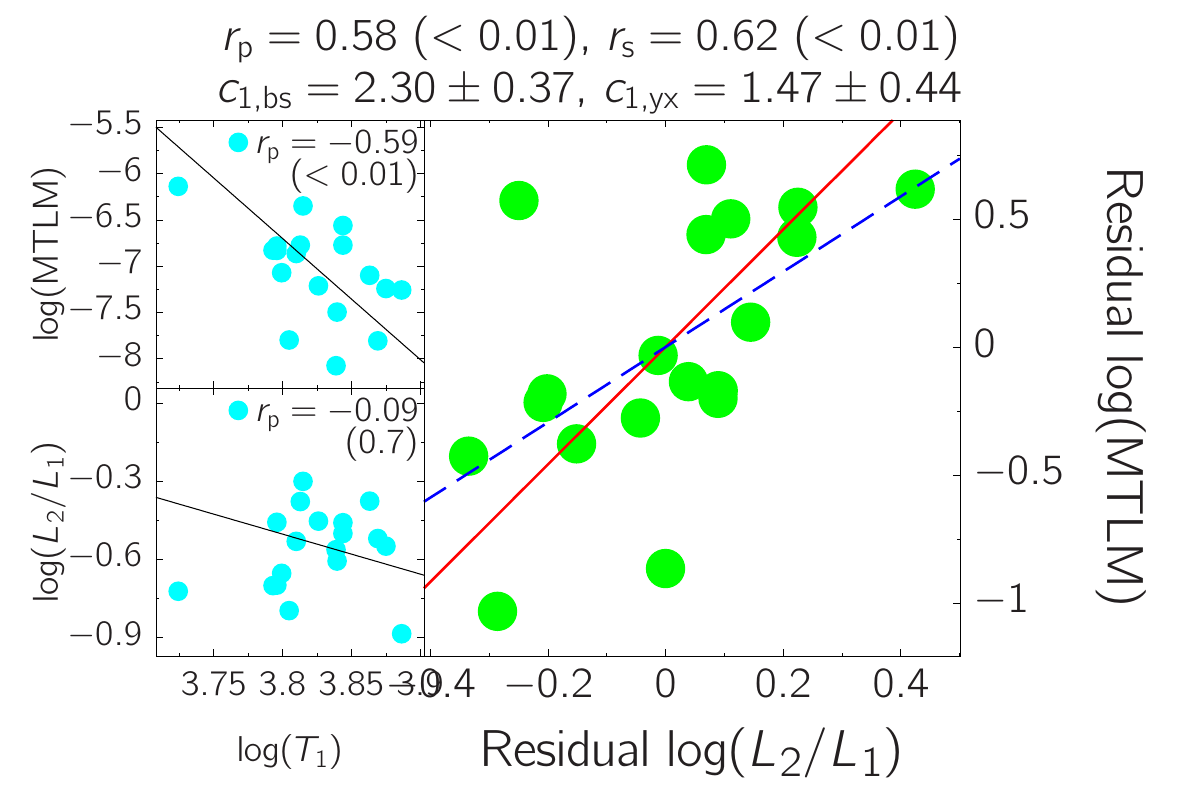}
	\centerline{(b)}
\end{minipage}
\begin{minipage}[b]{0.24\textwidth}
	\includegraphics[width=\textwidth]{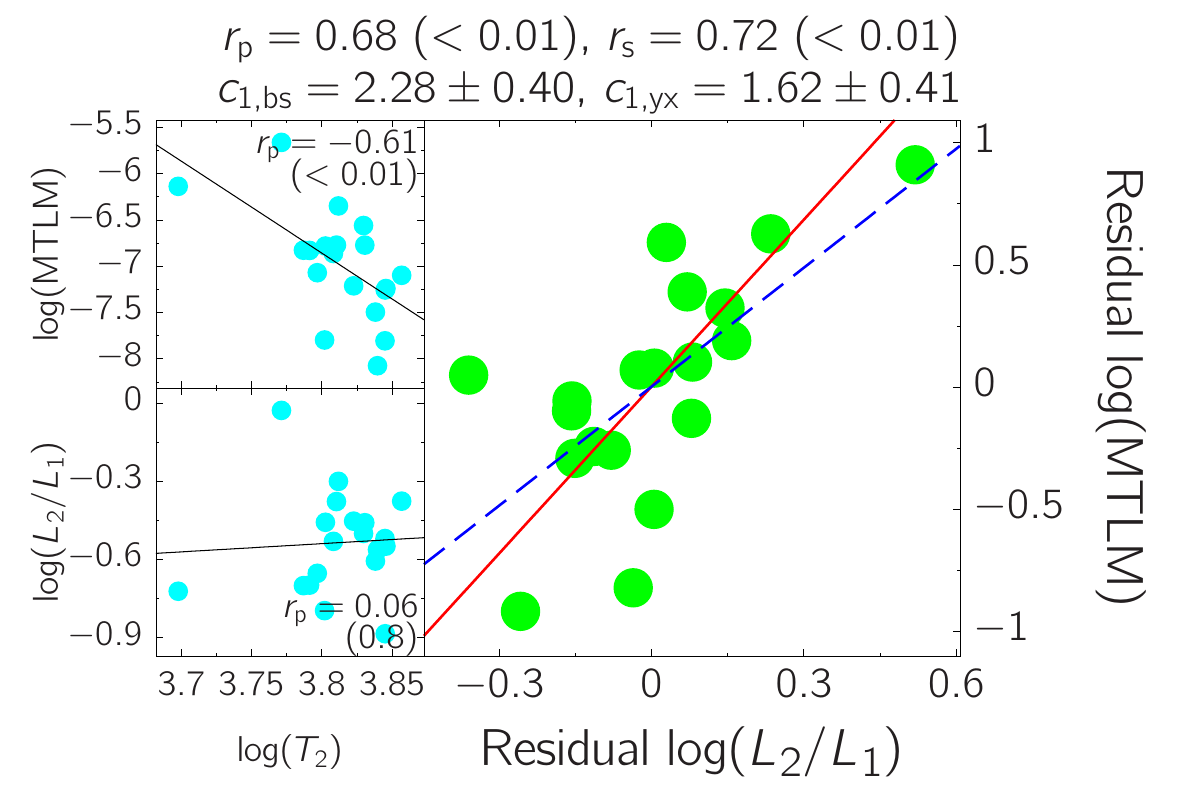}
	\centerline{(c)}
\end{minipage}
\begin{minipage}[b]{0.24\textwidth}
	\includegraphics[width=\textwidth]{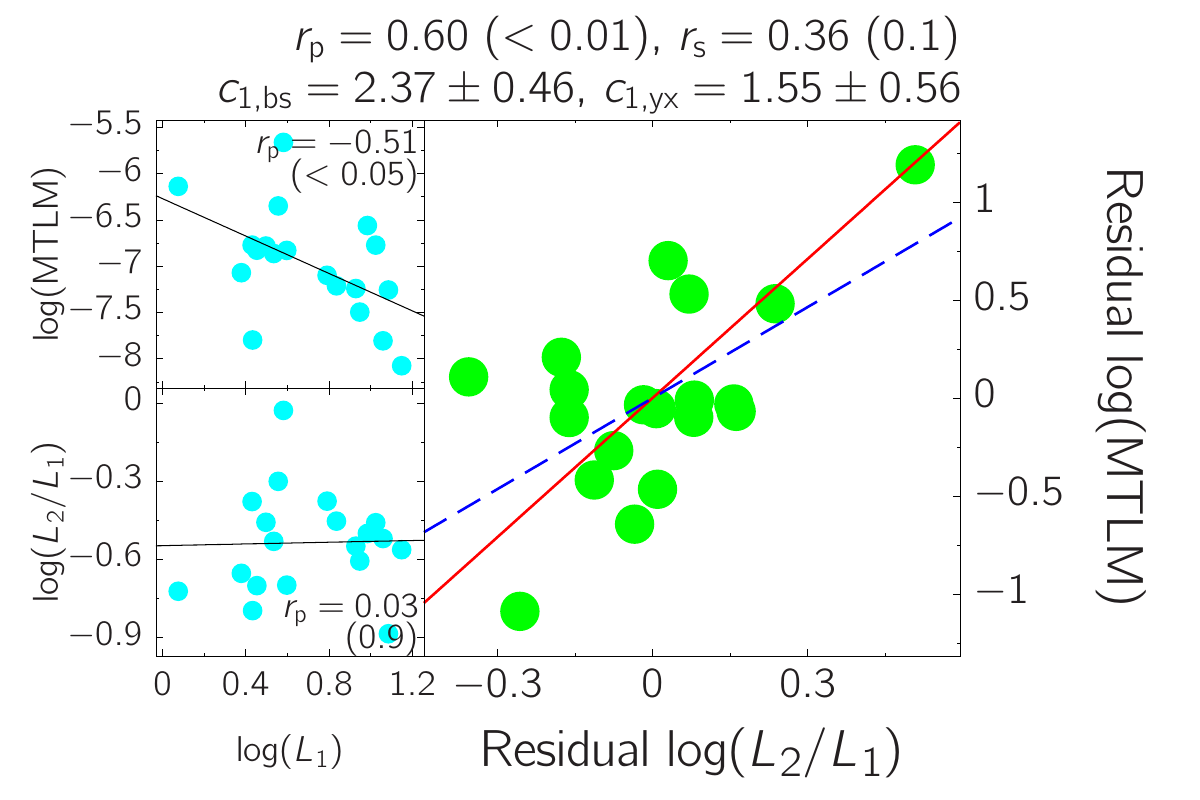}
	\centerline{(d)}
\end{minipage}
\begin{minipage}[b]{0.24\textwidth}
	\includegraphics[width=\textwidth]{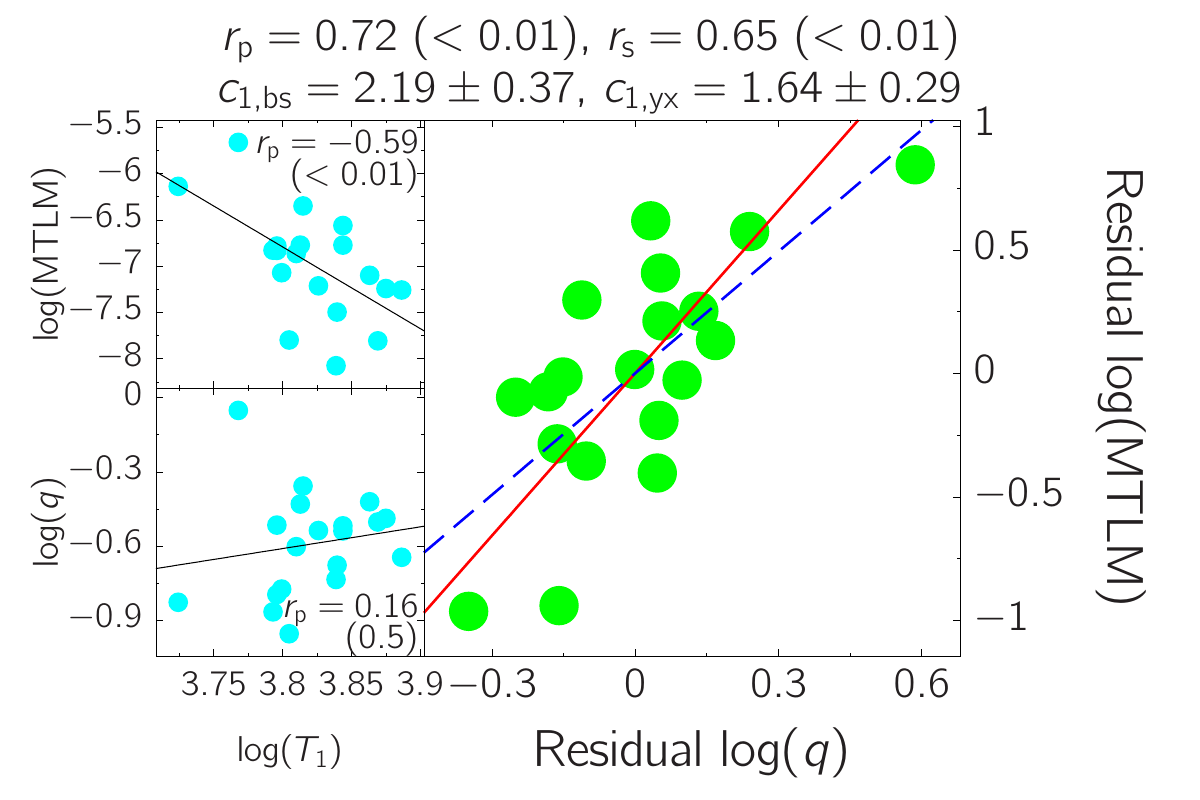}
	\centerline{(e)}
\end{minipage}
\begin{minipage}[b]{0.24\textwidth}
	\includegraphics[width=\textwidth]{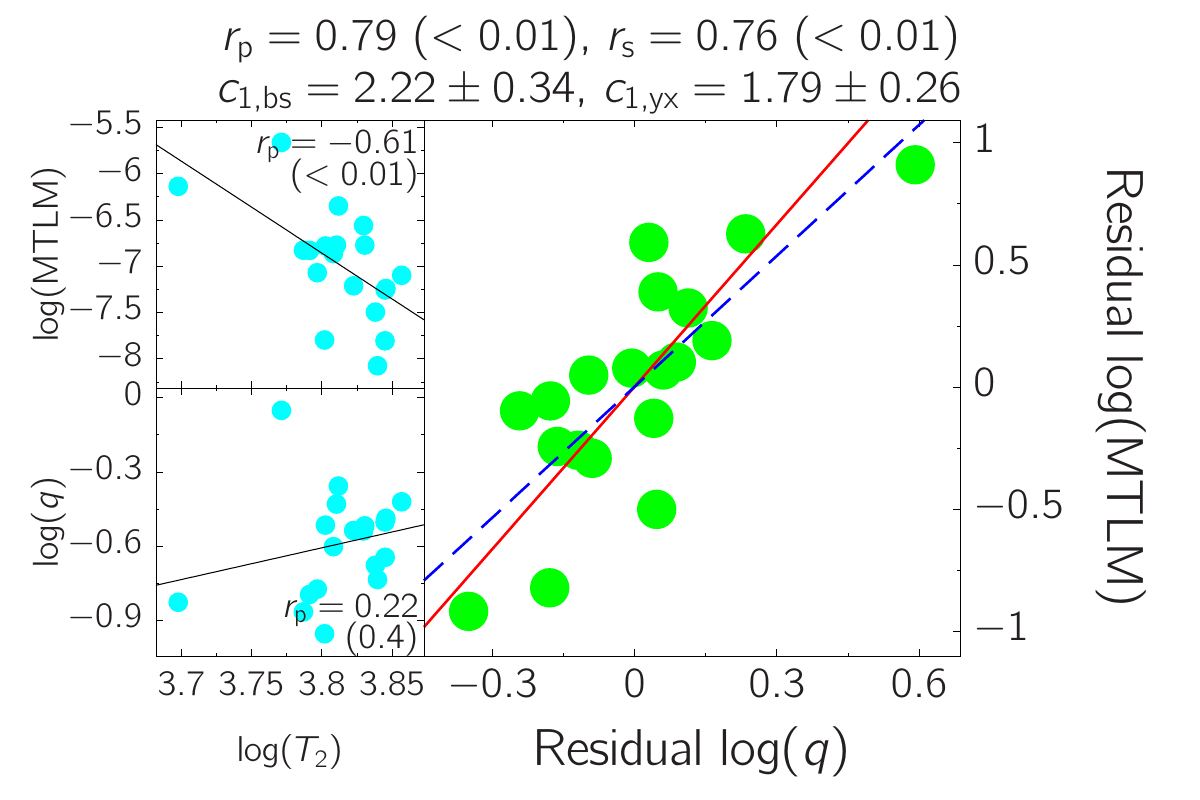}
	\centerline{(f)}
\end{minipage}
\begin{minipage}[b]{0.24\textwidth}
	\includegraphics[width=\textwidth]{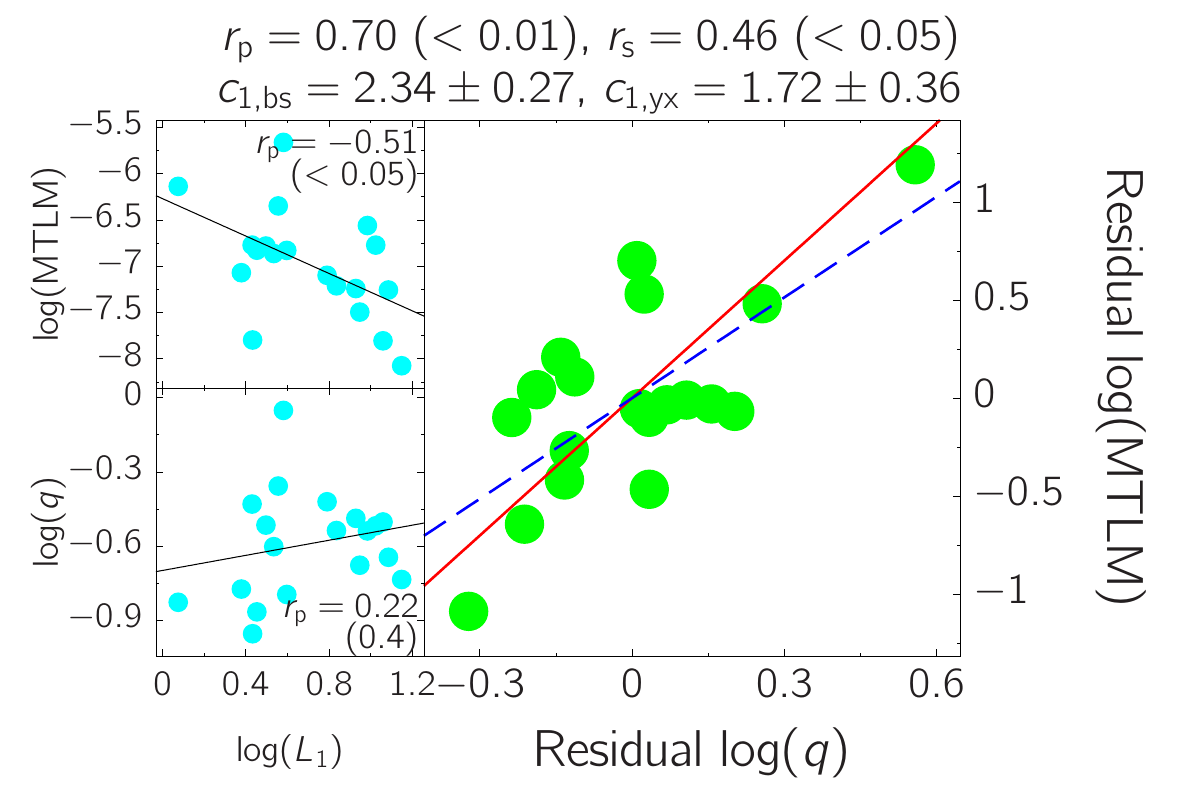}
	\centerline{(g)}
\end{minipage}
\begin{minipage}[b]{0.24\textwidth}
	\includegraphics[width=\textwidth]{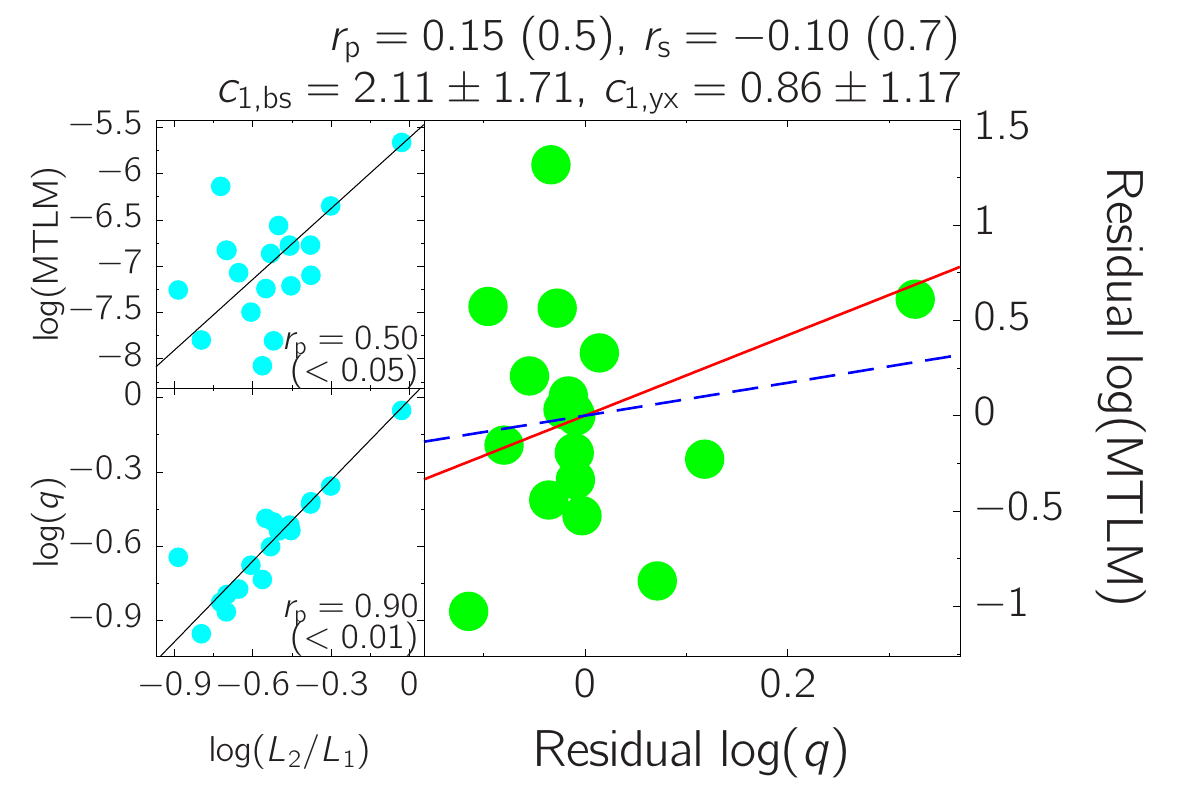}
	\centerline{(h)}
\end{minipage}
\caption{
{\it Top row}: Partial regression plots of the luminosity ratio vs. MTLM rate, controlling for the mass ratio (a), primary temperature (b), secondary temperature (c), and primary luminosity (d), as in Figure \ref{Fig_PRP_An_ME}. 
{\it Bottom row}: Partial regression plots of the mass ratio vs. MTLM rate, controlling for the primary temperature (e), secondary temperature (f), primary luminosity (g), and luminosity ratio (h). 
The residuals for panels (a,b,h) and (c-g) are based on the OLS bisector and OLS($Y|X$), respectively. 
\label{Fig_PRP_Ap_qLrat-ME}}
\end{figure*}
\subsection{Positive period variation}\label{Sec_Cor_PPC}
\subsubsection{MTLM}\label{Sec_Cor_MELM}
Figure \ref{Fig_RMSEP}c shows the result of cross-validation for a PLS analysis with the MTLM rate. 
The RMSEP values have a minimum at $N=2$, and those at $N=2$--$5$ are similar. 
The explained variance for Y-variables increases by only a few percent when $N\geq 4$ (Table \ref{Tab_EV}). 
Therefore, the model with $N=3$ is optimal. 

The VIP scores in Fig. \ref{Fig_PWVIP}c show that seven parameters have scores greater than one: $R_1$, $M_1$, $q$, $T_1$, $T_2$, $L_1$, and $L_2/L_1$. 
The first component is contributed by primary's parameters, in particular $T_1$ and $L_1$ have large contributions. 
On the other hand, though $L_2/L_1$ also contributes to the first component, its weights and loadings have the sign opposite to those of the primary's parameters. 
Accordingly, the first component should be related to the radiation energy of primaries relative to that of secondaries. 
In the second component, the secondary's ($R_2$ and $M_2$) and system's ($\Mtot$ and $J$) parameters have relatively large contributions. 
The third component is contributed by $|\Delta T|$ and $f$, and their weights and loadings indicate that the two contributions are opposite. 

Table \ref{Tab_Pcor_MELM} summarizes partial correlations coefficients for the seven parameters. 
The sign of each coefficient for five ($q$, $T_1$, $T_2$, $L_1$, and $L_2/L_1$) is the same as that of the corresponding $r_\mathrm{p}$. 
Of the five, $L_1$ is affected by $T_1$ and $T_2$, and $T_1$ is affected by $T_2$. 
As shown in the following section, $T_2$ is genuinely correlated with the ML rate. 
In the partial regression plot of $T_2$ and the MTLM rate controlling for the ML rate, it was confirmed that there was no negative correlation. 
Therefore, $L_1$, $T_1$, and $T_2$ are ruled out. 
The parameters $L_2/L_1$ and $q$ affect each other. 
Although the partial correlation coefficient of $L_2/L_1$ controlling for $q$ is slightly larger than that of $q$ controlling for $L_2/L_1$, the difference in the strength of the relationship is unclear in their partial regression plots. 
As a consequence, in this work, it is difficult to distinguish which is the most plausible. 
A possible reason is discussed in Section \ref{Sec_Disc_PPC}. 

Figure \ref{Fig_PRP_Ap_qLrat-ME} shows the partial regression plots for $L_2/L_1$ and $q$ with four control variables. 
From the regression lines in the plots, power-law relations for $L_2/L_1$ and $q$ are estimated as 
\begin{eqnarray}\label{Eq_MELM_depend}
	|\dot{M}_{12}| &\propto& \left(\frac{L_2}{L_1}\right)^{2.3 \pm 0.2} \hspace{3mm} \mathrm{and} \hspace{3mm} |\dot{M}_{12}| \propto \left(\frac{L_2}{L_1}\right)^{1.6 \pm 0.3},  \\
	|\dot{M}_{12}| &\propto& q^{2.3 \pm 0.2} \hspace{3mm} \mathrm{and} \hspace{3mm} |\dot{M}_{12}| \propto q^{1.7 \pm 0.2}
\end{eqnarray}
with the OLS bisector and OLS($Y|X$), respectively.

\begin{table*}
\centering
\caption{Partial correlation coefficients between the nine binary parameters ($X$) and  ML rate ($Y$) 
\label{Tab_Pcor_ML}}
\begin{tabular}{crrrrrrrrrrr}
\hline \hline
\multicolumn{1}{c}{$X$ $\backslash$ $Z$} & \multicolumn{1}{c}{$r_\mathrm{p}$} & \multicolumn{1}{c}{$p$-value} & \multicolumn{1}{c}{$P$} & \multicolumn{1}{c}{$a$} & \multicolumn{1}{c}{$R_1$} & \multicolumn{1}{c}{$M_1$} & \multicolumn{1}{c}{$T_1$} & \multicolumn{1}{c}{$T_2$} & \multicolumn{1}{c}{$|\Delta T|$} & \multicolumn{1}{c}{$L_1$} & \multicolumn{1}{c}{$L_2$} \\
\hline
$P$         & $-$0.51 & $<0.05$ & ---   & $-$0.14 & $-$0.09 & $-$0.49  & $-$0.33 & $-$0.34 &  $-$0.55$^*$ &  0.00 & $-$0.28  \\ 
$a$         & $-$0.51 & $<0.05$ & $-$0.07 & ---   & $-$0.07 & $-$0.46  & $-$0.25 & $-$0.22 &  $-$0.55$^*$ &  0.16 & $-$0.14  \\ 
$R_1$       & $-$0.53 & $<0.05$ & $-$0.20 & $-$0.22 & ---   & $-$0.48  & $-$0.28 & $-$0.28 &  $-$0.58$^*$ &  0.20 & $-$0.33  \\ 
$M_1$       & $-$0.37 & 0.12 & $-$0.32 & $-$0.22 & $-$0.14 & ---    &  0.17 &  0.28 &  $-$0.42$^*$ &  0.27 & $-$0.23  \\ 
$T_1$       & $-$0.70 & $<0.05$ & $-$0.65 & $-$0.64 & $-$0.63 & $-$0.72  & ---   &  0.07 &  $-$0.79$^*$ & $-$0.36 & $-$0.62  \\ 
$T_2$       & $-$0.76 & $<0.05$ & $-$0.72 & $-$0.71 & $-$0.71 & $-$0.80  & $-$0.47 & ---   &  $-$0.78$^*$ & $-$0.53 & $-$0.68  \\ 
$|\Delta T|$ &  0.13 & 0.58 &  0.26$^*$ &  0.28$^*$ &  0.29$^*$ &  0.16  &  0.40 &  0.27$^*$ & ---   &  0.17 & 0.21$^*$  \\ 
$L_1$       & $-$0.69 & $<0.05$ & $-$0.59 & $-$0.62 & $-$0.60 & $-$0.74  & $-$0.30 & $-$0.22 &  $-$0.75$^*$ & ---   & $-$0.52  \\ 
$L_2$       & $-$0.58 & $<0.05$ & $-$0.40 & $-$0.38 & $-$0.40 & $-$0.53  & $-$0.30 & $-$0.15 &  $-$0.59$^*$ & $-$0.09 & ---    \\ 
\hline
\end{tabular}
\parbox{\hsize}{\emph{Notes.}
Correlation coefficients are summarized as in Table \ref{Tab_Pcor_MEML}. }
\end{table*}
\subsubsection{ML}\label{Sec_Cor_ML}
In Fig. \ref{Fig_RMSEP}d, the RMSEP values have a minimum at $N=2$, and those within $N\leq 4$ are similar. 
The explained variance for Y-variables increases by only several percent when $N\geq 4$ (Table \ref{Tab_EV}). 
Therefore, the model with $N=3$ is optimal. 

The VIP scores in Fig. \ref{Fig_PWVIP}d show that nine parameters have scores greater than one: $P$, $a$, $R_1$, $M_1$, $T_1$, $T_2$, $|\Delta T|$, $L_1$, and $L_2$. 
The first component is mainly contributed by the temperature and luminosity of each star. 
In the second component, although $M_2$, $q$, and $T_2$ have relatively large contributions, $M_2$ and $T_2$ show opposite contributions. 
The third component is contributed by $M_1$ and $L_2/L_1$. 

Table \ref{Tab_Pcor_ML} summarizes the partial correlation coefficients for the nine parameters. 
Only four ($T_2$, $|\Delta T|$, $L_1$, and $L_2$) have coefficients of which signs are the same as those of the corresponding $r_\mathrm{p}$. 
Because $|\Delta T|$ have small correlation coefficients, it is ruled out. 
Of the other three, $T_2$ affects the other two whereas $T_2$ is less affected by the others. 
The parameter $L_2$ is also affected by $L_1$. 
Therefore, $T_2$ is the most plausible. 

\begin{figure}
\centering
\begin{minipage}[b]{0.23\textwidth}
	\includegraphics[width=\textwidth]{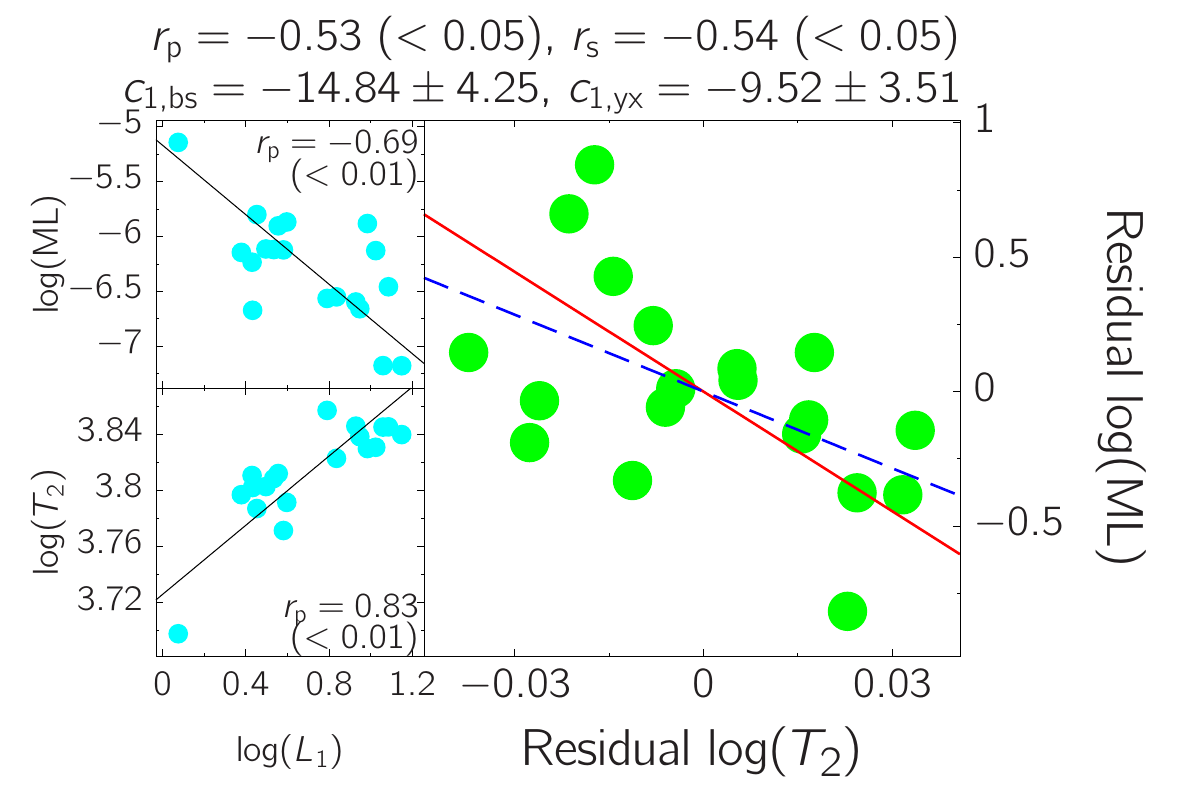}
	\centerline{(a)}
\end{minipage}
\begin{minipage}[b]{0.23\textwidth}
	\includegraphics[width=\textwidth]{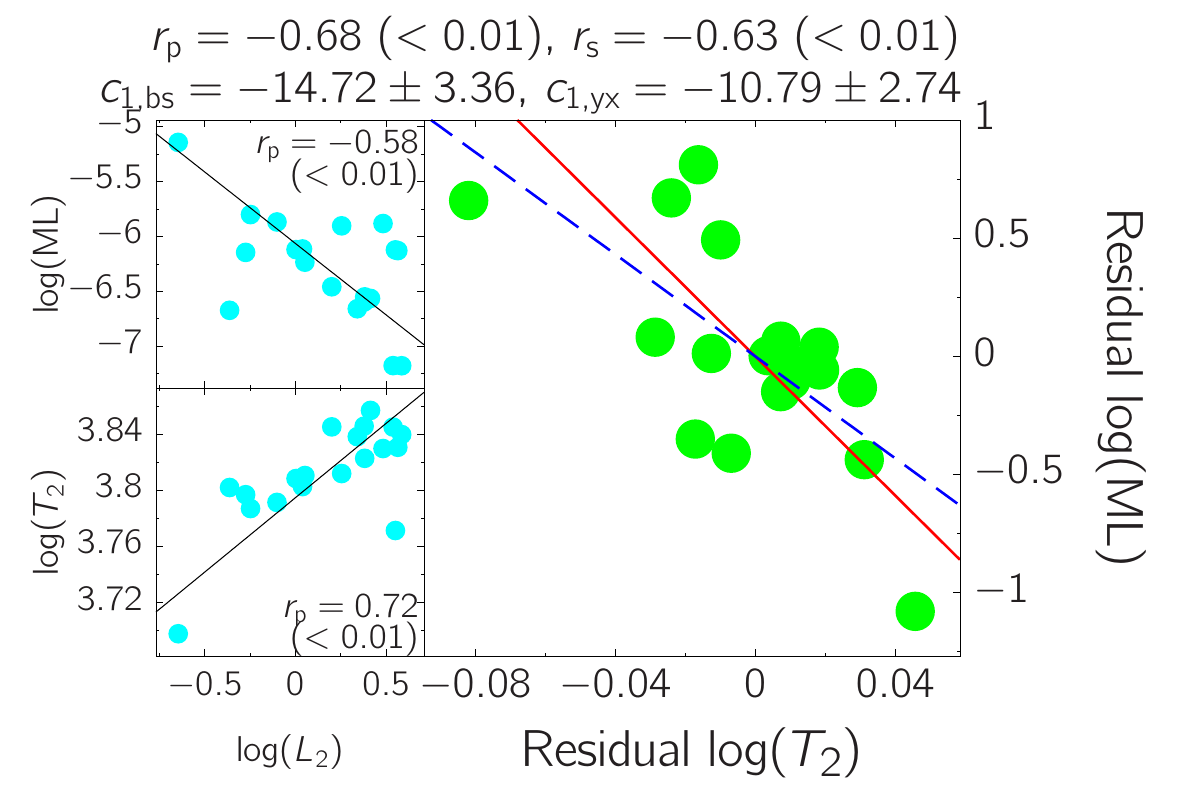}
	\centerline{(b)}
\end{minipage}
\caption{
Partial regression plots of the secondary temperature vs. ML rate, controlling for the primary (left) and secondary (right) luminosities, as in Fig. \ref{Fig_PRP_An_ME}. 
The residuals are based on the OLS bisector. 
\label{Fig_PRP_Ap_ML}}
\end{figure}

Figure \ref{Fig_PRP_Ap_ML} shows the partial regression plots of $T_2$ and the ML rate, controlling for $L_1$ and $L_2$. 
Negative correlations with similar slopes are found, and its power-law relations are 
\begin{eqnarray}\label{Eq_ML_depend}
	|\dot{\Mtot}| \propto T_2^{-14.8 \pm 2.6}  \hspace{3mm} \mathrm{and} \hspace{3mm}  |\dot{\Mtot}| \propto T_2^{-10.3 \pm 2.2} 
\end{eqnarray}
with the OLS bisector and OLS($Y|X$), respectively.

\section{Properties of sample binaries}\label{Sec_Property}
\begin{figure*}
	\centering
	\includegraphics[width=0.3\textwidth]{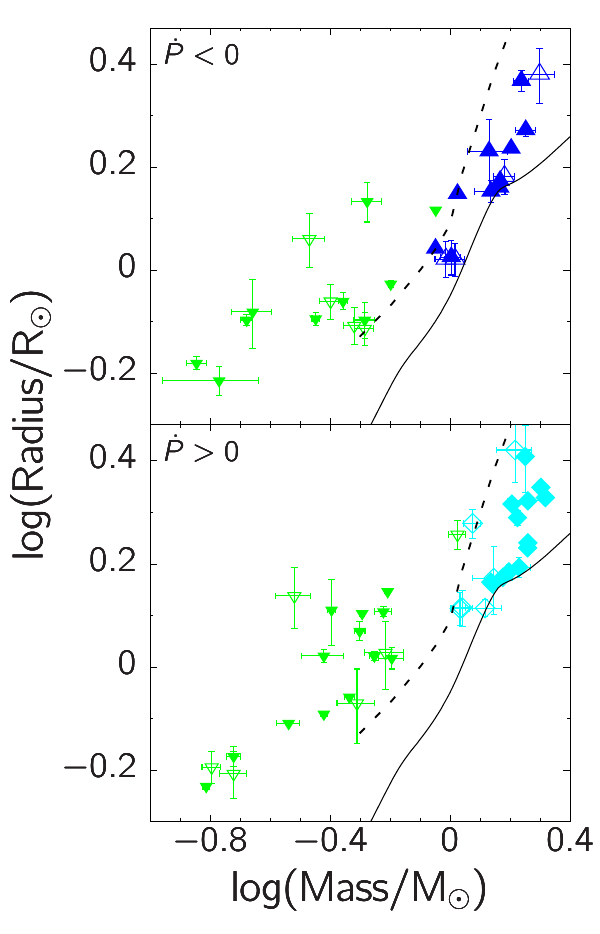}
	\includegraphics[width=0.3\textwidth]{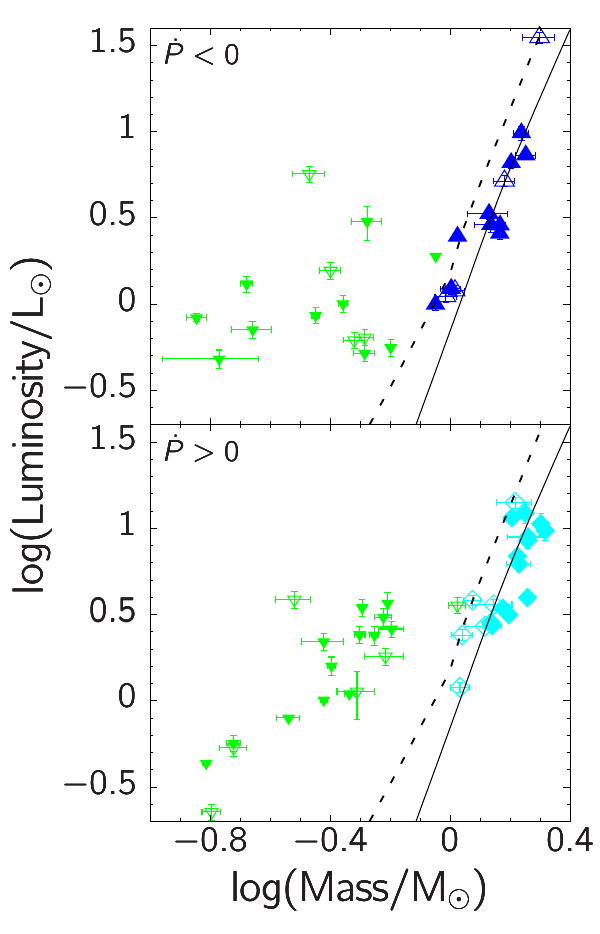}
	\includegraphics[width=0.3\textwidth]{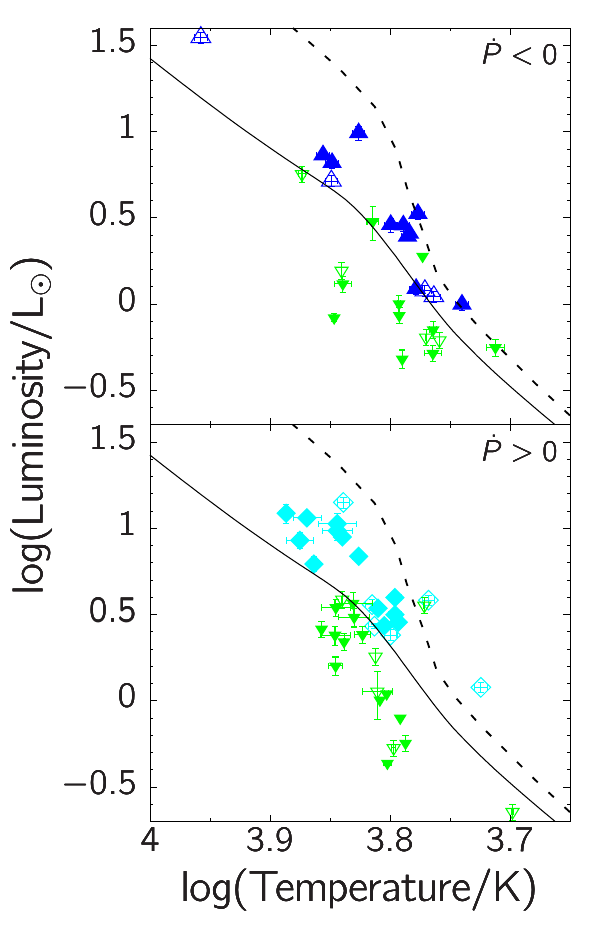}
\caption{Mass--radius, mass--luminosity, and H-R diagrams. 
The top and bottom panels are the diagrams for the An and Ap samples, respectively. 
The blue triangles and cyan diamonds represent primary stars and the smaller inverse triangles represent secondaries. 
The solid and open symbols represent binaries of which parameters are determined based on spectroscopic and photometric mass-ratios, respectively. 
The solid and dashed lines are the ZAMS and TAMS lines taken from \citet{Pols1998-MNRAS}, respectively. 
\label{Fig_EV}}
\end{figure*}
\subsection{Evolutionary status}\label{Sec_Evol}
Figure \ref{Fig_EV} shows $M$--$R$, $M$--$L$, and Hertzsprung-Russell (H-R) diagrams, together with the lines of zero-age main sequence (ZAMS) and terminal-age main sequence (TAMS) taken from \citet{Pols1998-MNRAS}. 
To estimate underlying functional relations, OLS-bisector lines are computed for the $M$--$R$ and $M$--$L$ relations. 
Table \ref{Tab_Slopes} summarizes their regression slopes and intercepts. 

In the $M$--$R$ diagrams, almost all primaries are located between the ZAMS and TAMS lines. 
Both regression lines for the primaries of the An and Ap samples have slopes larger than 1 (Table \ref{Tab_Slopes}), which is slightly steeper than the slope ($c_1=0.92 \pm 0.04$) reported by \citet{Gazeas2008-MNRAS}. 
The primaries of eight (42 percent) Ap systems with $R_1>1.8$ $R_\odot$ are relatively far from the ZAMS line, while those of the other Ap systems are closer to the ZAMS. 
In other words, these eight primaries are more highly evolved than those of the others. 
The eight also have $R_2>1$ $\Rsol$, which are also larger than the radii of almost all of the others. 
Furthermore, they are relatively far from the regression line for the secondaries. 
Thus, the eight Ap systems exhibit a tendency differing from that of the other Ap systems. 
As mentioned in Section \ref{Sec_Prop_Asso_Mass}, the eight systems have $a>3.5$ $\Rsol$. 

By contrast, in the $M$--$L$ diagrams, 4 An and 11 Ap primaries are below the ZAMS line and also have $M_1>1.3$ $\Msol$. 
The former indicates that these primaries are underluminous for their masses. 
Concerning the secondaries, the data points of the An sample are relatively scattered, whereas those of the Ap sample show a clear positive correlation. 
It is also found that only three An systems (IK Per, OO Aql, and V1073 Cyg) have $L_2>1.6$ $\Lsol$ while the majority of the Ap sample have such secondary luminosities. 

In the H-R diagrams, almost all primaries are located between the ZAMS and TAMS lines, as well as in the $M$--$R$ diagrams. 
Moreover, the component stars of the An sample tend to have lower temperatures than those of the Ap sample. 
In other words, the majority of the An (Ap) components have spectral types later (earlier) than F5. 

The secondary stars of the sample binaries are generally oversized and overluminous for their masses. 
This tendency is well known and has been considered by previous studies: 
e.g., energy transfer between the two components \citep{Lucy1968-ApJ1123,Webbink2003-ASPC} and the reversal of mass ratio \citep{Stepien2006-AcA199}.

\begin{table*}
\centering
\caption{The slopes and intercepts of regression lines between binary parameters \label{Tab_Slopes}}
\resizebox{\textwidth}{!}{%
\begin{tabular}{ccrrrrrrrr}
\hline \hline
\multicolumn{2}{c}{Parameters} & \multicolumn{8}{c}{$ \log Y = c_1 \log X + c_0 $} \\
\multicolumn{2}{c}{} & \multicolumn{4}{c}{An} & \multicolumn{4}{c}{Ap$^a$} \\
\multicolumn{2}{c}{} & \multicolumn{2}{c}{Primary} & \multicolumn{2}{c}{Secondary} & \multicolumn{2}{c}{Primary} & \multicolumn{2}{c}{Secondary} \\ 
\multicolumn{1}{c}{$X$} & \multicolumn{1}{c}{$Y$} & \multicolumn{1}{c}{$c_1$} & \multicolumn{1}{c}{$c_0$} & \multicolumn{1}{c}{$c_1$} & \multicolumn{1}{c}{$c_0$} & \multicolumn{1}{c}{$c_1$} & \multicolumn{1}{c}{$c_0$} & \multicolumn{1}{c}{$c_1$} & \multicolumn{1}{c}{$c_0$} \\
\hline
$M$ & $R$ & $1.08 \pm 0.11$ & $0.04 \pm 0.02$  &  $0.46 \pm 0.08$ & $0.14 \pm 0.05$  & $1.16 \pm 0.17$ & $0.02 \pm 0.03$  &  $0.60 \pm 0.04$ & $0.24 \pm 0.03$  \\ 
$M$ & $L$ & $3.92 \pm 0.52$ &  $0.03 \pm 0.06$ &  $1.11 \pm 0.12$ & $0.49 \pm 0.13$  & $3.55 \pm 0.42$ &  $0.03 \pm 0.08$ &  $1.56 \pm 0.16$ &  $0.80 \pm 0.09$ \\
$a$   & $R$ & $ 1.26 \pm 0.13 $ & $ -0.41 \pm 0.06 $	&  $ 1.06 \pm 0.17 $ &  $ -0.55 \pm 0.08 $ & $ 0.90 \pm 0.07 $ & $ -0.23 \pm 0.04 $  &  $ 1.27 \pm 0.12 $ &  $ -0.67 \pm 0.06 $ \\
$q$   & $R$ & $ -0.40 \pm 0.09 $ & $ -0.06 \pm 0.04 $ &  $ 0.38 \pm 0.11 $ &  $ 0.16 \pm 0.07 $ & $ 0.78 \pm 0.25 $ & $ 0.70 \pm 0.16 $ &  $ 0.66 \pm 0.08 $ &  $ 0.39 \pm 0.06 $ \\
$a$   & $M$ & $1.15 \pm 0.20 $ & $-0.41 \pm 0.09$	&  $1.15 \pm 0.43$ &  $-0.96 \pm 0.22$ & $1.30 \pm 0.12$ & $-0.41 \pm 0.05$   &  $4.06 \pm 0.50$ &  $-2.27 \pm 0.21$ \\
$q$   & $M$ & $-0.35 \pm 0.06$ & $-0.07 \pm 0.03$	&  $0.71 \pm 0.04$ &  $-0.03 \pm 0.03$ & $0.88 \pm 0.49$ & $0.71 \pm 0.26$   &  $1.08 \pm 0.12$ &  $0.24 \pm 0.07$ \\
\hline
\end{tabular}
}
\parbox{\hsize}{\emph{Notes.}
The regression lines are computed with the OLS bisector. \\
$^a$The coefficients of the $a$--$M$ relationships for the Ap sample is computed using the Ap sample binaries with $a<3.5$ $\Rsol$. 
}
\end{table*}

\begin{figure*}
\centering
\begin{minipage}[b]{0.3\textwidth}
	\includegraphics[width=\textwidth]{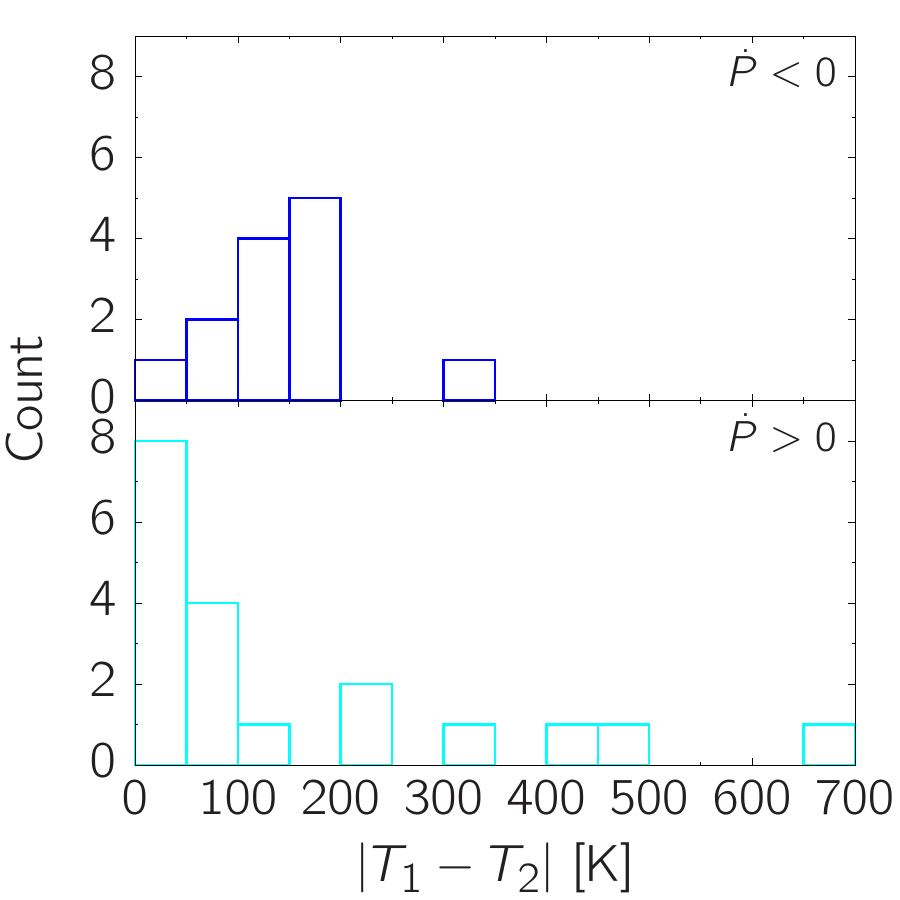}
	\centerline{(a)}
\end{minipage}
\begin{minipage}[b]{0.3\textwidth}
	\includegraphics[width=\textwidth]{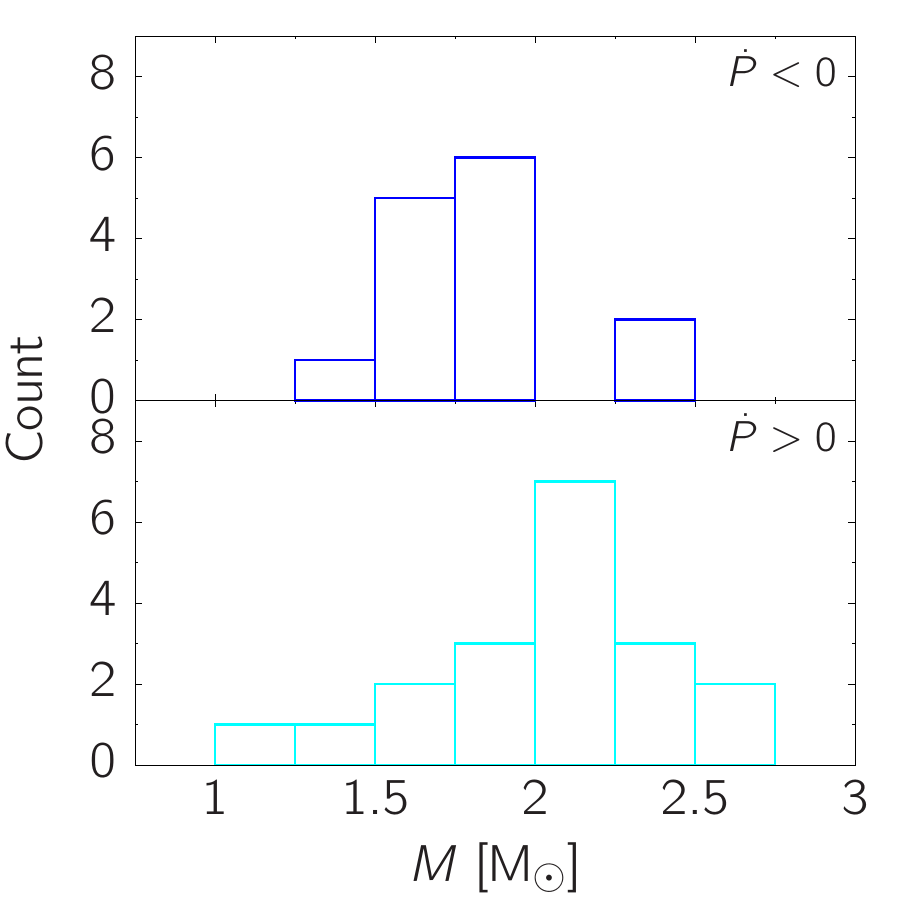}
	\centerline{(b)}
\end{minipage}
\begin{minipage}[b]{0.3\textwidth}
	\includegraphics[width=\textwidth]{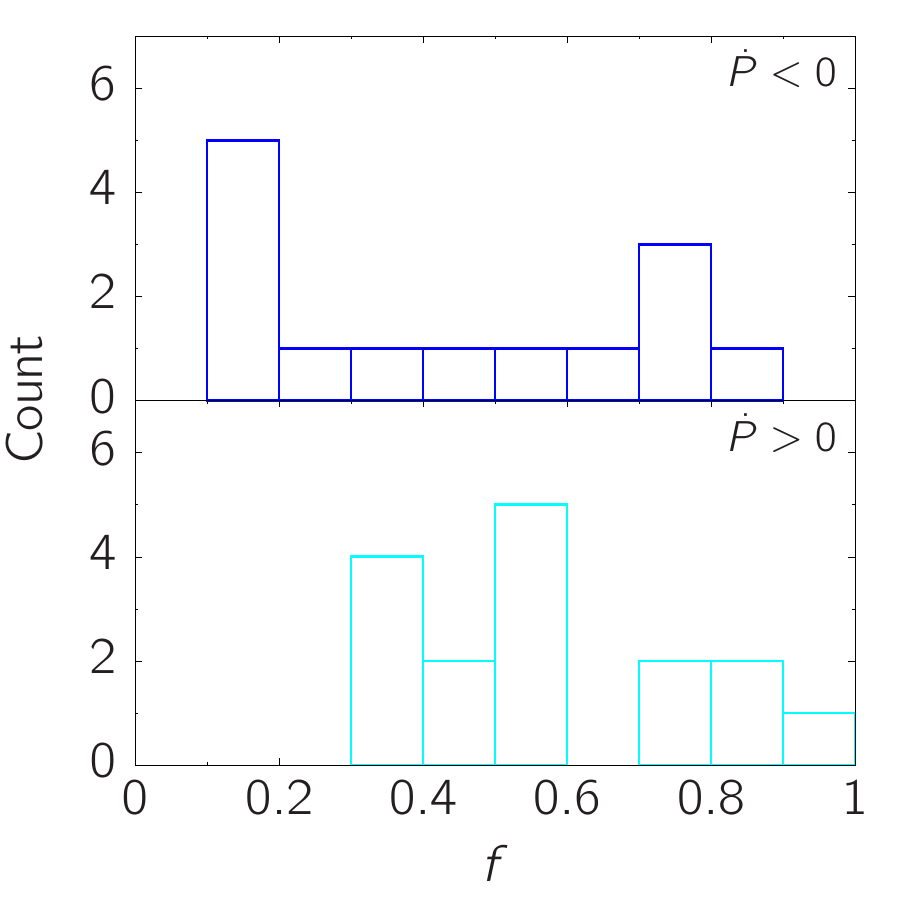}
	\centerline{(c)}
\end{minipage}
\caption{Histograms of the temperature difference (a), total mass (b), and fill-out factor (c) for the An (top) and Ap (bottom) samples. 
\label{Fig_Hist}}
\end{figure*}
\subsection{Histograms of binary parameters}
Notable features are found in the histograms of $|\Delta T|$, $\Mtot$, and $f$ (Fig. \ref{Fig_Hist}). 
In the histograms of $|\Delta T|$ (Fig. \ref{Fig_Hist}a), the An systems tend to have relatively large temperature-differences. 
The An sample shows a distribution with a peak around $|\Delta T|=150$--$200$ K, and only two (AU Ser and IK Per) have $|\Delta T|>200$ K (IK Per has $|\Delta T|=1600$ K). 
In contrast, the Ap sample shows a distribution with a peak around $|\Delta T|=0$--$50$ K and a long tail to the right. 
Although most Ap systems have $|\Delta T|<100$ K, a few systems distribute in a wide range of $|\Delta T|=100$--$700$ K. 
\citet{Rucinski1974-AcA} reported that a mean of temperature difference for A-type systems is $-80 \pm 25$ K, which is just the middle of the two peaks described above. 

In the histograms of $\Mtot$ (Fig. \ref{Fig_Hist}b), the An systems tend to have lower total masses than the Ap systems. 
All the An systems have $\Mtot<2$ $\Msol$ except for two (i.e., IK Per and V1073 Cyg). 
By contrast, the majority of the Ap systems have $\Mtot>2$ $\Msol$. 
The Ap sample shows a distribution with a peak at $\Mtot=2$--$2.25$ $\Msol$ and a longer tail to the left than to the right. 

The distributions of $f$ for the An and Ap samples may also differ (Fig. \ref{Fig_Hist}c). 
Although almost half of the An systems have $f<0.3$, no Ap systems have such small fill-out factors. 
The An sample shows a distribution concentrating at $f \sim 0.2$ and $0.8$, and a uniform distribution within $f=0.2$--$0.7$. 
In contrast, the Ap sample shows a distribution concentrating at $f \sim 0.5$ and 0.8.

\begin{figure*}
\centering
\begin{minipage}[b]{0.48\textwidth}
	\includegraphics[width=\textwidth]{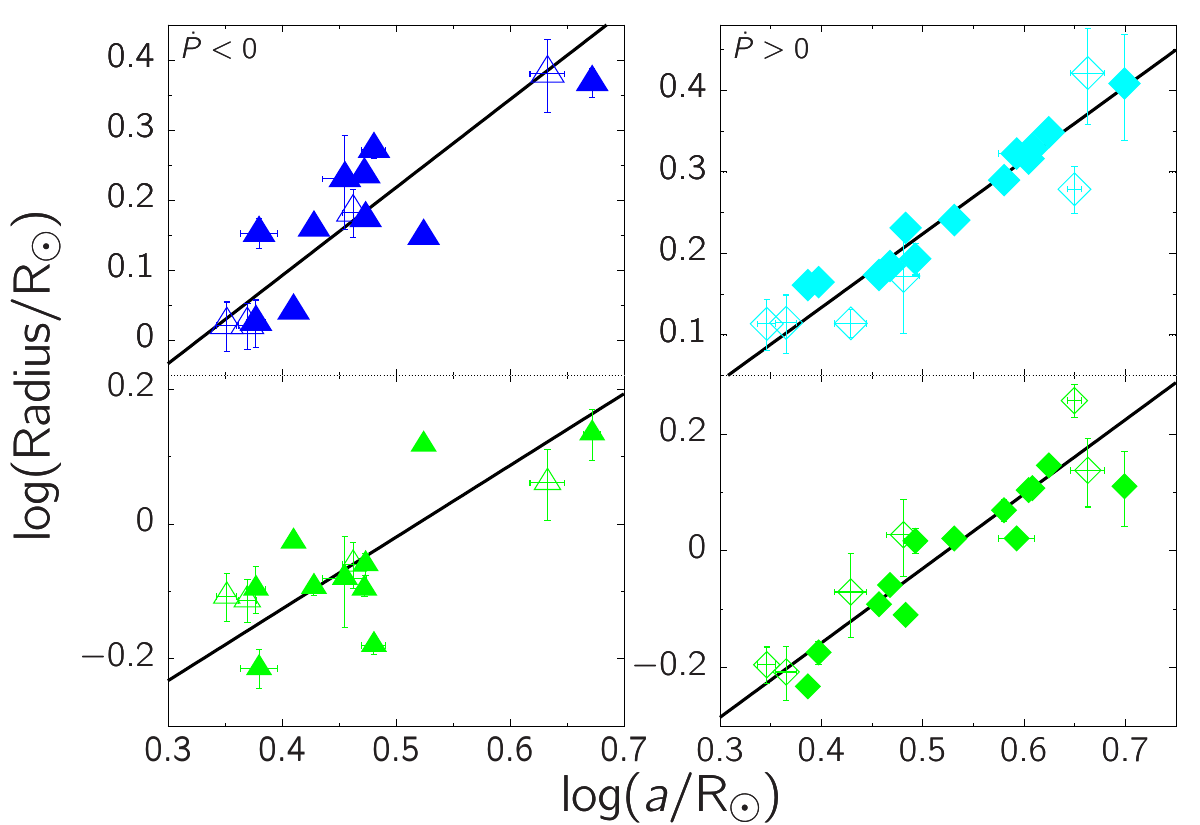}
	\centerline{(a)}
\end{minipage}
\begin{minipage}[b]{0.48\textwidth}
	\includegraphics[width=\textwidth]{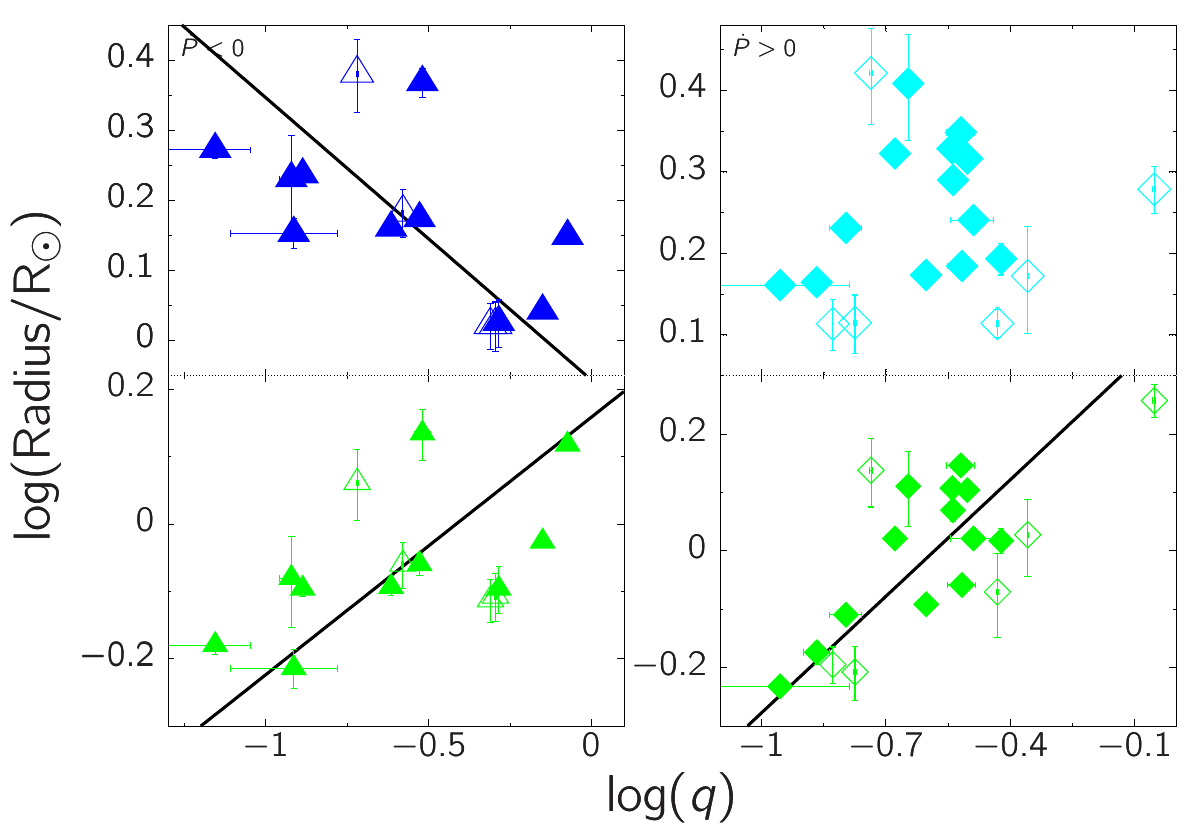}
	\centerline{(b)}
\end{minipage}
\caption{Primary (top panels) and secondary (bottom panels) radii as a function of separation (a) and mass ratio (b). 
The left and right panels are plots for the An and Ap samples, respectively. 
The regression lines are computed with the OLS bisector. 
\label{Fig_aq-R}}
\end{figure*}

\begin{figure*}
\centering
\begin{minipage}[b]{0.48\textwidth}
	\includegraphics[width=\textwidth]{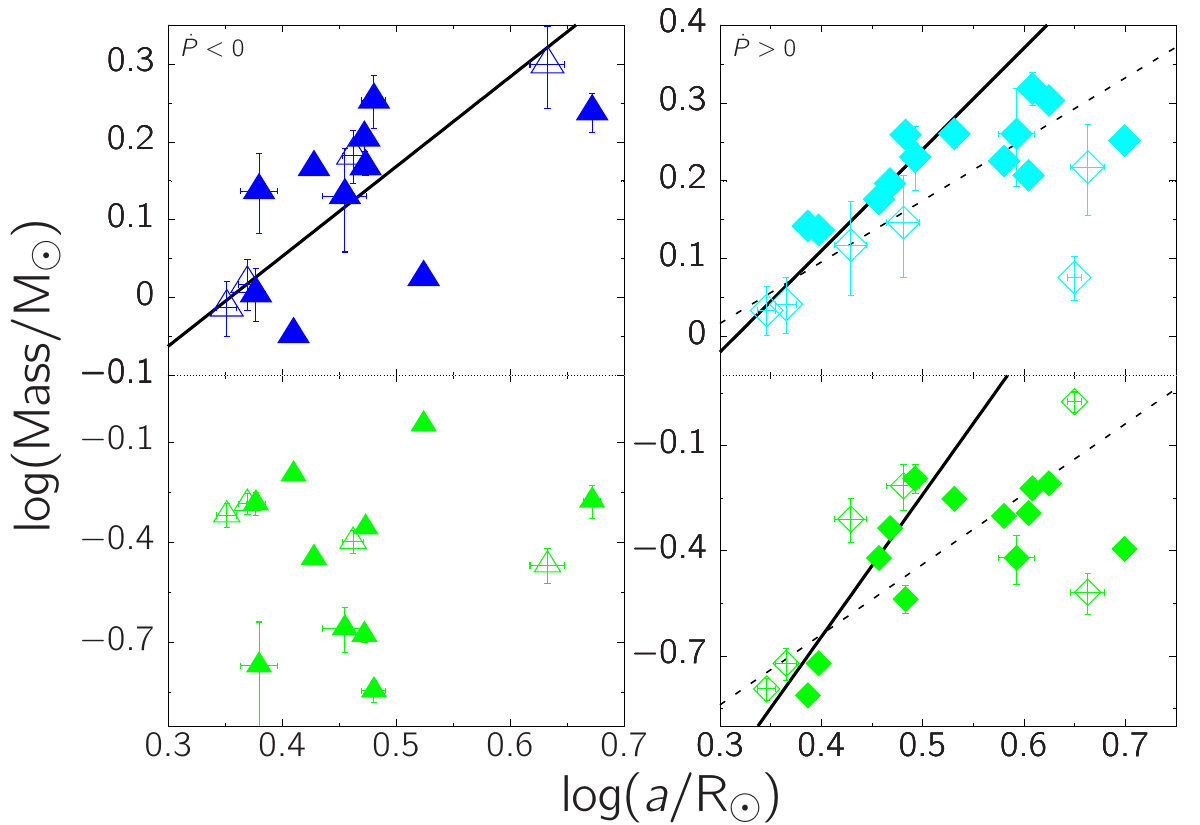}
	\centerline{(a)}
\end{minipage}
\begin{minipage}[b]{0.48\textwidth}
	\includegraphics[width=\textwidth]{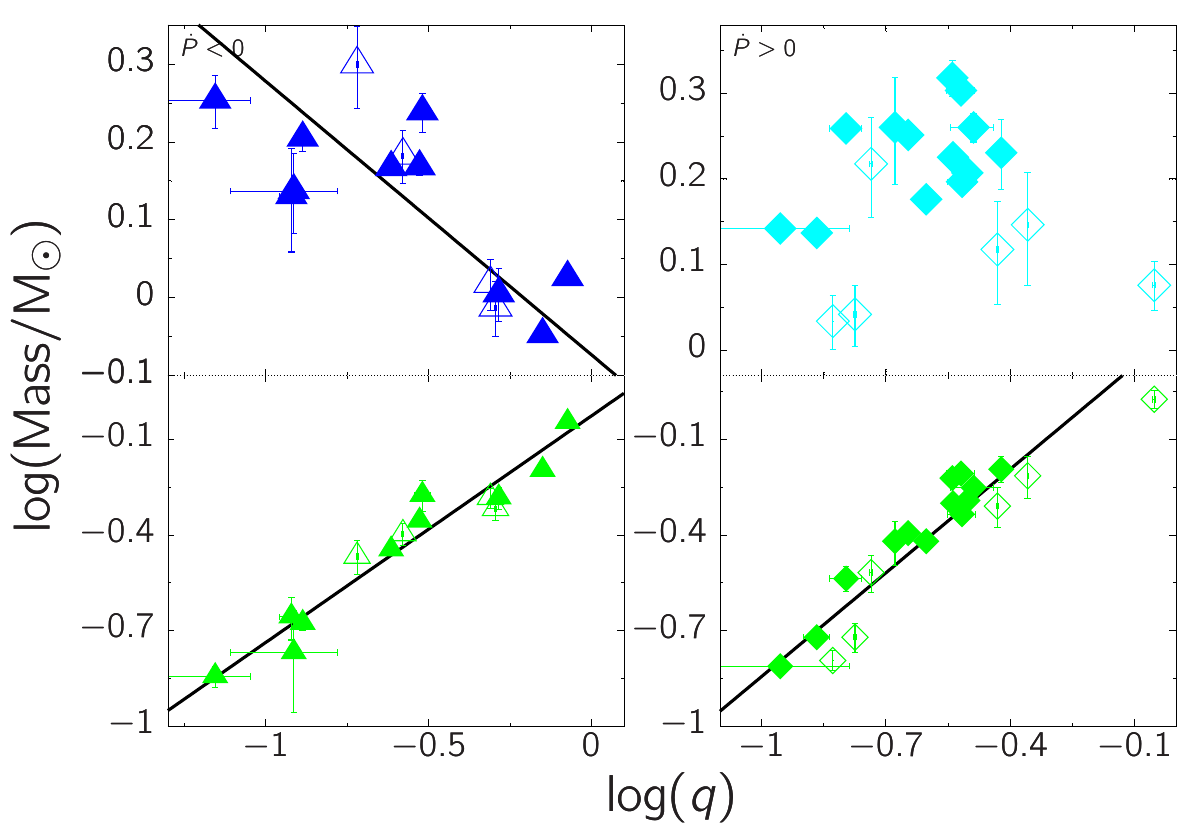}
	\centerline{(b)}
\end{minipage}
\caption{Primary (top panels) and secondary (bottom panels) masses as a function of separation (a) and mass ratio (b). 
The solid and dashed lines for the Ap sample are regression lines for the systems with $a<3.5$ $\Rsol$ and all systems, respectively. 
Symbols are the same as in Fig. \ref{Fig_aq-R}. 
\label{Fig_aq-M}}
\end{figure*}
\subsection{Associations between binary parameters}\label{Sec_Prop_Asso}
\subsubsection{Radius}
\citet{Eggleton1983-ApJ} provided an accurate formula for the effective radius of an inner Roche lobe, which is a function of the separation and mass ratio. 
The stellar radii in a contact binary are roughly approximated to the effective radii of the Roche lobes. 
Accordingly, the stellar radii of the sample binaries are expected to be correlated with the separation and mass ratio. 

Figure \ref{Fig_aq-R}a shows the scatter plots of $R_1$ and $R_2$ against $a$. 
All four plots display positive correlations as expected. 
Moreover, their OLS-bisector lines, shown in the plots, have slopes of $\sim 1$ (Table \ref{Tab_Slopes}). 
These slopes agree with the formula of \citet{Eggleton1983-ApJ}, in which the effective radius of a Roche lobe is proportional to the separation. 
Note that the Ap sample has distributions less dispersed than the An sample. 

Figure \ref{Fig_aq-R}b shows the scatter plots of $R_1$ and $R_2$ against $q$. 
On each plot, the data points are more dispersed than those on the corresponding plot in Fig. \ref{Fig_aq-R}a. 
A simple approximation by \citet{Paczynski1971-ARAA} indicates that the effective radius of a Roche lobe depends on the separation more than the mass ratio. 
Therefore, the dispersed distributions should be due to the more sensitive dependence of the lobe radii on the separation. 
In the An sample, the primaries and secondaries have negative and positive correlations, respectively. 
Their regression slopes in Table \ref{Tab_Slopes} are reasonably consistent with the Paczy\'{n}ski's approximation. 
By contrast, in the Ap sample, although the secondaries have a positive correlation, the primaries appear to have no clear correlation. 
Moreover, the regression slope for the Ap secondaries is approximately two times steeper than that for the An secondaries.

\begin{figure}
\centering
\includegraphics[width=0.48\textwidth]{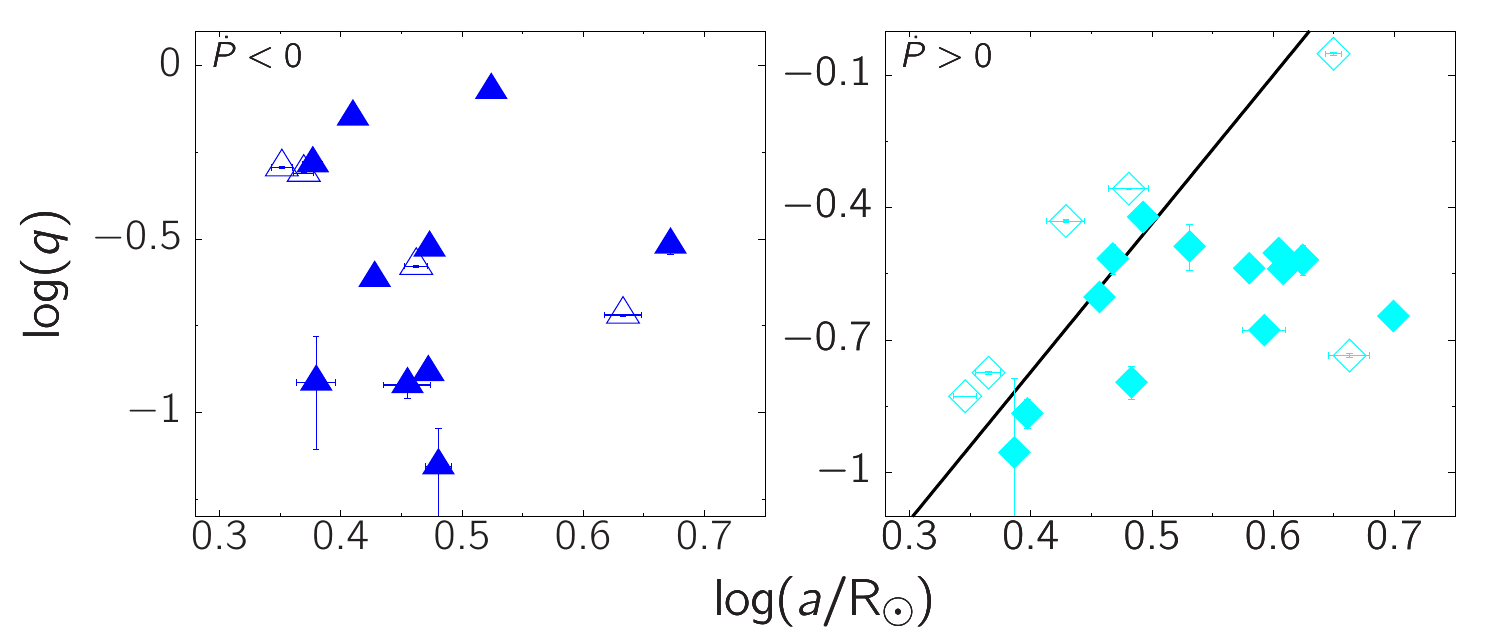}
\caption{Separation vs. mass ratio for the An (left) and {Ap} ({right}) systems. 
The regression line is for the systems with $a<3.5$ $\Rsol$. 
\label{Fig_a-q}}
\end{figure}
\subsubsection{Mass}\label{Sec_Prop_Asso_Mass}
We also examine the relations of $a$ and $q$ with $M_1$ and $M_2$ (Fig. \ref{Fig_aq-M}). 
In the $a$--$M$ relations for the An sample (Fig. \ref{Fig_aq-M}a), although the primaries show a positive correlation, the secondaries appear to show a very weak positive or no correlation. 
This unclear association may be due to the dispersed distribution in the $M$--$R$ diagram for the secondaries (Fig. \ref{Fig_EV}). 
By contrast, the Ap primaries and secondaries have positive correlations, especially below $a \sim 3.5$ $\Rsol$. 
Moreover, the regression slope for the Ap secondaries is about three times steeper than that for the Ap primaries. 
However, above this separation, the Ap components have a trend that differs from the linear relationship below $a \sim 3.5$ $\Rsol$. 
In the range $a > 3.5$ $\Rsol$, both components of the Ap systems tend to have masses lower than those predicted from the regression lines within $a < 3.5$ $\Rsol$. 

A similar trend is also found in the $a$--$q$ relation (Fig. \ref{Fig_a-q}). 
The Ap sample shows a positive correlation only below $a\sim3.5$ $\Rsol$, of which the regression line is 
\begin{equation}
	\log q = (3.36 \pm 0.60) \log a -(2.12 \pm 0.26). 
\end{equation}
The Ap systems with $a>3.5$ $\Rsol$ are far from the regression line. 
In contrast, the An sample shows no clear association. 
As mentioned in Section \ref{Sec_Evol}, the Ap systems with $a>3.5$ $\Rsol$ are more evolved in the $M$--$R$ diagram. 

Figure \ref{Fig_aq-M}b shows the scatter plots of $q$ and $M$. 
Trends similar to those in the $q$--$R$ relations are found. 
However, in both An and Ap samples, the regression slopes for the secondaries are 1.5--2 times steeper than those derived from the $q$--$R_2$ relationships. 
In each plot except for the Ap primaries, the data points are less dispersed than those on the corresponding plot in Fig. \ref{Fig_aq-M}a, unlike the $a$--$R$ and $a$--$M$ relations. 
This indicates that the stellar radius is closely associated with the separation while the mass is closely associated with the mass ratio. 

\begin{figure}
\centering
\includegraphics[width=0.48\textwidth]{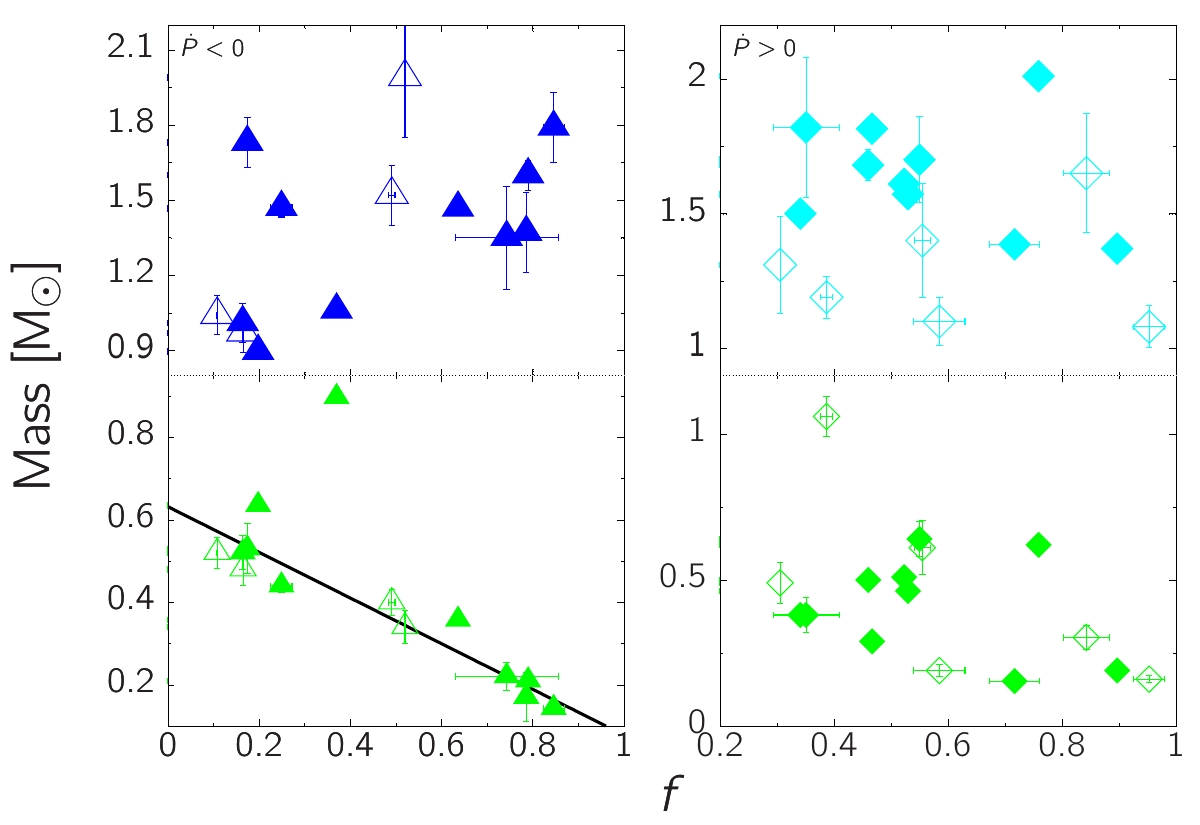}
\caption{Fill-out factor vs. primary (top) and secondary (bottom) masses. 
The regression line is computed without OO Aql. 
Symbols are the same as in Fig. \ref{Fig_aq-R}. 
\label{Fig_f-M}}
\end{figure}

A curious correlation is found in the scatter plots of $f$ and $M$ (Fig. \ref{Fig_f-M}). 
The An secondaries have a strong negative correlation, although the others show no clear correlations. 
The OLS-bisector line for the An secondaries except OO Aql is
\begin{equation}
	M_2=(-0.55 \pm 0.05)f+(0.63 \pm 0.03).
\end{equation}
OO Aql is the system that has the highest secondary mass in the An sample, and its secondary mass is extremely higher than that predicted from the regression line. 
Accordingly, the secondary mass of OO Aql seems to be an outlier. 

\begin{figure}
\centering
\begin{minipage}[b]{0.48\textwidth}
	\includegraphics[width=\textwidth]{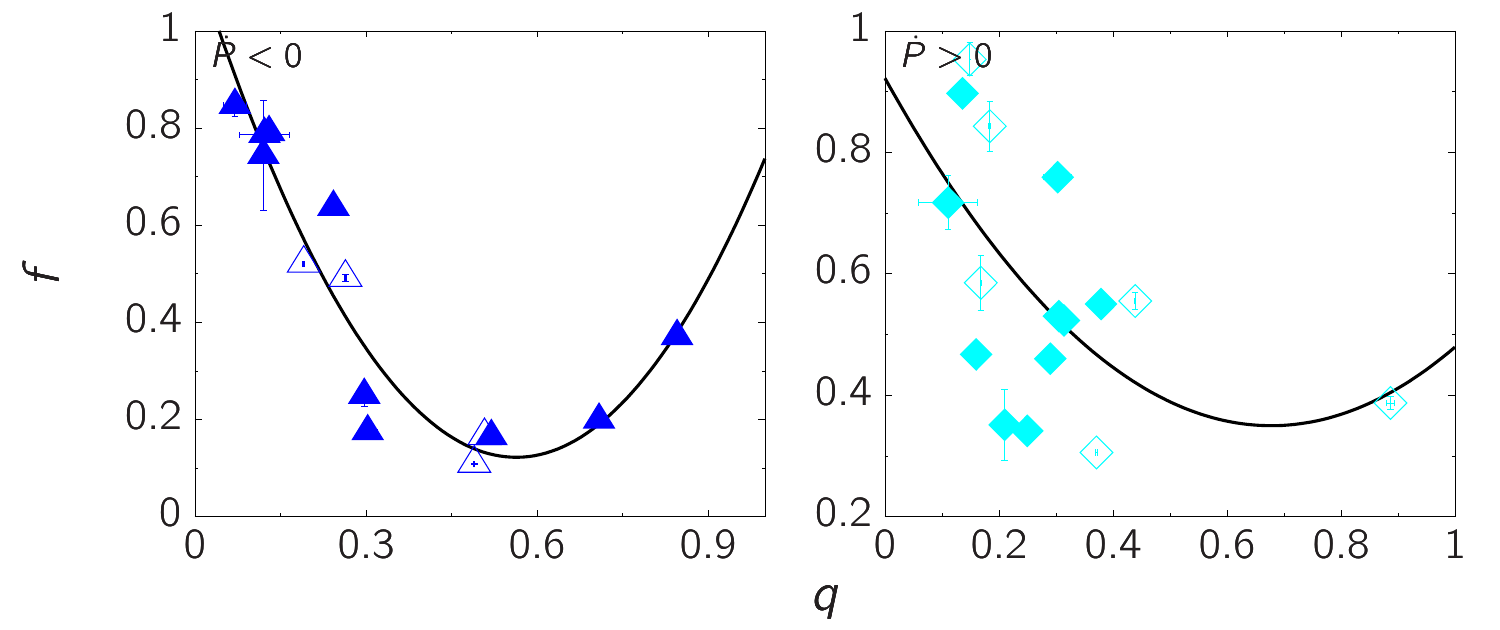}
	\centerline{(a)}
\end{minipage}
\begin{minipage}[b]{0.48\textwidth}
\includegraphics[width=\textwidth]{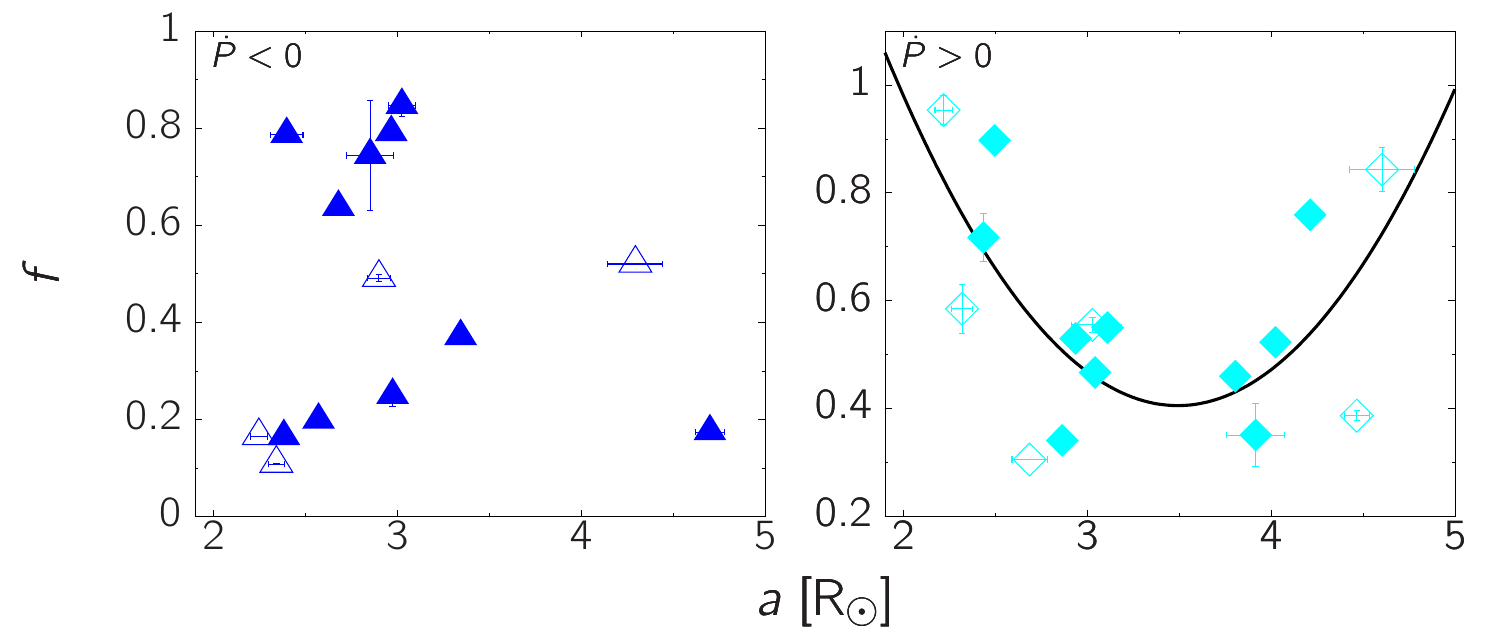}
	\centerline{(b)}
\end{minipage}
\caption{Scatter plots of fill-out factor against mass ratio (a) and separation (b). 
The quadratic regression curves are computed with least-squares fitting. 
Symbols are the same as in Fig. \ref{Fig_a-q}. 
\label{Fig_qa-f}}
\end{figure}
\subsubsection{Fill-out factor}
The fill-out factor also has associations with another two parameters: the mass ratio and separation. 
Figure \ref{Fig_qa-f}a shows the scatter plots of $q$ and $f$. 
Both An and Ap samples have positive quadratic associations. 
Their quadratic regression curves in the figure have minima at $q=0.56$ (for the An sample) and $0.68$ (for the Ap sample). 
Although the An systems are well fitted by the regression curve, the data points of the Ap sample are relatively dispersed from its regression curve. 

Another positive quadratic association is found in the scatter plot of $a$ and $f$ for the Ap sample (Fig. \ref{Fig_qa-f}b). 
Its quadratic regression curve has a minimum at $a=3.5$ $\Rsol$. 
However, the An sample shows no clear association, unlike the Ap sample.

\begin{figure}
\centering
\begin{minipage}[b]{0.48\textwidth}
	\includegraphics[width=\textwidth]{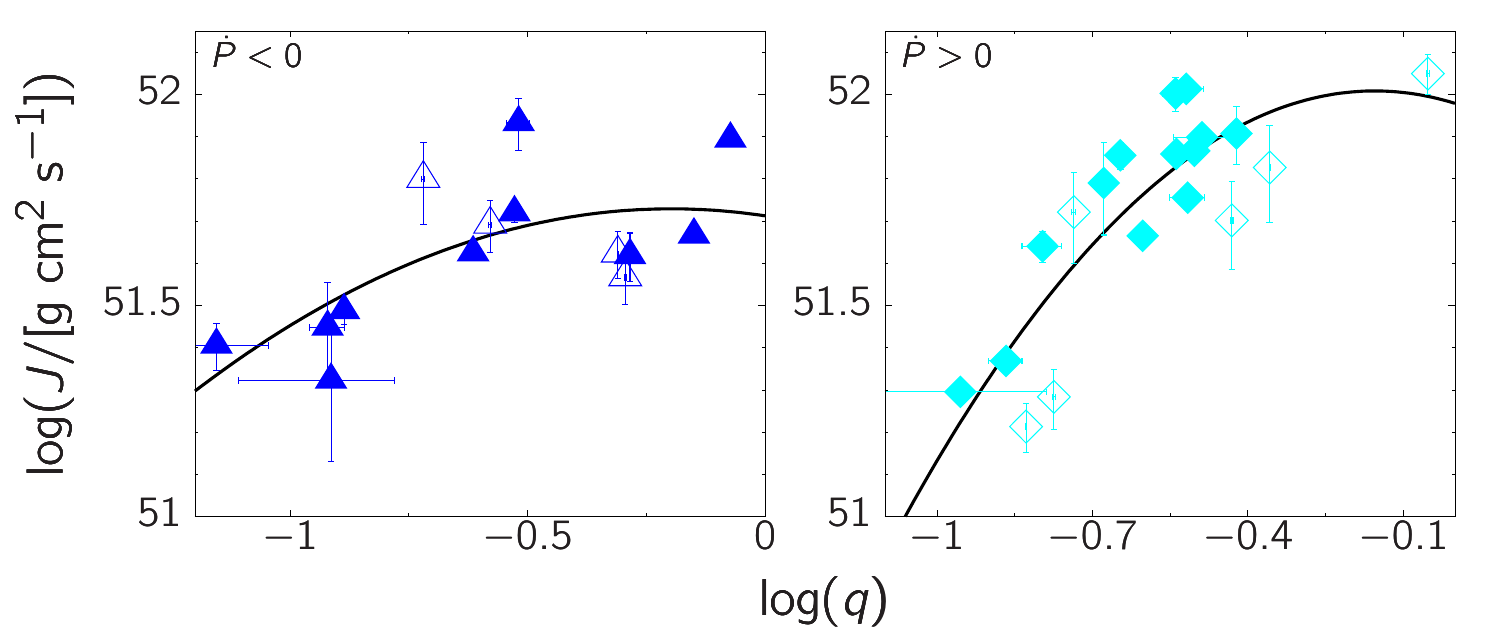}
	\centerline{(a)}
\end{minipage}
\begin{minipage}[b]{0.48\textwidth}
	\includegraphics[width=\textwidth]{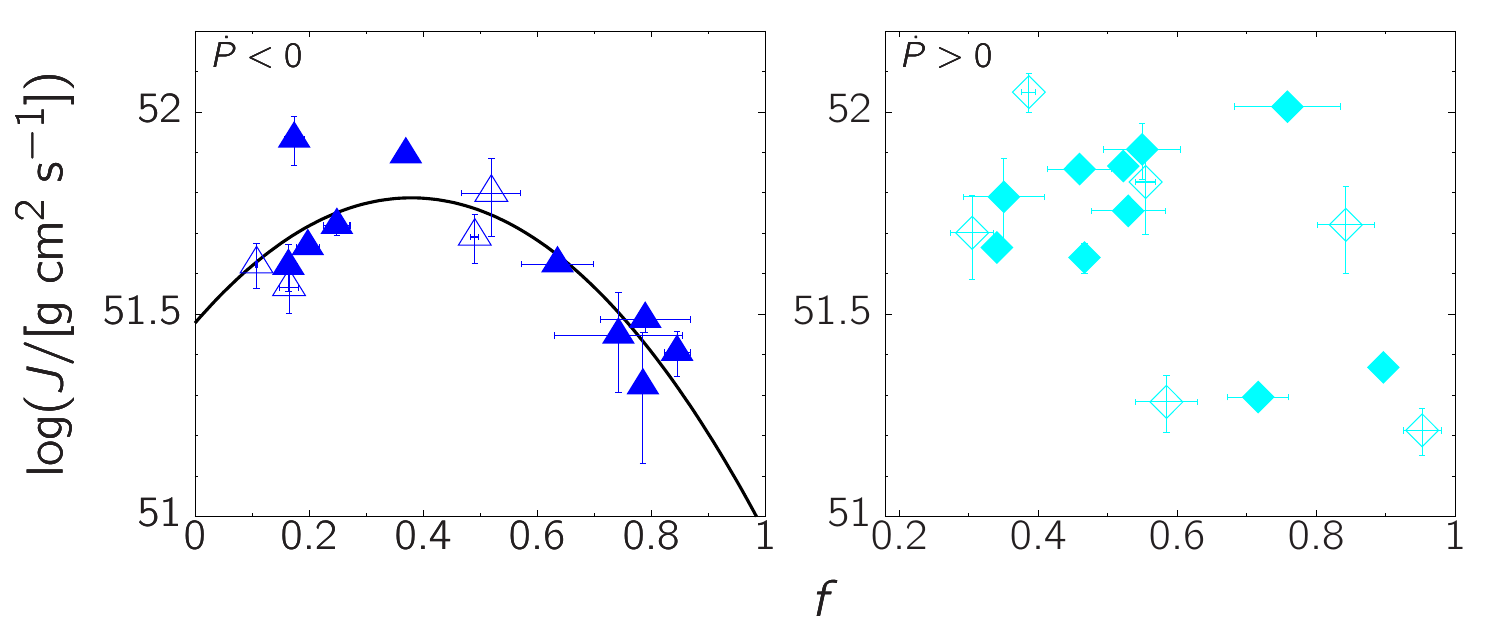}
	\centerline{(b)}
\end{minipage}
\begin{minipage}[b]{0.48\textwidth}
	\includegraphics[width=\textwidth]{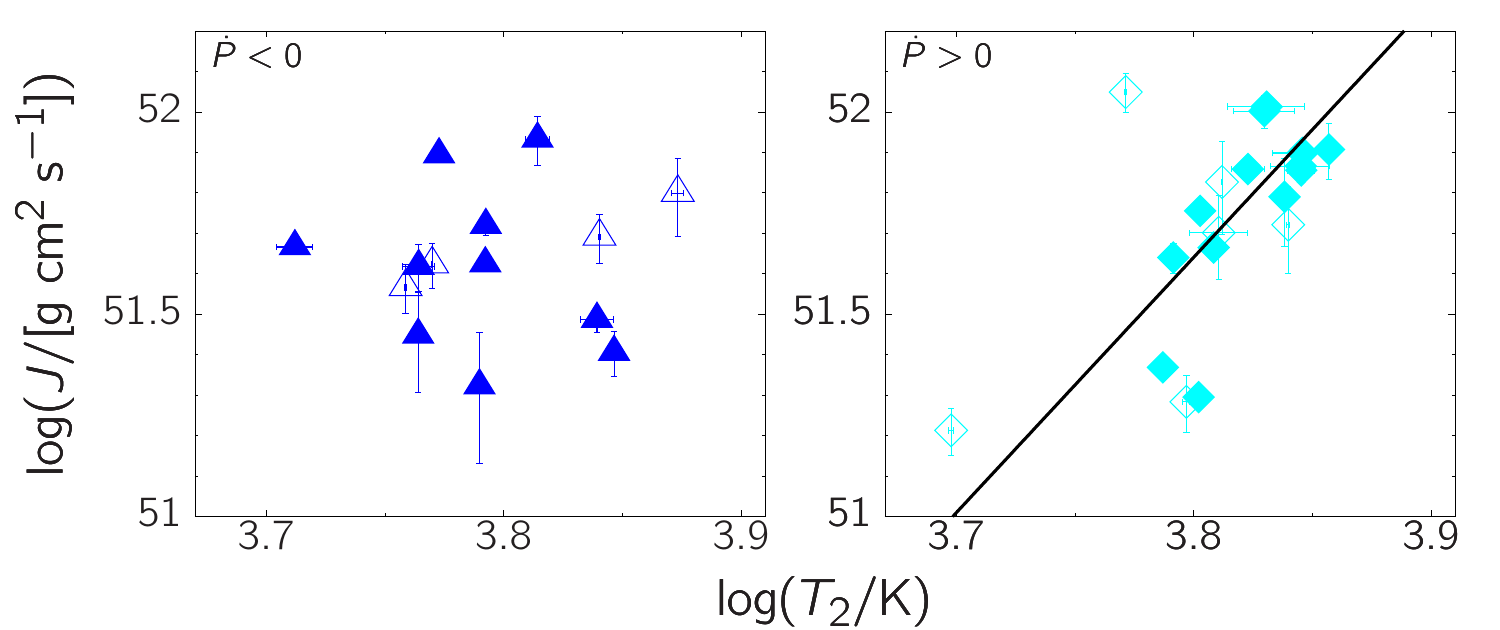}
	\centerline{(c)}
\end{minipage}
\caption{Scatter plots of total angular momentum against mass ratio (a), fill-out factor (b), and secondary temperature (c). 
Symbols are the same as in Fig. \ref{Fig_a-q}. 
\label{Fig_qft2-J}}
\end{figure}
\subsubsection{Angular momentum}
\citet{Vilhu1981-ApSS} reported that as the mass ratio of W UMa systems becomes small, its angular momentum also tends to become small. 
The $q$--$J$ relations in Fig. \ref{Fig_qft2-J}a show that the An and Ap samples have trends similar to that reported by \citet{Vilhu1981-ApSS}. 
Furthermore, their associations are a bit quadratic. 
However, the strength of dependence considerably differs between them. 
The OLS bisector yields the relationships of $J \propto q^{0.58 \pm 0.12}$ and $J \propto q^{1.20 \pm 0.16}$ for the An and Ap samples, respectively. 

Figure \ref{Fig_qft2-J}b shows the scatter plots of $f$ and $J$. 
Only the An sample has a negative quadratic association with a maximum at $f=0.38$. 
The angular momentum clearly decreases with increasing fill-out factor above $f \sim 0.4$. 

By contrast, only the Ap sample shows a positive correlation in the $T_2$--$J$ relation (Fig. \ref{Fig_qft2-J}c). 
Its OLS-bisector line is 
\begin{equation}
	\log J=(6.3 \pm 1.1) \log T_2 + (27.6 \pm 4.0). 
\end{equation}

\section{Discussion}\label{Sec_Discussion}
\subsection{Negative period variation}\label{Sec_Disc_NPC}
\subsubsection{The ML rate relative to the MTML rate deduced from the $q$--$M$ relations}\label{Sec_Disc_ratio_ME_ML}
We first focus on the correlations found in the $q$--$M$ relations of the An sample (Fig. \ref{Fig_aq-M}). 
The $q$--$M$ relations indicate that the primary mass is negatively correlated with the mass ratio while the secondary mass is positively correlated. 
These opposite correlations can be interpreted as a result of mass transfer.  
Therefore, we here assume the correlations arise from the mass transfer. 

Let us consider the following model. 
The primary star of a contact system transfers mass at a rate of $\dot{M}_1$. 
When a fraction $\beta$ of $\dot{M}_1$ is lost from the system and the rest is transferred to the secondary, we can write 
\begin{eqnarray}
	\dot{\Mtot}&=&\beta \dot{M}_1, \\
	\dot{M}_2&=&-(1-\beta) \dot{M}_1. 
\end{eqnarray}
With these equations and $q=M_2/M_1$, we obtain
\begin{equation}\label{Eq_qM1_qM2}
	\frac{\dot{q}}{q} = -\left(1+\frac{1-\beta}{q} \right)\frac{\dot{M}_1}{M_1} \hspace{3mm} \mathrm{ and } \hspace{3mm} 
	\frac{\dot{q}}{q} = \left(1+\frac{q}{1-\beta} \right)\frac{\dot{M}_2}{M_2}. 
\end{equation}
When the masses of the two component stars have power-law relations with the mass ratio, i.e., $M_1 \propto q^{\alpha_1}$ and $M_2 \propto q^{\alpha_2}$, 
equation (\ref{Eq_qM1_qM2}) is transformed into 
\begin{eqnarray}\label{Eq_Beta_Gen}
	\beta=1+\left(1+\frac{1}{\alpha_1} \right)q \hspace{3mm} \mathrm{ and } \hspace{3mm}
	\beta=1-\frac{\alpha_2}{1-\alpha_2}q.
\end{eqnarray}
Using $\alpha_1=-0.35\pm 0.06$ and $\alpha_2=0.71 \pm 0.04$ in Table \ref{Tab_Slopes},  
equation (\ref{Eq_Beta_Gen}) becomes 
\begin{equation}
	\beta = 1-(1.86 \pm 0.49) q \hspace{3mm} \mathrm{ and } \hspace{3mm}
	\beta = 1-(2.45 \pm 0.48) q. 
\end{equation}
These two relations agree within the limits of errors. 
Finally, by taking the weighted mean of the two relations, we find 
\begin{equation}\label{Eq_Beta_Final}
\beta = 1-(2.16 \pm 0.34) q. 
\end{equation}

Equation (\ref{Eq_Beta_Final}) indicates that the ML rate relative to the MTML rate decreases with increasing mass ratio. 
When a binary has $q=0.05$, which is comparable with the minimum mass ratio of W UMa systems, 89 percent of the mass transferred from the primary should be lost. 
In addition, the equation predicts that the ML in the An systems occurs within $q<0.46 \pm 0.08$. 

The ML in the An systems is likely to occur from Lagrange point \Ls  because of the following reasons. 
First, a combination of the MTML and ML accounts for the opposite correlations in the $q$--$M$ relations (Fig. \ref{Fig_aq-M}); 
the ML rate relative to the MTML rate is a function of the mass ratio [equation (\ref{Eq_Beta_Final})]. 
These indicate that the ML is closely associated with the MTML. 
Second, as the mass ratio becomes small, the layer between the inner and outer Roche lobes also becomes thin, and therefore much more matter is likely to be lost from the \Ls  point. 
\citet{Kuiper1941-ApJ} surmised that if matter flow from primary to secondary via the \Lf  point is sufficiently large a fraction of the matter will fly off in the vicinity of the \Ls  point. 
Thus, these pictures are consistent. 

Furthermore, the results in this section are consistent with previous studies which investigated the ML from the \Ls  point. 
\citet{Pribulla1998-CoSka} presented that the mass outflow from the \Ls  point is very efficient for low mass-ratio systems, of which trend agrees with equation (\ref{Eq_Beta_Final}). 
\citet{Yildiz2013-MNRAS} estimated the initial masses of the components of W UMa systems on the basis of luminosity excess. 
They concluded that, in A-type W UMa systems, 34 percent of the mass is transferred to another component and the remainder is lost from the system. 
With the mean mass-ratio of the An sample (i.e., $q=0.34 \pm 0.12$), equation (\ref{Eq_Beta_Final}) yields that $(73 \pm 28)$ percent of the mass is transferred to another component. 
These two ratios are of the same order, and thus reasonably agree with each other. 
These studies also support the above discussion.

\subsubsection{Associations between binary parameters}\label{Sec_Disc_associations}
Several associations in Section \ref{Sec_Prop_Asso} are also consistent with the discussion in Section \ref{Sec_Disc_ratio_ME_ML}. 
Below $q=0.5$--$0.6$, the $q$--$f$ relation (Fig. \ref{Fig_qa-f}a) indicates that contact systems with smaller mass-ratios tend to have higher fill-out factors. 
In a contact system with a higher fill-out factor, the stellar surface of the secondary is closer to its outer Roche lobe. 
Such a system is likely to lose more matter from the \Ls  point. 
Accordingly, this picture is consistent with the result that as the mass ratio becomes small, the ML rate becomes large [equation (\ref{Eq_Beta_Final})]. 
Furthermore, the $q$--$f$ relation also shows that $f$ slightly increases with increasing $q$ above $q\sim 0.5$. 
In this range of mass ratio, the ML should barely occur in systems, according to equation (\ref{Eq_Beta_Final}). 
Thus, the fill-out factor is expected to be constant or to slightly change. 

In the $q$--$J$ relation (Fig. \ref{Fig_qft2-J}a), below $q \sim 0.4$, as the mass ratio becomes small, the total angular momentum also becomes small. 
As mentioned above, the results indicate that a system with a small mass-ratio exhibits a large ML rate. 
Therefore, it is surmised that the ML carries away a large amount of angular momentum and the total angular momentum becomes small. 
A similar situation is also found in the $f$--$J$ relation (Fig. \ref{Fig_qft2-J}b), in which systems with higher fill-out factors have smaller total angular momenta above $f\sim0.4$. 

Moreover, only for the secondaries of the An sample, the mass is negatively and strongly correlated with the fill-out factor (Fig. \ref{Fig_f-M}). 
This trend can be interpreted as the result that a system with a higher fill-out factor loses more matter from the \Ls  point; the secondary mass tends to be small. 

\subsubsection{Plausible processes for interpreting the negative period variations}\label{Sec_Disc_Processes}
The MTML should be affected by the primary's parameters relative to the secondary's (Section \ref{Sec_Cor_MEML}). 
Furthermore, the primary radius has the closest correlation with the MTML rate; its power-law exponent is $-3.2$ or $-2.0$ [equation (\ref{Eq_MEML_depend})]. 
\citet{Kouzuma2018-PASJ} found that the MTML rate is negatively correlated with the primary mass of contact binaries within $M_1 \gtrsim1.2$ $\Msol$. 
This trend is consistent with the derived inverse relationship between $R_1$ and $|\dot{M}_{12}|$, assuming that $M_1$ is roughly approximated to $R_1$ in W UMa systems. 

The $R_1$--$\dot{M}_{12}$ relationships indicate that as the primary radius becomes large, the MTML rate becomes small. 
The $q$--$R_1$ relation (Fig. \ref{Fig_aq-R}b) shows that primaries with larger radii have smaller mass ratios. 
According to equation (\ref{Eq_Beta_Final}), An systems with smaller mass ratios tend to have larger ML rates relative to the MTML rates. 
Therefore, it is surmised that a large amount of ML leads to a small amount of MTML. 
This tendency also agrees with the positive correlation in the $q$--$\dot{M}_{12}$ relation (Table \ref{Tab_Pcor_MEML}), in which the MTML rate decreases with decreasing mass ratio. 

The rate of extra AML ($K$) should be affected by the fill-out factor and secondary's parameters (Section \ref{Sec_Cor_aAML}). 
Furthermore, the fill-out factor has the closest correlation with $K$; its power-law exponents are 1.1 or 0.7 [equation (\ref{Eq_K_depend})]. 
As shown in Section \ref{Sec_Cor_aAML}, the fill-out factor has a contribution opposite to the seondary's parameters. 
Figure \ref{Fig_aq-M}b shows that as $M_2$ increases, $q$ becomes high. 
In general, a binary with a higher mass-ratio tends to have a thicker layer between the inner and outer Roche lobes. 
Such a binary with a thick layer is likely to have a small fill-out factor. 
This situation is also directly verified in the negative $f$--$M_2$ relationship (Fig. \ref{Fig_f-M}). 
Therefore, it is deduced that the first PLS component represents the thickness between the inner lobe and stellar surface. 

Magnetic braking via stellar wind is a typical mechanism for the extra AML. 
Most component stars in the An sample have spectral types later than F5 (Section \ref{Sec_Evol}). 
Accordingly, they are expected to have, at least weak, magnetic activities; the magnetic braking can occur. 
However, \citet{Rucinski1982-AA} deduced that the AML via magnetic wind is expected to have weak dependence on the degree of contact. 
If it is true, the magnetic braking via stellar wind is implausible for describing the derived $f$--$K$ relationships and cannot be a dominant process for the AML. 
Moreover, the magnetic braking is generally expressed as a function of the stellar mass, stellar radius, and angular velocity \citep[e.g., ][]{Verbunt1981-AA,Stepien2006-AcA347}. 
According to such formulae, the primary should contribute to the extra AML more than the secondary.
However, our results suggest that the secondary, rather than the primary, affects the extra AML. 
Thus, it is difficult to describe the results by the magnetic braking. 

As shown in Section \ref{Sec_Disc_associations}, the ML from the \Ls point should cause a large amount of AML. 
If matter lost from the system carries away the specific angular momentum of the system, extra AML occurs and results in losing a large amount of AML. 
Theoretical studies have considered such a scenario for unraveling the evolution of binary systems \citep[e.g.,][]{Shu1979-ApJ,Pejcha2016-MNRAS2527,MacLeod2018-ApJ5}. 
Therefore, for the An sample, it is deduced that the extra AML is closely associated with the ML from the \Ls  point. 

The histograms in Fig. \ref{Fig_Hist} indicate that the above situation tends to arise when an A-type system has a relatively large temperature-difference (i.e., $|\Delta T| \gtrsim100$ K) and a relatively small total mass (i.e., $\Mtot \lesssim 2$ $\Msol$). 
The small amount of total mass may be due to a large amount of the ML described in this section. 
These characteristics may be a hint for interpreting the properties of the mass transfer and AML.

\subsection{Positive period variation}\label{Sec_Disc_PPC}
The MTLM is likely to be affected by the radiation of stars (Section \ref{Sec_Cor_MELM}). 
Furthermore, the luminosity ratio and mass ratio have the closest correlations with the MTLM rate; their power-law exponents are approximately equal. 
Both parameters equally affect each other, and in this work it is difficult to identify which parameter is more essential. 
In W UMa systems, the luminosity ratio is roughly proportional to the mass ratio \citep{Osaki1965-PASJ, Lucy1968-ApJ877}, which presumably causes the above situation. 
\citet{Kouzuma2018-PASJ} also derived a positive association between $q$ and the MTLM rate, which is consistent with the positive relationship in this study. 
A large luminosity ratio indicates that the secondary is brighter than the primary. 
Therefore, the relationships between $L_2/L_1$ and $\dot{M}_{12}$ indicate that radiation pressure should drive the MTLM process. 
This may be associated with the finding that the temperatures of two components tend to be relatively close (typically $|\Delta T|<100$ K), as shown in Fig. \ref{Fig_Hist}a. 

The ML should be affected by the temperature and luminosity of each component (Section \ref{Sec_Cor_ML}). 
Furthermore, the secondary temperature has the closest correlation with the ML rate; 
its power-law exponent is $-14.8$ or $-10.3$ [equation (\ref{Eq_ML_depend})]. 
According to these relationships, more rapid ML tends to occur in systems with lower temperatures, which results in a large amount of AML. 
In the $T_2$--$J$ relation (Fig. \ref{Fig_qft2-J}c), as the secondary temperature decreases, the total angular momentum also tends to decrease. 
Therefore, the positive correlation in the $T_2$--$J$ relation is consistent with the $T_2$--$\dot{M}_\mathrm{b}$ relationships. 

\citet{Cranmer2011-ApJ} theoretically predicted the dependence of the rate of ML via stellar wind on the effective temperature ($T_\mathrm{eff}$) and rotation period for main-sequence stars. 
According to their predictions, the ML rate decreases with increasing effective temperature above $T_\mathrm{eff}\sim$ 6000--6500 K. 
The $T_2$--$\dot{\Mtot}$ relationship in this study agrees with their inverse association. 
This agreement supports that the ML in the Ap systems should be predominantly due to the stellar wind. 

As discussed by \citet{Mochnacki1981-ApJ}, the orbital period decreases when the relative AML rate is at least $5/3$ times larger than the relative ML rate (i.e., $|\dot{J}/J|>5/3|\dot{\Mtot}/\Mtot|$). 
If the ML actually occurs in the Ap systems, the AML due to the ML should be sufficiently small, that is, the relative AML rate is $5/3$ times smaller than the relative ML one. 
For instance, a spherical wind leads to widening the separation \citep{MacLeod2018-ApJ5}. 
As a consequence, extra AML, such as the magnetic braking via stellar winds and carrying away the specific angular momentum of the system, should work ineffectively in the Ap systems. 

The histograms in Fig. \ref{Fig_Hist} show that the Ap systems have relatively small temperature-difference (i.e., $|\Delta T|\lesssim 100$ K) and high total mass (i.e., $\Mtot \gtrsim 2$ $\Msol$). 
Furthermore, Ap systems tend to have higher temperatures than An systems, that is, roughly earlier than F5 (Section \ref{Sec_Evol}). 
It suggests that the Ap systems can have weaker magnetic activity than the An systems. 
This agrees with the above prediction that the extra AML is ineffective. 
The differences between the An and Ap samples mentioned here may be a clue for unraveling the properties of the MTLM and ML. 

Another notable feature is that several associations in Section \ref{Sec_Prop_Asso} show trends differing between below and above $a \sim 3.5$ $\Rsol$: in the relations of $a$ with $M_1$, $M_2$, $q$, and $f$ (Figs. \ref{Fig_aq-M}a, \ref{Fig_a-q}, and \ref{Fig_qa-f}b). 
However, such a trend is not found in the $a$--$R$ relations (Fig. \ref{Fig_aq-R}a). 
Moreover, the An sample shows no such trend. 
The Ap systems with $a>3.5$ $\Rsol$ have masses higher than 1.8 $\Msol$, and they are more evolved than the other Ap systems (Section \ref{Sec_Evol}).
These differences need to be further investigated.

\section{Summary and conclusions}\label{Sec_Summary}
Sample A-type W UMa systems with monotonic orbital-period variations were collected from the literature. 
Using their well-defined binary parameters, we have investigated which parameters are genuinely correlated with the physical processes causing the period variations. 
The properties of the sample systems have also been examined by analyzing associations between parameters. 
The main results of this study are summarized as follows. 

For A-type W UMa systems with negative period variations:  
\begin{enumerate}
\item The mass transfer from more- to less-massive stars (MTML) is mainly affected by primary's parameters, particularly those relative to secondary's (Section \ref{Sec_Cor_MEML}). 
\item Subsequently, the temperature difference and the thickness between the inner Roche lobe and stellar surface may also affect the MTML (Section \ref{Sec_Cor_MEML}). 
\item The MTML rate is proportional to $R_1^{-3.2 \pm 0.4}$ or $R_1^{-2.0 \pm 0.4}$ (Section \ref{Sec_Cor_MEML}). 
\item The ML rate relative to the MTML rate decreases with increasing mass ratio: $\beta=1-(2.16 \pm 0.34) q$ (Section \ref{Sec_Disc_ratio_ME_ML}). 
\item The ML should occur below $q=0.46 \pm 0.08$ (Section \ref{Sec_Disc_ratio_ME_ML}). 
\item The ML is likely to occur from the \Ls  point (Sections \ref{Sec_Disc_ratio_ME_ML} and \ref{Sec_Disc_associations}). 
\item The extra AML is mainly affected by the thickness between the inner Roche lobe and stellar surface (Sections \ref{Sec_Cor_aAML} and \ref{Sec_Disc_Processes}). 
\item Subsequently, the temperature difference may also affect the extra AML (Section \ref{Sec_Cor_aAML}). 
\item The rate of extra AML is proportional to $f^{1.1 \pm 0.1}$ or $f^{0.7 \pm 0.1}$ (Section \ref{Sec_Cor_aAML}). 
\item The AML should predominantly occur by the ML from the \Ls  point, which also carries away specific angular momentum from the system (Section \ref{Sec_Disc_Processes}). 
\end{enumerate}

For A-type W UMa systems with positive period variations: 
\begin{enumerate}
\item The mass transfer from less- to more-massive stars (MTLM) is mainly affected by the radiation of the primary relative to the secondary (Section \ref{Sec_Cor_MELM}). 
\item Subsequently, secondary's and system's parameters may also affect the MTLM. 
\item The MTLM rate is proportional to $(L_2/L_1)^{2.3 \pm 0.2}$ and/or $q^{2.3 \pm 0.2}$ or $(L_2/L_1)^{1.6 \pm 0.3}$ and/or $q^{1.7 \pm 0.2}$ (Section \ref{Sec_Cor_MELM}). 
\item Radiation pressure should drive the MTLM process (Section \ref{Sec_Disc_PPC}). 
\item The ML is mainly affected by the temperature and luminosity of each component star (Section \ref{Sec_Cor_ML}). 
\item The ML rate is proportional to $T_2^{-14.8 \pm 2.6}$ or $T_2^{-10.3 \pm 2.2}$ (Section \ref{Sec_Cor_ML}). 
\item The ML should occur via stellar wind, which differs from the dominant ML process in A-type systems with negative period variations (Section \ref{Sec_Disc_PPC}). 
\item The trends in the relations of $a$ with $M_1$, $M_2$, $q$, and $f$ differ between below and above $a\sim3.5$ $\Rsol$ (Figs. \ref{Fig_aq-M}a, \ref{Fig_a-q}, and \ref{Fig_qa-f}b). 
\end{enumerate}

Several differences between A-type systems with negative (An) and positive (Ap) period variations are found: 
\begin{enumerate}
\item The An sample systems tend to have higher temperature-differences than the Ap ones; the histograms of $|\Delta T|$ for the An and Ap display distributions with peaks around $|\Delta T|=150$--$200$ K and $|\Delta T|=0$--$50$ K, respectively (Fig. \ref{Fig_Hist}a). 
\item In general, most of the An and Ap systems have $\Mtot<2$ $\Msol$ and $\Mtot>2$ $\Msol$, respectively (Fig. \ref{Fig_Hist}b). 
\item The histograms of the fill-out factor for the An and Ap samples appear to display different distributions (Fig. \ref{Fig_Hist}c). 
\item Only the An sample shows clear associations in the $f$--$M_2$ and $f$--$J$ relations (Figs. \ref{Fig_f-M} and \ref{Fig_qft2-J}). 
\item Only the Ap sample shows clear associations in the $a$--$q$, $a$--$f$, and $T_2$--$J$ relations (Figs. \ref{Fig_a-q}, \ref{Fig_qa-f}b, and \ref{Fig_qft2-J}c). 
\end{enumerate}

These results can help to examine and construct theoretical models on the evolution of W UMa systems. 
This paper focuses only on A-type W UMa contact systems. 
Other types of close binaries, such as W-type W UMa and semi-detached systems, need to be also investigated. 
Comparative studies should clarify the characteristics of mass transfer and AML, and provide clues about the evolution of close binary systems.

\section*{Acknowledgements}
This work was supported by the Chukyo University Grant for Overseas Research Program, 2018. 
The author is deeply grateful to David Montes for supporting my stay in Spain. 
The author would like to thank the anonymous referee for constructive comments and suggestions that helped improve the manuscript.

\newcommand*\aap{A\&A}
\let\astap=\aap
\newcommand*\aapr{A\&A~Rev.}
\newcommand*\aaps{A\&AS}
\newcommand*\actaa{Acta Astron.}
\newcommand*\aj{AJ}
\newcommand*\ao{Appl.~Opt.}
\let\applopt\ao
\newcommand*\apj{ApJ}
\newcommand*\apjl{ApJ}
\let\apjlett\apjl
\newcommand*\apjs{ApJS}
\let\apjsupp\apjs
\newcommand*\aplett{Astrophys.~Lett.}
\newcommand*\apspr{Astrophys.~Space~Phys.~Res.}
\newcommand*\apss{Ap\&SS}
\newcommand*\araa{ARA\&A}
\newcommand*\azh{AZh}
\newcommand*\baas{BAAS}
\newcommand*\bac{Bull. astr. Inst. Czechosl.}
\newcommand*\bain{Bull.~Astron.~Inst.~Netherlands}
\newcommand*\caa{Chinese Astron. Astrophys.}
\newcommand*\cjaa{Chinese J. Astron. Astrophys.}
\newcommand*\fcp{Fund.~Cosmic~Phys.}
\newcommand*\gca{Geochim.~Cosmochim.~Acta}
\newcommand*\grl{Geophys.~Res.~Lett.}
\newcommand*\iaucirc{IAU~Circ.}
\newcommand*\icarus{Icarus}
\newcommand*\jcap{J. Cosmology Astropart. Phys.}
\newcommand*\jcp{J.~Chem.~Phys.}
\newcommand*\jgr{J.~Geophys.~Res.}
\newcommand*\jqsrt{J.~Quant.~Spectr.~Rad.~Transf.}
\newcommand*\jrasc{JRASC}
\newcommand*\memras{MmRAS}
\newcommand*\memsai{Mem.~Soc.~Astron.~Italiana}
\newcommand*\mnras{MNRAS}
\newcommand*\na{New A}
\newcommand*\nar{New A Rev.}
\newcommand*\nat{Nature}
\newcommand*\nphysa{Nucl.~Phys.~A}
\newcommand*\pasa{PASA}
\newcommand*\pasj{PASJ}
\newcommand*\pasp{PASP}
\newcommand*\physrep{Phys.~Rep.}
\newcommand*\physscr{Phys.~Scr}
\newcommand*\planss{Planet.~Space~Sci.}
\newcommand*\pra{Phys.~Rev.~A}
\newcommand*\prb{Phys.~Rev.~B}
\newcommand*\prc{Phys.~Rev.~C}
\newcommand*\prd{Phys.~Rev.~D}
\newcommand*\pre{Phys.~Rev.~E}
\newcommand*\prl{Phys.~Rev.~Lett.}
\newcommand*\procspie{Proc.~SPIE}
\newcommand*\qjras{QJRAS}
\newcommand*\rmxaa{Rev. Mexicana Astron. Astrofis.}
\newcommand*\skytel{S\&T}
\newcommand*\solphys{Sol.~Phys.}
\newcommand*\sovast{Soviet~Ast.}
\newcommand*\ssr{Space~Sci.~Rev.}
\newcommand*\zap{ZAp}

\bibliographystyle{elsarticle-harv}

\begin{thebibliography}{106}
\expandafter\ifx\csname natexlab\endcsname\relax\def\natexlab#1{#1}\fi
\providecommand{\url}[1]{\texttt{#1}}
\providecommand{\href}[2]{#2}
\providecommand{\path}[1]{#1}
\providecommand{\DOIprefix}{doi:}
\providecommand{\ArXivprefix}{arXiv:}
\providecommand{\URLprefix}{URL: }
\providecommand{\Pubmedprefix}{pmid:}
\providecommand{\doi}[1]{\href{http://dx.doi.org/#1}{\path{#1}}}
\providecommand{\Pubmed}[1]{\href{pmid:#1}{\path{#1}}}
\providecommand{\bibinfo}[2]{#2}
\ifx\xfnm\relax \def\xfnm[#1]{\unskip,\space#1}\fi
\bibitem[{Andersen and Bro(2010)}]{Anderson2010-JC}
\bibinfo{author}{Andersen, C.M.}, \bibinfo{author}{Bro, R.},
  \bibinfo{year}{2010}.
\newblock \bibinfo{title}{Variable selection in regression—a tutorial}.
\newblock \bibinfo{journal}{Journal of Chemometrics} \bibinfo{volume}{24},
  \bibinfo{pages}{728--737}.
\newblock \URLprefix
  \url{https://analyticalsciencejournals.onlinelibrary.wiley.com/doi/abs/10.1002/cem.1360},
  \DOIprefix\doi{https://doi.org/10.1002/cem.1360},
  \href{http://arxiv.org/abs/https://analyticalsciencejournals.onlinelibrary.wiley.com/doi/pdf/10.1002/cem.1360}{{\tt
  arXiv:https://analyticalsciencejournals.onlinelibrary.wiley.com/doi/pdf/10.1002/cem.1360}}.
\bibitem[{{Applegate}(1992)}]{Applegate1992-ApJ}
\bibinfo{author}{{Applegate}, J.H.}, \bibinfo{year}{1992}.
\newblock \bibinfo{title}{{A mechanism for orbital period modulation in close
  binaries}}.
\newblock \bibinfo{journal}{\apj} \bibinfo{volume}{385},
  \bibinfo{pages}{621--629}.
\newblock \DOIprefix\doi{10.1086/170967}.
\bibitem[{{Bell} et~al.(1990){Bell}, {Rainger} and
  {Hilditch}}]{Bell1990-MNRAS632}
\bibinfo{author}{{Bell}, S.A.}, \bibinfo{author}{{Rainger}, P.P.},
  \bibinfo{author}{{Hilditch}, R.W.}, \bibinfo{year}{1990}.
\newblock \bibinfo{title}{{Spots on AG Virginis - Paradigm or panacea?}}
\newblock \bibinfo{journal}{\mnras} \bibinfo{volume}{247},
  \bibinfo{pages}{632--646}.
\bibitem[{{Binnendijk}(1970)}]{Binnendijk1970-VA}
\bibinfo{author}{{Binnendijk}, L.}, \bibinfo{year}{1970}.
\newblock \bibinfo{title}{{The orbital elements of W Ursae Majoris systems}}.
\newblock \bibinfo{journal}{Vistas in Astronomy} \bibinfo{volume}{12},
  \bibinfo{pages}{217--256}.
\newblock \DOIprefix\doi{10.1016/0083-6656(70)90041-3}.
\bibitem[{Chong and Jun(2005)}]{Chong2005-CILS}
\bibinfo{author}{Chong, I.G.}, \bibinfo{author}{Jun, C.H.},
  \bibinfo{year}{2005}.
\newblock \bibinfo{title}{Performance of some variable selection methods when
  multicollinearity is present}.
\newblock \bibinfo{journal}{Chemometrics and intelligent laboratory systems}
  \bibinfo{volume}{78}, \bibinfo{pages}{103--112}.
\bibitem[{{Cox}(2000)}]{Cox2000-book}
\bibinfo{author}{{Cox}, A.N.}, \bibinfo{year}{2000}.
\newblock \bibinfo{title}{{Allen's astrophysical quantities}}.
\bibitem[{{Cranmer} and {Saar}(2011)}]{Cranmer2011-ApJ}
\bibinfo{author}{{Cranmer}, S.R.}, \bibinfo{author}{{Saar}, S.H.},
  \bibinfo{year}{2011}.
\newblock \bibinfo{title}{{Testing a Predictive Theoretical Model for the Mass
  Loss Rates of Cool Stars}}.
\newblock \bibinfo{journal}{\apj} \bibinfo{volume}{741}, \bibinfo{pages}{54}.
\newblock \DOIprefix\doi{10.1088/0004-637X/741/1/54},
  \href{http://arxiv.org/abs/1108.4369}{{\tt arXiv:1108.4369}}.
\bibitem[{{Djurasevic}(1992)}]{Djurasevic1992-ApSS}
\bibinfo{author}{{Djurasevic}, G.}, \bibinfo{year}{1992}.
\newblock \bibinfo{title}{{An analysis of active close binaries (CB) based on
  photometric measurements. I - A model of active CB with spots on the
  components. II - Active CB with accretion discs}}.
\newblock \bibinfo{journal}{\apss} \bibinfo{volume}{196},
  \bibinfo{pages}{241--265}.
\newblock \DOIprefix\doi{10.1007/BF00692893}.
\bibitem[{{Eggleton}(1983)}]{Eggleton1983-ApJ}
\bibinfo{author}{{Eggleton}, P.P.}, \bibinfo{year}{1983}.
\newblock \bibinfo{title}{{Aproximations to the radii of Roche lobes.}}
\newblock \bibinfo{journal}{\apj} \bibinfo{volume}{268},
  \bibinfo{pages}{368--369}.
\newblock \DOIprefix\doi{10.1086/160960}.
\bibitem[{{Ekmek{\c c}i} et~al.(2012){Ekmek{\c c}i}, {Elmasl{\i}}, {Y{\i}lmaz},
  {K{\i}l{\i}{\c c}o{\u g}lu}, {Tanr{\i}verdi}, {Ba{\c s}t{\"u}rk}, {{\c
  S}enavc{\i}}, {{\c C}al{\i}{\c s}kan}, {Albayrak} and
  {Selam}}]{Ekmekci2012-NewA}
\bibinfo{author}{{Ekmek{\c c}i}, F.}, \bibinfo{author}{{Elmasl{\i}}, A.},
  \bibinfo{author}{{Y{\i}lmaz}, M.}, \bibinfo{author}{{K{\i}l{\i}{\c c}o{\u
  g}lu}, T.}, \bibinfo{author}{{Tanr{\i}verdi}, T.}, \bibinfo{author}{{Ba{\c
  s}t{\"u}rk}, {\"O}.}, \bibinfo{author}{{{\c S}enavc{\i}}, H.V.},
  \bibinfo{author}{{{\c C}al{\i}{\c s}kan}, {\c S}.},
  \bibinfo{author}{{Albayrak}, B.}, \bibinfo{author}{{Selam}, S.O.},
  \bibinfo{year}{2012}.
\newblock \bibinfo{title}{{Physical parameters of some close binaries: ET Boo,
  V1123 Tau, V1191 Cyg, V1073 Cyg and V357 Peg}}.
\newblock \bibinfo{journal}{\na} \bibinfo{volume}{17},
  \bibinfo{pages}{603--609}.
\newblock \DOIprefix\doi{10.1016/j.newast.2012.03.001},
  \href{http://arxiv.org/abs/1203.6189}{{\tt arXiv:1203.6189}}.
\bibitem[{{Essam} et~al.(2010){Essam}, {Saad}, {Nouh}, {Dumitrescu},
  {El-Khateeb} and {Haroon}}]{Essam2010-NewA}
\bibinfo{author}{{Essam}, A.}, \bibinfo{author}{{Saad}, S.M.},
  \bibinfo{author}{{Nouh}, M.I.}, \bibinfo{author}{{Dumitrescu}, A.},
  \bibinfo{author}{{El-Khateeb}, M.M.}, \bibinfo{author}{{Haroon}, A.},
  \bibinfo{year}{2010}.
\newblock \bibinfo{title}{{Photometric and spectroscopic analysis of YY CrB}}.
\newblock \bibinfo{journal}{\na} \bibinfo{volume}{15},
  \bibinfo{pages}{227--233}.
\newblock \DOIprefix\doi{10.1016/j.newast.2009.07.006}.
\bibitem[{{Gazeas} and {St{\k{e}}pie{\'n}}(2008)}]{Gazeas2008-MNRAS}
\bibinfo{author}{{Gazeas}, K.}, \bibinfo{author}{{St{\k{e}}pie{\'n}}, K.},
  \bibinfo{year}{2008}.
\newblock \bibinfo{title}{{Angular momentum and mass evolution of contact
  binaries}}.
\newblock \bibinfo{journal}{\mnras} \bibinfo{volume}{390},
  \bibinfo{pages}{1577--1586}.
\newblock \DOIprefix\doi{10.1111/j.1365-2966.2008.13844.x},
  \href{http://arxiv.org/abs/0803.0212}{{\tt arXiv:0803.0212}}.
\bibitem[{{Gazeas} et~al.(2006){Gazeas}, {Niarchos}, {Zola}, {Kreiner} and
  {Rucinski}}]{Gazeas2006-AcA}
\bibinfo{author}{{Gazeas}, K.D.}, \bibinfo{author}{{Niarchos}, P.G.},
  \bibinfo{author}{{Zola}, S.}, \bibinfo{author}{{Kreiner}, J.M.},
  \bibinfo{author}{{Rucinski}, S.M.}, \bibinfo{year}{2006}.
\newblock \bibinfo{title}{{Physical Parameters of Components in Close Binary
  Systems: VI}}.
\newblock \bibinfo{journal}{\actaa} \bibinfo{volume}{56},
  \bibinfo{pages}{127--143}.
\newblock \href{http://arxiv.org/abs/0903.1364}{{\tt arXiv:0903.1364}}.
\bibitem[{{G{\"u}rol}(2005)}]{Gurol2005-NewA}
\bibinfo{author}{{G{\"u}rol}, B.}, \bibinfo{year}{2005}.
\newblock \bibinfo{title}{{Long term photometric and period study of AU
  Serpentis}}.
\newblock \bibinfo{journal}{\na} \bibinfo{volume}{10},
  \bibinfo{pages}{653--675}.
\newblock \DOIprefix\doi{10.1016/j.newast.2005.04.004}.
\bibitem[{{G{\"u}rol} and {M{\"u}yessero{\u g}lu}(2005)}]{Gurol2005-AN}
\bibinfo{author}{{G{\"u}rol}, B.}, \bibinfo{author}{{M{\"u}yessero{\u g}lu},
  Z.}, \bibinfo{year}{2005}.
\newblock \bibinfo{title}{{First light curve and period study of LO
  Andromedae}}.
\newblock \bibinfo{journal}{Astronomische Nachrichten} \bibinfo{volume}{326},
  \bibinfo{pages}{43}.
\newblock \DOIprefix\doi{10.1002/asna.200410338}.
\bibitem[{{Hilditch} et~al.(1989){Hilditch}, {King} and
  {McFarlane}}]{Hilditch1989-MNRAS}
\bibinfo{author}{{Hilditch}, R.W.}, \bibinfo{author}{{King}, D.J.},
  \bibinfo{author}{{McFarlane}, T.M.}, \bibinfo{year}{1989}.
\newblock \bibinfo{title}{{Contact and near-contact binary systems. X - The
  contact system TV MUSCAE}}.
\newblock \bibinfo{journal}{\mnras} \bibinfo{volume}{237},
  \bibinfo{pages}{447--459}.
\newblock \DOIprefix\doi{10.1093/mnras/237.2.447}.
\bibitem[{{Irwin}(1959)}]{Irwin1959-AJ}
\bibinfo{author}{{Irwin}, J.B.}, \bibinfo{year}{1959}.
\newblock \bibinfo{title}{{Standard light-time curves}}.
\newblock \bibinfo{journal}{\aj} \bibinfo{volume}{64}, \bibinfo{pages}{149}.
\newblock \DOIprefix\doi{10.1086/107913}.
\bibitem[{{Isobe} et~al.(1990){Isobe}, {Feigelson}, {Akritas} and
  {Babu}}]{Isobe1990-ApJ}
\bibinfo{author}{{Isobe}, T.}, \bibinfo{author}{{Feigelson}, E.D.},
  \bibinfo{author}{{Akritas}, M.G.}, \bibinfo{author}{{Babu}, G.J.},
  \bibinfo{year}{1990}.
\newblock \bibinfo{title}{{Linear Regression in Astronomy. I.}}
\newblock \bibinfo{journal}{\apj} \bibinfo{volume}{364}, \bibinfo{pages}{104}.
\newblock \DOIprefix\doi{10.1086/169390}.
\bibitem[{{King} and {Hilditch}(1984)}]{King1984-MNRAS}
\bibinfo{author}{{King}, D.J.}, \bibinfo{author}{{Hilditch}, R.W.},
  \bibinfo{year}{1984}.
\newblock \bibinfo{title}{{Contact and near-contact binary systems. II - RR
  Cen, EZ Hya, V502 OPH and RS SCT}}.
\newblock \bibinfo{journal}{\mnras} \bibinfo{volume}{209},
  \bibinfo{pages}{645--653}.
\newblock \DOIprefix\doi{10.1093/mnras/209.3.645}.
\bibitem[{{Kouzuma}(2018)}]{Kouzuma2018-PASJ}
\bibinfo{author}{{Kouzuma}, S.}, \bibinfo{year}{2018}.
\newblock \bibinfo{title}{{Mass-transfer properties of overcontact systems in
  the Kepler eclipsing binary catalog}}.
\newblock \bibinfo{journal}{\pasj} \bibinfo{volume}{70}, \bibinfo{pages}{90}.
\newblock \DOIprefix\doi{10.1093/pasj/psy086},
  \href{http://arxiv.org/abs/1808.06401}{{\tt arXiv:1808.06401}}.
\bibitem[{{Kuiper}(1941)}]{Kuiper1941-ApJ}
\bibinfo{author}{{Kuiper}, G.P.}, \bibinfo{year}{1941}.
\newblock \bibinfo{title}{{On the Interpretation of {\ensuremath{\beta}} Lyrae
  and Other Close Binaries.}}
\newblock \bibinfo{journal}{\apj} \bibinfo{volume}{93}, \bibinfo{pages}{133}.
\newblock \DOIprefix\doi{10.1086/144252}.
\bibitem[{{Kwee}(1958)}]{Kwee1958-BAN}
\bibinfo{author}{{Kwee}, K.K.}, \bibinfo{year}{1958}.
\newblock \bibinfo{title}{{Investigation of variations in the period of sixteen
  bright short-period eclipsing binary stars}}.
\newblock \bibinfo{journal}{\bain} \bibinfo{volume}{14}, \bibinfo{pages}{131}.
\bibitem[{{Lapasset} and {Gomez}(1990)}]{Lapasset1990-AA}
\bibinfo{author}{{Lapasset}, E.}, \bibinfo{author}{{Gomez}, M.},
  \bibinfo{year}{1990}.
\newblock \bibinfo{title}{{Simultaneous analysis of light and radial velocity
  curves of the peculiar contact system V 508 Ophiuchi}}.
\newblock \bibinfo{journal}{\aap} \bibinfo{volume}{231},
  \bibinfo{pages}{365--374}.
\bibitem[{{Lee} et~al.(2007){Lee}, {Kim} and {Kim}}]{Lee2007-PASP}
\bibinfo{author}{{Lee}, J.W.}, \bibinfo{author}{{Kim}, H.I.},
  \bibinfo{author}{{Kim}, S.L.}, \bibinfo{year}{2007}.
\newblock \bibinfo{title}{{A Period Study and Spot Model for the Eclipsing
  Binary TU Bootis}}.
\newblock \bibinfo{journal}{\pasp} \bibinfo{volume}{119},
  \bibinfo{pages}{1099--1107}.
\newblock \DOIprefix\doi{10.1086/521605}.
\bibitem[{{Lee} and {Park}(2018)}]{Lee2018-PASP}
\bibinfo{author}{{Lee}, J.W.}, \bibinfo{author}{{Park}, J.H.},
  \bibinfo{year}{2018}.
\newblock \bibinfo{title}{{Physical Nature and Orbital Behavior of the
  Eclipsing System UZ Leonis}}.
\newblock \bibinfo{journal}{\pasp} \bibinfo{volume}{130},
  \bibinfo{pages}{034201}.
\newblock \DOIprefix\doi{10.1088/1538-3873/aaa390},
  \href{http://arxiv.org/abs/1712.07864}{{\tt arXiv:1712.07864}}.
\bibitem[{{Lee} et~al.(2015){Lee}, {Youn}, {Park} and {Wolf}}]{Lee2015-AJ}
\bibinfo{author}{{Lee}, J.W.}, \bibinfo{author}{{Youn}, J.H.},
  \bibinfo{author}{{Park}, J.H.}, \bibinfo{author}{{Wolf}, M.},
  \bibinfo{year}{2015}.
\newblock \bibinfo{title}{{The Physical Nature and Orbital Behavior of the
  Eclipsing System DK Cygni}}.
\newblock \bibinfo{journal}{\aj} \bibinfo{volume}{149}, \bibinfo{pages}{194}.
\newblock \DOIprefix\doi{10.1088/0004-6256/149/6/194},
  \href{http://arxiv.org/abs/1504.03752}{{\tt arXiv:1504.03752}}.
\bibitem[{{Li} et~al.(2016){Li}, {Wei}, {Yang} and {Dai}}]{Li2016-RAA}
\bibinfo{author}{{Li}, H.L.}, \bibinfo{author}{{Wei}, J.Y.},
  \bibinfo{author}{{Yang}, Y.G.}, \bibinfo{author}{{Dai}, H.F.},
  \bibinfo{year}{2016}.
\newblock \bibinfo{title}{{OO Aquilae: a solar-type contact binary with
  intrinsic light curve changes}}.
\newblock \bibinfo{journal}{Research in Astronomy and Astrophysics}
  \bibinfo{volume}{16}, \bibinfo{pages}{2}.
\newblock \DOIprefix\doi{10.1088/1674-4527/16/1/002}.
\bibitem[{{Liu}(2021)}]{Liu2021-PASP}
\bibinfo{author}{{Liu}, L.}, \bibinfo{year}{2021}.
\newblock \bibinfo{title}{{Error Analysis of the Light Curve Solution of
  Contact Binaries Based on the W-D Code}}.
\newblock \bibinfo{journal}{\pasp} \bibinfo{volume}{133},
  \bibinfo{pages}{084202}.
\newblock \DOIprefix\doi{10.1088/1538-3873/ac1ac1},
  \href{http://arxiv.org/abs/2109.02807}{{\tt arXiv:2109.02807}}.
\bibitem[{{Lu}(1986)}]{Lu1986-PASP}
\bibinfo{author}{{Lu}, W.}, \bibinfo{year}{1986}.
\newblock \bibinfo{title}{{The spectroscopic orbit of the W Ursae Majoris
  system V508 Ophiuchi}}.
\newblock \bibinfo{journal}{\pasp} \bibinfo{volume}{98},
  \bibinfo{pages}{577--580}.
\newblock \DOIprefix\doi{10.1086/131797}.
\bibitem[{{Lu} and {Rucinski}(1999)}]{Lu1999-AJ}
\bibinfo{author}{{Lu}, W.}, \bibinfo{author}{{Rucinski}, S.M.},
  \bibinfo{year}{1999}.
\newblock \bibinfo{title}{{Radial Velocity Studies of Close Binary Stars. I.}}
\newblock \bibinfo{journal}{\aj} \bibinfo{volume}{118},
  \bibinfo{pages}{515--526}.
\newblock \DOIprefix\doi{10.1086/300933},
  \href{http://arxiv.org/abs/astro-ph/9902168}{{\tt arXiv:astro-ph/9902168}}.
\bibitem[{{Lucy}(1968a)}]{Lucy1968-ApJ877}
\bibinfo{author}{{Lucy}, L.B.}, \bibinfo{year}{1968}a.
\newblock \bibinfo{title}{{The Light Curves of W Ursae Majoris Stars}}.
\newblock \bibinfo{journal}{\apj} \bibinfo{volume}{153}, \bibinfo{pages}{877}.
\newblock \DOIprefix\doi{10.1086/149712}.
\bibitem[{{Lucy}(1968b)}]{Lucy1968-ApJ1123}
\bibinfo{author}{{Lucy}, L.B.}, \bibinfo{year}{1968}b.
\newblock \bibinfo{title}{{The Structure of Contact Binaries}}.
\newblock \bibinfo{journal}{\apj} \bibinfo{volume}{151}, \bibinfo{pages}{1123}.
\newblock \DOIprefix\doi{10.1086/149510}.
\bibitem[{{MacLeod} et~al.(2018){MacLeod}, {Ostriker} and
  {Stone}}]{MacLeod2018-ApJ5}
\bibinfo{author}{{MacLeod}, M.}, \bibinfo{author}{{Ostriker}, E.C.},
  \bibinfo{author}{{Stone}, J.M.}, \bibinfo{year}{2018}.
\newblock \bibinfo{title}{{Runaway Coalescence at the Onset of Common Envelope
  Episodes}}.
\newblock \bibinfo{journal}{\apj} \bibinfo{volume}{863}, \bibinfo{pages}{5}.
\newblock \DOIprefix\doi{10.3847/1538-4357/aacf08},
  \href{http://arxiv.org/abs/1803.03261}{{\tt arXiv:1803.03261}}.
\bibitem[{{McLean}(1981)}]{McLean1981-MNRAS}
\bibinfo{author}{{McLean}, B.J.}, \bibinfo{year}{1981}.
\newblock \bibinfo{title}{{Radial velocities for contact binary systems. I - W
  Ursae Majoris and AW Ursae Majoris}}.
\newblock \bibinfo{journal}{\mnras} \bibinfo{volume}{195},
  \bibinfo{pages}{931--938}.
\newblock \DOIprefix\doi{10.1093/mnras/195.4.931}.
\bibitem[{{Mestel}(1968)}]{Mestel1968-MNRAS}
\bibinfo{author}{{Mestel}, L.}, \bibinfo{year}{1968}.
\newblock \bibinfo{title}{{Magnetic braking by a stellar wind-I}}.
\newblock \bibinfo{journal}{\mnras} \bibinfo{volume}{138},
  \bibinfo{pages}{359}.
\newblock \DOIprefix\doi{10.1093/mnras/138.3.359}.
\bibitem[{{Mochnacki}(1981)}]{Mochnacki1981-ApJ}
\bibinfo{author}{{Mochnacki}, S.W.}, \bibinfo{year}{1981}.
\newblock \bibinfo{title}{{Contact binary stars.}}
\newblock \bibinfo{journal}{\apj} \bibinfo{volume}{245},
  \bibinfo{pages}{650--670}.
\newblock \DOIprefix\doi{10.1086/158841}.
\bibitem[{{Morton}(1960)}]{Morton1960-ApJ}
\bibinfo{author}{{Morton}, D.C.}, \bibinfo{year}{1960}.
\newblock \bibinfo{title}{{Evolutionary Mass Exchange in Close Binary
  Systems.}}
\newblock \bibinfo{journal}{\apj} \bibinfo{volume}{132}, \bibinfo{pages}{146}.
\newblock \DOIprefix\doi{10.1086/146908}.
\bibitem[{{Niarchos} et~al.(1996){Niarchos}, {Hoffmann} and
  {Duerbeck}}]{Niarchos1996-AAS}
\bibinfo{author}{{Niarchos}, P.G.}, \bibinfo{author}{{Hoffmann}, M.},
  \bibinfo{author}{{Duerbeck}, H.W.}, \bibinfo{year}{1996}.
\newblock \bibinfo{title}{{TU Bootis: an ambiguous W Ursae Majoris system.}}
\newblock \bibinfo{journal}{\aaps} \bibinfo{volume}{117},
  \bibinfo{pages}{105--112}.
\bibitem[{{Noori} and {Abedi}(2017)}]{Noori2017-NewA}
\bibinfo{author}{{Noori}, H.R.}, \bibinfo{author}{{Abedi}, A.},
  \bibinfo{year}{2017}.
\newblock \bibinfo{title}{{First period investigation and light-curve study of
  the eclipsing contact binary V776 Cas}}.
\newblock \bibinfo{journal}{\na} \bibinfo{volume}{56}, \bibinfo{pages}{5--9}.
\newblock \DOIprefix\doi{10.1016/j.newast.2017.04.007}.
\bibitem[{{Osaki}(1965)}]{Osaki1965-PASJ}
\bibinfo{author}{{Osaki}, Y.}, \bibinfo{year}{1965}.
\newblock \bibinfo{title}{{Mass-Luminosity Relationship in Close Binary Systems
  of W Ursae Majoris Type}}.
\newblock \bibinfo{journal}{\pasj} \bibinfo{volume}{17}, \bibinfo{pages}{97}.
\bibitem[{{Paczy{\'n}ski}(1971)}]{Paczynski1971-ARAA}
\bibinfo{author}{{Paczy{\'n}ski}, B.}, \bibinfo{year}{1971}.
\newblock \bibinfo{title}{{Evolutionary Processes in Close Binary Systems}}.
\newblock \bibinfo{journal}{\araa} \bibinfo{volume}{9}, \bibinfo{pages}{183}.
\newblock \DOIprefix\doi{10.1146/annurev.aa.09.090171.001151}.
\bibitem[{{Paczy{\'n}ski} and {Sienkiewicz}(1972)}]{Paczynski1972-AcA}
\bibinfo{author}{{Paczy{\'n}ski}, B.}, \bibinfo{author}{{Sienkiewicz}, R.},
  \bibinfo{year}{1972}.
\newblock \bibinfo{title}{{Evolution of Close Binaries VIII. Mass Exchange on
  the Dynamical Time Scale}}.
\newblock \bibinfo{journal}{\actaa} \bibinfo{volume}{22},
  \bibinfo{pages}{73--91}.
\bibitem[{{Park} et~al.(2013){Park}, {Lee}, {Kim}, {Lee} and
  {Jeon}}]{Park2013-PASJ}
\bibinfo{author}{{Park}, J.H.}, \bibinfo{author}{{Lee}, J.W.},
  \bibinfo{author}{{Kim}, S.L.}, \bibinfo{author}{{Lee}, C.U.},
  \bibinfo{author}{{Jeon}, Y.B.}, \bibinfo{year}{2013}.
\newblock \bibinfo{title}{{The Light and Period Variations of the Eclipsing
  Binary BX Draconis}}.
\newblock \bibinfo{journal}{\pasj} \bibinfo{volume}{65}, \bibinfo{pages}{1}.
\newblock \DOIprefix\doi{10.1093/pasj/65.1.1},
  \href{http://arxiv.org/abs/1207.6974}{{\tt arXiv:1207.6974}}.
\bibitem[{{Pejcha} et~al.(2016){Pejcha}, {Metzger} and
  {Tomida}}]{Pejcha2016-MNRAS2527}
\bibinfo{author}{{Pejcha}, O.}, \bibinfo{author}{{Metzger}, B.D.},
  \bibinfo{author}{{Tomida}, K.}, \bibinfo{year}{2016}.
\newblock \bibinfo{title}{{Binary stellar mergers with marginally bound ejecta:
  excretion discs, inflated envelopes, outflows, and their luminous
  transients}}.
\newblock \bibinfo{journal}{\mnras} \bibinfo{volume}{461},
  \bibinfo{pages}{2527--2539}.
\newblock \DOIprefix\doi{10.1093/mnras/stw1481},
  \href{http://arxiv.org/abs/1604.07414}{{\tt arXiv:1604.07414}}.
\bibitem[{{Pols} et~al.(1998){Pols}, {Schr{\"o}der}, {Hurley}, {Tout} and
  {Eggleton}}]{Pols1998-MNRAS}
\bibinfo{author}{{Pols}, O.R.}, \bibinfo{author}{{Schr{\"o}der}, K.P.},
  \bibinfo{author}{{Hurley}, J.R.}, \bibinfo{author}{{Tout}, C.A.},
  \bibinfo{author}{{Eggleton}, P.P.}, \bibinfo{year}{1998}.
\newblock \bibinfo{title}{{Stellar evolution models for Z = 0.0001 to 0.03}}.
\newblock \bibinfo{journal}{\mnras} \bibinfo{volume}{298},
  \bibinfo{pages}{525--536}.
\newblock \DOIprefix\doi{10.1046/j.1365-8711.1998.01658.x}.
\bibitem[{{Pribulla}(1998)}]{Pribulla1998-CoSka}
\bibinfo{author}{{Pribulla}, T.}, \bibinfo{year}{1998}.
\newblock \bibinfo{title}{{Efficiency of mass transfer and outflow in close
  binaries}}.
\newblock \bibinfo{journal}{Contributions of the Astronomical Observatory
  Skalnate Pleso} \bibinfo{volume}{28}, \bibinfo{pages}{101--108}.
\bibitem[{{Pribulla} et~al.(1999){Pribulla}, {Chochol}, {Rovithis-Livaniou} and
  {Rovithis}}]{Pribulla1999-AA}
\bibinfo{author}{{Pribulla}, T.}, \bibinfo{author}{{Chochol}, D.},
  \bibinfo{author}{{Rovithis-Livaniou}, H.}, \bibinfo{author}{{Rovithis}, P.},
  \bibinfo{year}{1999}.
\newblock \bibinfo{title}{{The contact binary AW Ursae Majoris as a member of a
  multiple system}}.
\newblock \bibinfo{journal}{\aap} \bibinfo{volume}{345},
  \bibinfo{pages}{137--148}.
\bibitem[{{Pribulla} et~al.(2007){Pribulla}, {Rucinski}, {Conidis}, {DeBond},
  {Thomson}, {Gazeas} and {Og{\l}oza}}]{Pribulla2007-AJ}
\bibinfo{author}{{Pribulla}, T.}, \bibinfo{author}{{Rucinski}, S.M.},
  \bibinfo{author}{{Conidis}, G.}, \bibinfo{author}{{DeBond}, H.},
  \bibinfo{author}{{Thomson}, J.R.}, \bibinfo{author}{{Gazeas}, K.},
  \bibinfo{author}{{Og{\l}oza}, W.}, \bibinfo{year}{2007}.
\newblock \bibinfo{title}{{Radial Velocity Studies of Close Binary Stars.
  XII.}}
\newblock \bibinfo{journal}{\aj} \bibinfo{volume}{133},
  \bibinfo{pages}{1977--1987}.
\newblock \DOIprefix\doi{10.1086/512772},
  \href{http://arxiv.org/abs/astro-ph/0611875}{{\tt arXiv:astro-ph/0611875}}.
\bibitem[{{Pribulla} et~al.(2009){Pribulla}, {Rucinski}, {DeBond}, {De Ridder},
  {Karmo}, {Thomson}, {Croll}, {Og{\l}oza}, {Pilecki} and
  {Siwak}}]{Pribulla2009-AJ3646}
\bibinfo{author}{{Pribulla}, T.}, \bibinfo{author}{{Rucinski}, S.M.},
  \bibinfo{author}{{DeBond}, H.}, \bibinfo{author}{{De Ridder}, A.},
  \bibinfo{author}{{Karmo}, T.}, \bibinfo{author}{{Thomson}, J.R.},
  \bibinfo{author}{{Croll}, B.}, \bibinfo{author}{{Og{\l}oza}, W.},
  \bibinfo{author}{{Pilecki}, B.}, \bibinfo{author}{{Siwak}, M.},
  \bibinfo{year}{2009}.
\newblock \bibinfo{title}{{Radial Velocity Studies of Close Binary Stars.
  XIV}}.
\newblock \bibinfo{journal}{\aj} \bibinfo{volume}{137},
  \bibinfo{pages}{3646--3654}.
\newblock \DOIprefix\doi{10.1088/0004-6256/137/3/3646},
  \href{http://arxiv.org/abs/0810.1658}{{\tt arXiv:0810.1658}}.
\bibitem[{{Pribulla} et~al.(2006){Pribulla}, {Rucinski}, {Lu}, {Mochnacki},
  {Conidis}, {Blake}, {DeBond}, {Thomson}, {Pych}, {Og{\l}oza} and
  {Siwak}}]{Pribulla2006-AJ}
\bibinfo{author}{{Pribulla}, T.}, \bibinfo{author}{{Rucinski}, S.M.},
  \bibinfo{author}{{Lu}, W.}, \bibinfo{author}{{Mochnacki}, S.W.},
  \bibinfo{author}{{Conidis}, G.}, \bibinfo{author}{{Blake}, R.M.},
  \bibinfo{author}{{DeBond}, H.}, \bibinfo{author}{{Thomson}, J.R.},
  \bibinfo{author}{{Pych}, W.}, \bibinfo{author}{{Og{\l}oza}, W.},
  \bibinfo{author}{{Siwak}, M.}, \bibinfo{year}{2006}.
\newblock \bibinfo{title}{{Radial Velocity Studies of Close Binary Stars. XI.}}
\newblock \bibinfo{journal}{\aj} \bibinfo{volume}{132},
  \bibinfo{pages}{769--780}.
\newblock \DOIprefix\doi{10.1086/505536},
  \href{http://arxiv.org/abs/astro-ph/0605357}{{\tt arXiv:astro-ph/0605357}}.
\bibitem[{{Pr{\v s}a} et~al.(2011){Pr{\v s}a}, {Batalha}, {Slawson}, {Doyle},
  {Welsh}, {Orosz}, {Seager}, {Rucker}, {Mjaseth}, {Engle}, {Conroy},
  {Jenkins}, {Caldwell}, {Koch} and {Borucki}}]{Prsa2011-AJ}
\bibinfo{author}{{Pr{\v s}a}, A.}, \bibinfo{author}{{Batalha}, N.},
  \bibinfo{author}{{Slawson}, R.W.}, \bibinfo{author}{{Doyle}, L.R.},
  \bibinfo{author}{{Welsh}, W.F.}, \bibinfo{author}{{Orosz}, J.A.},
  \bibinfo{author}{{Seager}, S.}, \bibinfo{author}{{Rucker}, M.},
  \bibinfo{author}{{Mjaseth}, K.}, \bibinfo{author}{{Engle}, S.G.},
  \bibinfo{author}{{Conroy}, K.}, \bibinfo{author}{{Jenkins}, J.},
  \bibinfo{author}{{Caldwell}, D.}, \bibinfo{author}{{Koch}, D.},
  \bibinfo{author}{{Borucki}, W.}, \bibinfo{year}{2011}.
\newblock \bibinfo{title}{{Kepler Eclipsing Binary Stars. I. Catalog and
  Principal Characterization of 1879 Eclipsing Binaries in the First Data
  Release}}.
\newblock \bibinfo{journal}{\aj} \bibinfo{volume}{141}, \bibinfo{pages}{83}.
\newblock \DOIprefix\doi{10.1088/0004-6256/141/3/83},
  \href{http://arxiv.org/abs/1006.2815}{{\tt arXiv:1006.2815}}.
\bibitem[{{Pych} et~al.(2004){Pych}, {Rucinski}, {DeBond}, {Thomson},
  {Capobianco}, {Blake}, {Og{\l}oza}, {Stachowski}, {Rogoziecki}, {Ligeza} and
  {Gazeas}}]{Pych2004-AJ}
\bibinfo{author}{{Pych}, W.}, \bibinfo{author}{{Rucinski}, S.M.},
  \bibinfo{author}{{DeBond}, H.}, \bibinfo{author}{{Thomson}, J.R.},
  \bibinfo{author}{{Capobianco}, C.C.}, \bibinfo{author}{{Blake}, R.M.},
  \bibinfo{author}{{Og{\l}oza}, W.}, \bibinfo{author}{{Stachowski}, G.},
  \bibinfo{author}{{Rogoziecki}, P.}, \bibinfo{author}{{Ligeza}, P.},
  \bibinfo{author}{{Gazeas}, K.}, \bibinfo{year}{2004}.
\newblock \bibinfo{title}{{Radial Velocity Studies of Close Binary Stars. IX.}}
\newblock \bibinfo{journal}{\aj} \bibinfo{volume}{127},
  \bibinfo{pages}{1712--1719}.
\newblock \DOIprefix\doi{10.1086/382105},
  \href{http://arxiv.org/abs/astro-ph/0311350}{{\tt arXiv:astro-ph/0311350}}.
\bibitem[{{Qian}(2001)}]{Qian2001-MNRAS914}
\bibinfo{author}{{Qian}, S.}, \bibinfo{year}{2001}.
\newblock \bibinfo{title}{{Orbital period changes of contact binary systems:
  direct evidence for thermal relaxation oscillation theory}}.
\newblock \bibinfo{journal}{\mnras} \bibinfo{volume}{328},
  \bibinfo{pages}{914--924}.
\newblock \DOIprefix\doi{10.1046/j.1365-8711.2001.04921.x}.
\bibitem[{{Qian} et~al.(2008){Qian}, {He}, {Soonthornthum}, {Liu}, {Zhu}, {Li},
  {Liao} and {Dai}}]{Qian2008-AJ1940}
\bibinfo{author}{{Qian}, S.B.}, \bibinfo{author}{{He}, J.J.},
  \bibinfo{author}{{Soonthornthum}, B.}, \bibinfo{author}{{Liu}, L.},
  \bibinfo{author}{{Zhu}, L.Y.}, \bibinfo{author}{{Li}, L.J.},
  \bibinfo{author}{{Liao}, W.P.}, \bibinfo{author}{{Dai}, Z.B.},
  \bibinfo{year}{2008}.
\newblock \bibinfo{title}{{High Fill-Out, Extreme Mass Ratio Overcontact Binary
  Systems. VIII. EM Piscium}}.
\newblock \bibinfo{journal}{\aj} \bibinfo{volume}{136},
  \bibinfo{pages}{1940--1946}.
\newblock \DOIprefix\doi{10.1088/0004-6256/136/5/1940}.
\bibitem[{{Qian} et~al.(2006){Qian}, {Liu}, {Soonthornthum}, {Zhu} and
  {He}}]{Qian2006-AJ131}
\bibinfo{author}{{Qian}, S.B.}, \bibinfo{author}{{Liu}, L.},
  \bibinfo{author}{{Soonthornthum}, B.}, \bibinfo{author}{{Zhu}, L.Y.},
  \bibinfo{author}{{He}, J.J.}, \bibinfo{year}{2006}.
\newblock \bibinfo{title}{{Deep, Low Mass Ratio Overcontact Binary Systems. VI.
  AH Cancri in the Old Open Cluster M67}}.
\newblock \bibinfo{journal}{\aj} \bibinfo{volume}{131},
  \bibinfo{pages}{3028--3039}.
\newblock \DOIprefix\doi{10.1086/503561}.
\bibitem[{{Qian} et~al.(2007){Qian}, {Xiang}, {Zhu}, {Dai}, {He} and
  {Yuan}}]{Qian2007-AJ357}
\bibinfo{author}{{Qian}, S.B.}, \bibinfo{author}{{Xiang}, F.Y.},
  \bibinfo{author}{{Zhu}, L.Y.}, \bibinfo{author}{{Dai}, Z.B.},
  \bibinfo{author}{{He}, J.J.}, \bibinfo{author}{{Yuan}, J.Z.},
  \bibinfo{year}{2007}.
\newblock \bibinfo{title}{{A New CCD Photometric Investigation of the
  Short-Period Close Binary AP Leonis}}.
\newblock \bibinfo{journal}{\aj} \bibinfo{volume}{133},
  \bibinfo{pages}{357--363}.
\newblock \DOIprefix\doi{10.1086/509499}.
\bibitem[{{Qian} and {Yang}(2004)}]{Qian2004-AJ}
\bibinfo{author}{{Qian}, S.B.}, \bibinfo{author}{{Yang}, Y.G.},
  \bibinfo{year}{2004}.
\newblock \bibinfo{title}{{GR Virginis: A Deep Overcontact Binary}}.
\newblock \bibinfo{journal}{\aj} \bibinfo{volume}{128},
  \bibinfo{pages}{2430--2434}.
\newblock \DOIprefix\doi{10.1086/425051}.
\bibitem[{{Qian} et~al.(2005){Qian}, {Yang}, {Soonthornthum}, {Zhu}, {He} and
  {Yuan}}]{Qian2005-AJ224}
\bibinfo{author}{{Qian}, S.B.}, \bibinfo{author}{{Yang}, Y.G.},
  \bibinfo{author}{{Soonthornthum}, B.}, \bibinfo{author}{{Zhu}, L.Y.},
  \bibinfo{author}{{He}, J.J.}, \bibinfo{author}{{Yuan}, J.Z.},
  \bibinfo{year}{2005}.
\newblock \bibinfo{title}{{Deep, Low Mass Ratio Overcontact Binary Systems.
  III. CU Tauri and TV Muscae}}.
\newblock \bibinfo{journal}{\aj} \bibinfo{volume}{130},
  \bibinfo{pages}{224--233}.
\newblock \DOIprefix\doi{10.1086/430673}.
\bibitem[{{Raghavan} et~al.(2010){Raghavan}, {McAlister}, {Henry}, {Latham},
  {Marcy}, {Mason}, {Gies}, {White} and {ten Brummelaar}}]{Raghavan2010-ApJS}
\bibinfo{author}{{Raghavan}, D.}, \bibinfo{author}{{McAlister}, H.A.},
  \bibinfo{author}{{Henry}, T.J.}, \bibinfo{author}{{Latham}, D.W.},
  \bibinfo{author}{{Marcy}, G.W.}, \bibinfo{author}{{Mason}, B.D.},
  \bibinfo{author}{{Gies}, D.R.}, \bibinfo{author}{{White}, R.J.},
  \bibinfo{author}{{ten Brummelaar}, T.A.}, \bibinfo{year}{2010}.
\newblock \bibinfo{title}{{A Survey of Stellar Families: Multiplicity of
  Solar-type Stars}}.
\newblock \bibinfo{journal}{\apjs} \bibinfo{volume}{190},
  \bibinfo{pages}{1--42}.
\newblock \DOIprefix\doi{10.1088/0067-0049/190/1/1},
  \href{http://arxiv.org/abs/1007.0414}{{\tt arXiv:1007.0414}}.
\bibitem[{{Rahunen}(1981)}]{Rahunen1981-AA}
\bibinfo{author}{{Rahunen}, T.}, \bibinfo{year}{1981}.
\newblock \bibinfo{title}{{Evolution of W UMa systems and angular momentum
  loss.}}
\newblock \bibinfo{journal}{\aap} \bibinfo{volume}{102},
  \bibinfo{pages}{81--90}.
\bibitem[{{Rasio}(1995)}]{Rasio1995-ApJ}
\bibinfo{author}{{Rasio}, F.A.}, \bibinfo{year}{1995}.
\newblock \bibinfo{title}{{The Minimum Mass Ratio of W Ursae Majoris
  Binaries}}.
\newblock \bibinfo{journal}{\apjl} \bibinfo{volume}{444}, \bibinfo{pages}{L41}.
\newblock \DOIprefix\doi{10.1086/187855},
  \href{http://arxiv.org/abs/astro-ph/9502028}{{\tt arXiv:astro-ph/9502028}}.
\bibitem[{{Rucinski}(1974)}]{Rucinski1974-AcA}
\bibinfo{author}{{Rucinski}, S.M.}, \bibinfo{year}{1974}.
\newblock \bibinfo{title}{{Binaries. II. A- and W-type Systems. The W UMa-type
  Systems as Contact}}.
\newblock \bibinfo{journal}{\actaa} \bibinfo{volume}{24}, \bibinfo{pages}{119}.
\bibitem[{{Rucinski}(1982)}]{Rucinski1982-AA}
\bibinfo{author}{{Rucinski}, S.M.}, \bibinfo{year}{1982}.
\newblock \bibinfo{title}{{Contact binaries - Angular momentum loss in and out
  of contact}}.
\newblock \bibinfo{journal}{\aap} \bibinfo{volume}{112},
  \bibinfo{pages}{273--276}.
\bibitem[{{Rucinski} et~al.(2003){Rucinski}, {Capobianco}, {Lu}, {DeBond},
  {Thomson}, {Mochnacki}, {Blake}, {Og{\l}oza}, {Stachowski} and
  {Rogoziecki}}]{Rucinski2003-AJ}
\bibinfo{author}{{Rucinski}, S.M.}, \bibinfo{author}{{Capobianco}, C.C.},
  \bibinfo{author}{{Lu}, W.}, \bibinfo{author}{{DeBond}, H.},
  \bibinfo{author}{{Thomson}, J.R.}, \bibinfo{author}{{Mochnacki}, S.W.},
  \bibinfo{author}{{Blake}, R.M.}, \bibinfo{author}{{Og{\l}oza}, W.},
  \bibinfo{author}{{Stachowski}, G.}, \bibinfo{author}{{Rogoziecki}, P.},
  \bibinfo{year}{2003}.
\newblock \bibinfo{title}{{Radial Velocity Studies of Close Binary Stars.
  VIII.}}
\newblock \bibinfo{journal}{\aj} \bibinfo{volume}{125},
  \bibinfo{pages}{3258--3264}.
\newblock \DOIprefix\doi{10.1086/374949},
  \href{http://arxiv.org/abs/astro-ph/0302399}{{\tt arXiv:astro-ph/0302399}}.
\bibitem[{{Rucinski} and {Lu}(1999)}]{Rucinski1999-AJ}
\bibinfo{author}{{Rucinski}, S.M.}, \bibinfo{author}{{Lu}, W.},
  \bibinfo{year}{1999}.
\newblock \bibinfo{title}{{Radial Velocity Studies of Close Binary Stars. II.}}
\newblock \bibinfo{journal}{\aj} \bibinfo{volume}{118},
  \bibinfo{pages}{2451--2459}.
\newblock \DOIprefix\doi{10.1086/301101},
  \href{http://arxiv.org/abs/astro-ph/9906314}{{\tt arXiv:astro-ph/9906314}}.
\bibitem[{{Rucinski} et~al.(2002){Rucinski}, {Lu}, {Capobianco}, {Mochnacki},
  {Blake}, {Thomson}, {Og{\l}oza} and {Stachowski}}]{Rucinski2002-AJ}
\bibinfo{author}{{Rucinski}, S.M.}, \bibinfo{author}{{Lu}, W.},
  \bibinfo{author}{{Capobianco}, C.C.}, \bibinfo{author}{{Mochnacki}, S.W.},
  \bibinfo{author}{{Blake}, R.M.}, \bibinfo{author}{{Thomson}, J.R.},
  \bibinfo{author}{{Og{\l}oza}, W.}, \bibinfo{author}{{Stachowski}, G.},
  \bibinfo{year}{2002}.
\newblock \bibinfo{title}{{Radial Velocity Studies of Close Binary Stars. VI.}}
\newblock \bibinfo{journal}{\aj} \bibinfo{volume}{124},
  \bibinfo{pages}{1738--1745}.
\newblock \DOIprefix\doi{10.1086/342341},
  \href{http://arxiv.org/abs/astro-ph/0201213}{{\tt arXiv:astro-ph/0201213}}.
\bibitem[{{Rucinski} et~al.(2001){Rucinski}, {Lu}, {Mochnacki}, {Og{\l}oza} and
  {Stachowski}}]{Rucinski2001-AJ}
\bibinfo{author}{{Rucinski}, S.M.}, \bibinfo{author}{{Lu}, W.},
  \bibinfo{author}{{Mochnacki}, S.W.}, \bibinfo{author}{{Og{\l}oza}, W.},
  \bibinfo{author}{{Stachowski}, G.}, \bibinfo{year}{2001}.
\newblock \bibinfo{title}{{Radial Velocity Studies of Close Binary Stars. V.}}
\newblock \bibinfo{journal}{\aj} \bibinfo{volume}{122},
  \bibinfo{pages}{1974--1980}.
\newblock \DOIprefix\doi{10.1086/323106},
  \href{http://arxiv.org/abs/astro-ph/0106160}{{\tt arXiv:astro-ph/0106160}}.
\bibitem[{{Sarotsakulchai} et~al.(2018){Sarotsakulchai}, {Qian},
  {Soonthornthum}, {Zhou}, {Zhang}, {Reichart}, {Haislip}, {Kouprianov} and
  {Poshyachinda}}]{Sarotsakulchai2018-AJ}
\bibinfo{author}{{Sarotsakulchai}, T.}, \bibinfo{author}{{Qian}, S.B.},
  \bibinfo{author}{{Soonthornthum}, B.}, \bibinfo{author}{{Zhou}, X.},
  \bibinfo{author}{{Zhang}, J.}, \bibinfo{author}{{Reichart}, D.E.},
  \bibinfo{author}{{Haislip}, J.B.}, \bibinfo{author}{{Kouprianov}, V.V.},
  \bibinfo{author}{{Poshyachinda}, S.}, \bibinfo{year}{2018}.
\newblock \bibinfo{title}{{TY Pup: A Low-mass-ratio and Deep Contact Binary as
  a Progenitor Candidate of Luminous Red Novae}}.
\newblock \bibinfo{journal}{\aj} \bibinfo{volume}{156}, \bibinfo{pages}{199}.
\newblock \DOIprefix\doi{10.3847/1538-3881/aadcfa},
  \href{http://arxiv.org/abs/1807.00478}{{\tt arXiv:1807.00478}}.
\bibitem[{{Schatzman}(1962)}]{Schatzman1962-AnAp}
\bibinfo{author}{{Schatzman}, E.}, \bibinfo{year}{1962}.
\newblock \bibinfo{title}{{A theory of the role of magnetic activity during
  star formation}}.
\newblock \bibinfo{journal}{Annales d'Astrophysique} \bibinfo{volume}{25},
  \bibinfo{pages}{18}.
\bibitem[{{Selam} et~al.(2018){Selam}, {Esmer}, {{\c S}enavc{\i}}, {Bahar},
  {Y{\"o}r{\"u}ko{\u g}lu}, {Y{\i}lmaz} and {Ba{\c
  s}t{\"u}rk}}]{Selam2018-ApSS}
\bibinfo{author}{{Selam}, S.O.}, \bibinfo{author}{{Esmer}, E.M.},
  \bibinfo{author}{{{\c S}enavc{\i}}, H.V.}, \bibinfo{author}{{Bahar}, E.},
  \bibinfo{author}{{Y{\"o}r{\"u}ko{\u g}lu}, O.}, \bibinfo{author}{{Y{\i}lmaz},
  M.}, \bibinfo{author}{{Ba{\c s}t{\"u}rk}, {\"O}.}, \bibinfo{year}{2018}.
\newblock \bibinfo{title}{{A simultaneous spectroscopic and photometric study
  of two eclipsing binaries: V566 Oph and V972 Her}}.
\newblock \bibinfo{journal}{\apss} \bibinfo{volume}{363}, \bibinfo{pages}{34}.
\newblock \DOIprefix\doi{10.1007/s10509-018-3252-y}.
\bibitem[{{Shu} et~al.(1979){Shu}, {Lubow} and {Anderson}}]{Shu1979-ApJ}
\bibinfo{author}{{Shu}, F.H.}, \bibinfo{author}{{Lubow}, S.H.},
  \bibinfo{author}{{Anderson}, L.}, \bibinfo{year}{1979}.
\newblock \bibinfo{title}{{On the structure of contact binaries. III. Mass and
  energy flow.}}
\newblock \bibinfo{journal}{\apj} \bibinfo{volume}{229},
  \bibinfo{pages}{223--241}.
\newblock \DOIprefix\doi{10.1086/156948}.
\bibitem[{{St{\k{e}}pie{\'n}}(2006a)}]{Stepien2006-AcA199}
\bibinfo{author}{{St{\k{e}}pie{\'n}}, K.}, \bibinfo{year}{2006}a.
\newblock \bibinfo{title}{{Evolutionary Status of Late-Type Contact Binaries}}.
\newblock \bibinfo{journal}{\actaa} \bibinfo{volume}{56},
  \bibinfo{pages}{199--218}.
\newblock \href{http://arxiv.org/abs/astro-ph/0510464}{{\tt
  arXiv:astro-ph/0510464}}.
\bibitem[{{St{\k{e}}pie{\'n}}(2006b)}]{Stepien2006-AcA347}
\bibinfo{author}{{St{\k{e}}pie{\'n}}, K.}, \bibinfo{year}{2006}b.
\newblock \bibinfo{title}{{The Low-Mass Limit for Total Mass of W UMa-type
  Binaries}}.
\newblock \bibinfo{journal}{\actaa} \bibinfo{volume}{56},
  \bibinfo{pages}{347--364}.
\newblock \href{http://arxiv.org/abs/astro-ph/0701529}{{\tt
  arXiv:astro-ph/0701529}}.
\bibitem[{{St{\k{e}}pie{\'n}} and {Kiraga}(2015)}]{Stepien2015-AA}
\bibinfo{author}{{St{\k{e}}pie{\'n}}, K.}, \bibinfo{author}{{Kiraga}, M.},
  \bibinfo{year}{2015}.
\newblock \bibinfo{title}{{Model computations of blue stragglers and W UMa-type
  stars in globular clusters}}.
\newblock \bibinfo{journal}{\aap} \bibinfo{volume}{577}, \bibinfo{pages}{A117}.
\newblock \DOIprefix\doi{10.1051/0004-6361/201425550},
  \href{http://arxiv.org/abs/1503.07758}{{\tt arXiv:1503.07758}}.
\bibitem[{{Tian} et~al.(2018){Tian}, {Zhu}, {Qian}, {Li} and
  {Jiang}}]{Tian2018-RAA}
\bibinfo{author}{{Tian}, X.M.}, \bibinfo{author}{{Zhu}, L.Y.},
  \bibinfo{author}{{Qian}, S.B.}, \bibinfo{author}{{Li}, L.J.},
  \bibinfo{author}{{Jiang}, L.Q.}, \bibinfo{year}{2018}.
\newblock \bibinfo{title}{{Multi-color light curves and orbital period research
  of eclipsing binary V1073 Cyg}}.
\newblock \bibinfo{journal}{Research in Astronomy and Astrophysics}
  \bibinfo{volume}{18}, \bibinfo{pages}{020}.
\newblock \DOIprefix\doi{10.1088/1674-4527/18/2/20},
  \href{http://arxiv.org/abs/1712.01072}{{\tt arXiv:1712.01072}}.
\bibitem[{{Tokovinin} et~al.(2006){Tokovinin}, {Thomas}, {Sterzik} and
  {Udry}}]{Tokovinin2006-AA}
\bibinfo{author}{{Tokovinin}, A.}, \bibinfo{author}{{Thomas}, S.},
  \bibinfo{author}{{Sterzik}, M.}, \bibinfo{author}{{Udry}, S.},
  \bibinfo{year}{2006}.
\newblock \bibinfo{title}{{Tertiary companions to close spectroscopic
  binaries}}.
\newblock \bibinfo{journal}{\aap} \bibinfo{volume}{450},
  \bibinfo{pages}{681--693}.
\newblock \DOIprefix\doi{10.1051/0004-6361:20054427},
  \href{http://arxiv.org/abs/astro-ph/0601518}{{\tt arXiv:astro-ph/0601518}}.
\bibitem[{{Tout} and {Hall}(1991)}]{Tout1991-MNRAS}
\bibinfo{author}{{Tout}, C.A.}, \bibinfo{author}{{Hall}, D.S.},
  \bibinfo{year}{1991}.
\newblock \bibinfo{title}{{Wind driven mass transfer in interacting binary
  systems.}}
\newblock \bibinfo{journal}{\mnras} \bibinfo{volume}{253},
  \bibinfo{pages}{9--18}.
\newblock \DOIprefix\doi{10.1093/mnras/253.1.9}.
\bibitem[{{vant Veer}(1979)}]{vantVeer1979-AA}
\bibinfo{author}{{vant Veer}, F.}, \bibinfo{year}{1979}.
\newblock \bibinfo{title}{{The angular momentum controlled evolution of solar
  type contact binaries.}}
\newblock \bibinfo{journal}{\aap} \bibinfo{volume}{80},
  \bibinfo{pages}{287--295}.
\bibitem[{{Verbunt} and {Zwaan}(1981)}]{Verbunt1981-AA}
\bibinfo{author}{{Verbunt}, F.}, \bibinfo{author}{{Zwaan}, C.},
  \bibinfo{year}{1981}.
\newblock \bibinfo{title}{{Magnetic braking in low-mass X-ray binaries.}}
\newblock \bibinfo{journal}{\aap} \bibinfo{volume}{100},
  \bibinfo{pages}{L7--L9}.
\bibitem[{{Vilhu}(1981)}]{Vilhu1981-ApSS}
\bibinfo{author}{{Vilhu}, O.}, \bibinfo{year}{1981}.
\newblock \bibinfo{title}{{Problems of Low Mass Binary Evolution}}.
\newblock \bibinfo{journal}{\apss} \bibinfo{volume}{78},
  \bibinfo{pages}{401--418}.
\newblock \DOIprefix\doi{10.1007/BF00648946}.
\bibitem[{{Vilhu}(1982)}]{Vilhu1982-AA}
\bibinfo{author}{{Vilhu}, O.}, \bibinfo{year}{1982}.
\newblock \bibinfo{title}{{Detached to contact scenario for the origin of W UMa
  stars}}.
\newblock \bibinfo{journal}{\aap} \bibinfo{volume}{109},
  \bibinfo{pages}{17--22}.
\bibitem[{{Webbink}(2003)}]{Webbink2003-ASPC}
\bibinfo{author}{{Webbink}, R.F.}, \bibinfo{year}{2003}.
\newblock \bibinfo{title}{{Contact Binaries}}, in: \bibinfo{editor}{{Turcotte},
  S.}, \bibinfo{editor}{{Keller}, S.C.}, \bibinfo{editor}{{Cavallo}, R.M.}
  (Eds.), \bibinfo{booktitle}{3D Stellar Evolution}, p.~\bibinfo{pages}{76}.
\newblock \href{http://arxiv.org/abs/astro-ph/0304420}{{\tt
  arXiv:astro-ph/0304420}}.
\bibitem[{{Wilson} and {Devinney}(1971)}]{Wilson1971-ApJ}
\bibinfo{author}{{Wilson}, R.E.}, \bibinfo{author}{{Devinney}, E.J.},
  \bibinfo{year}{1971}.
\newblock \bibinfo{title}{{Realization of Accurate Close-Binary Light Curves:
  Application to MR Cygni}}.
\newblock \bibinfo{journal}{\apj} \bibinfo{volume}{166}, \bibinfo{pages}{605}.
\newblock \DOIprefix\doi{10.1086/150986}.
\bibitem[{Wold(1975)}]{Wold1975-QS}
\bibinfo{author}{Wold, H.}, \bibinfo{year}{1975}.
\newblock \bibinfo{title}{Path models with latent variables: The nipals
  approach}, in: \bibinfo{booktitle}{Quantitative sociology}.
  \bibinfo{publisher}{Elsevier}, pp. \bibinfo{pages}{307--357}.
\bibitem[{Wold et~al.(2004)Wold, Eriksson, Trygg and Kettaneh}]{Wold2004-UU}
\bibinfo{author}{Wold, S.}, \bibinfo{author}{Eriksson, L.},
  \bibinfo{author}{Trygg, J.}, \bibinfo{author}{Kettaneh, N.},
  \bibinfo{year}{2004}.
\newblock \bibinfo{title}{The pls method--partial least squares projections to
  latent structures--and its applications in industrial rdp (research,
  development, and production)}.
\newblock \bibinfo{journal}{Unea University} .
\bibitem[{Wold et~al.(2001)Wold, Sj{\"o}str{\"o}m and Eriksson}]{Wold2001-CILS}
\bibinfo{author}{Wold, S.}, \bibinfo{author}{Sj{\"o}str{\"o}m, M.},
  \bibinfo{author}{Eriksson, L.}, \bibinfo{year}{2001}.
\newblock \bibinfo{title}{Pls-regression: a basic tool of chemometrics}.
\newblock \bibinfo{journal}{Chemometrics and intelligent laboratory systems}
  \bibinfo{volume}{58}, \bibinfo{pages}{109--130}.
\bibitem[{{Wolf} et~al.(2000){Wolf}, {Mol{\'{\i}}k}, {Hornoch} and {{\v
  S}arounov{\'a}}}]{Wolf2000-AAS}
\bibinfo{author}{{Wolf}, M.}, \bibinfo{author}{{Mol{\'{\i}}k}, P.},
  \bibinfo{author}{{Hornoch}, K.}, \bibinfo{author}{{{\v S}arounov{\'a}}, L.},
  \bibinfo{year}{2000}.
\newblock \bibinfo{title}{{Period changes in W UMa-type eclipsing binaries: DK
  Cygni, V401 Cygni, AD Phoenicis and Y Sextantis}}.
\newblock \bibinfo{journal}{\aaps} \bibinfo{volume}{147},
  \bibinfo{pages}{243--249}.
\newblock \DOIprefix\doi{10.1051/aas:2000300}.
\bibitem[{{Wolf} et~al.(1996){Wolf}, {Sarounova} and {Molik}}]{Wolf1996-IBVS}
\bibinfo{author}{{Wolf}, M.}, \bibinfo{author}{{Sarounova}, L.},
  \bibinfo{author}{{Molik}, P.}, \bibinfo{year}{1996}.
\newblock \bibinfo{title}{{Period Changes in V839 Ophiuchi}}.
\newblock \bibinfo{journal}{Information Bulletin on Variable Stars}
  \bibinfo{volume}{4304}.
\bibitem[{{Wood}(1950)}]{Wood1950-ApJ}
\bibinfo{author}{{Wood}, F.B.}, \bibinfo{year}{1950}.
\newblock \bibinfo{title}{{On the Change of Period of Eclipsing Variables
  Stars.}}
\newblock \bibinfo{journal}{\apj} \bibinfo{volume}{112}, \bibinfo{pages}{196}.
\newblock \DOIprefix\doi{10.1086/145328}.
\bibitem[{{Xiang} et~al.(2015){Xiang}, {Yu} and {Xiao}}]{Xiang2015-AJ62}
\bibinfo{author}{{Xiang}, F.Y.}, \bibinfo{author}{{Yu}, Y.X.},
  \bibinfo{author}{{Xiao}, T.Y.}, \bibinfo{year}{2015}.
\newblock \bibinfo{title}{{CCD Photometric Study and Period Investigation of
  V508 Oph}}.
\newblock \bibinfo{journal}{\aj} \bibinfo{volume}{149}, \bibinfo{pages}{62}.
\newblock \DOIprefix\doi{10.1088/0004-6256/149/2/62}.
\bibitem[{{Yakut} and {Eggleton}(2005)}]{Yakut2005-ApJ1055}
\bibinfo{author}{{Yakut}, K.}, \bibinfo{author}{{Eggleton}, P.P.},
  \bibinfo{year}{2005}.
\newblock \bibinfo{title}{{Evolution of Close Binary Systems}}.
\newblock \bibinfo{journal}{\apj} \bibinfo{volume}{629},
  \bibinfo{pages}{1055--1074}.
\newblock \DOIprefix\doi{10.1086/431300}.
\bibitem[{{Yang} and {Liu}(2003)}]{Yang2003-AJ}
\bibinfo{author}{{Yang}, Y.}, \bibinfo{author}{{Liu}, Q.},
  \bibinfo{year}{2003}.
\newblock \bibinfo{title}{{RZ Tauri: An Unstable W Ursae Majoris Binary with a
  Magnetically Active Component}}.
\newblock \bibinfo{journal}{\aj} \bibinfo{volume}{126},
  \bibinfo{pages}{1960--1966}.
\newblock \DOIprefix\doi{10.1086/377019}.
\bibitem[{{Yang}(2012)}]{Yang2012-RAA}
\bibinfo{author}{{Yang}, Y.G.}, \bibinfo{year}{2012}.
\newblock \bibinfo{title}{{A new photometric study of the triple star system EF
  Draconis}}.
\newblock \bibinfo{journal}{Research in Astronomy and Astrophysics}
  \bibinfo{volume}{12}, \bibinfo{pages}{419--432}.
\newblock \DOIprefix\doi{10.1088/1674-4527/12/4/006}.
\bibitem[{{Yang} et~al.(2013a){Yang}, {Dai} and {Zhang}}]{Yang2013-NewA}
\bibinfo{author}{{Yang}, Y.G.}, \bibinfo{author}{{Dai}, H.F.},
  \bibinfo{author}{{Zhang}, J.F.}, \bibinfo{year}{2013}a.
\newblock \bibinfo{title}{{New photometric studies of two contact binaries CE
  Leo and V366 Cas with possible tertiary companions}}.
\newblock \bibinfo{journal}{\na} \bibinfo{volume}{19}, \bibinfo{pages}{27--33}.
\newblock \DOIprefix\doi{10.1016/j.newast.2012.07.002}.
\bibitem[{{Yang} et~al.(2013b){Yang}, {Qian} and {Dai}}]{Yang2013-AJ60}
\bibinfo{author}{{Yang}, Y.G.}, \bibinfo{author}{{Qian}, S.B.},
  \bibinfo{author}{{Dai}, H.F.}, \bibinfo{year}{2013}b.
\newblock \bibinfo{title}{{Photometric Studies of Three Neglected Short-period
  Contact Binaries GN Bootis, BL Leonis, and V1918 Cygni}}.
\newblock \bibinfo{journal}{\aj} \bibinfo{volume}{145}, \bibinfo{pages}{60}.
\newblock \DOIprefix\doi{10.1088/0004-6256/145/3/60}.
\bibitem[{{Yang} et~al.(2005a){Yang}, {Qian}, {Gonzalez-Rojas} and
  {Yuan}}]{Yang2005-ApSS}
\bibinfo{author}{{Yang}, Y.G.}, \bibinfo{author}{{Qian}, S.B.},
  \bibinfo{author}{{Gonzalez-Rojas}, D.J.}, \bibinfo{author}{{Yuan}, J.Z.},
  \bibinfo{year}{2005}a.
\newblock \bibinfo{title}{{The First Photometric Analyses Of Four New
  Discovered Ew-Type Eclipsing Binaries: GSC 1848-1264, GSC 0804-0118, GSC
  0619-0232 And GSC 2936-0478}}.
\newblock \bibinfo{journal}{\apss} \bibinfo{volume}{300},
  \bibinfo{pages}{337--356}.
\newblock \DOIprefix\doi{10.1007/s10509-005-4161-4}.
\bibitem[{{Yang} et~al.(2012){Yang}, {Qian} and
  {Soonthornthum}}]{Yang2012-AJ122}
\bibinfo{author}{{Yang}, Y.G.}, \bibinfo{author}{{Qian}, S.B.},
  \bibinfo{author}{{Soonthornthum}, B.}, \bibinfo{year}{2012}.
\newblock \bibinfo{title}{{Deep, Low-mass Ratio Overcontact Binary Systems.
  XII. CK Bootis with Possible Cyclic Magnetic Activity and Additional
  Companion}}.
\newblock \bibinfo{journal}{\aj} \bibinfo{volume}{143}, \bibinfo{pages}{122}.
\newblock \DOIprefix\doi{10.1088/0004-6256/143/5/122}.
\bibitem[{{Yang} et~al.(2013c){Yang}, {Qian}, {Zhang}, {Dai} and
  {Soonthornthum}}]{Yang2013-AJ}
\bibinfo{author}{{Yang}, Y.G.}, \bibinfo{author}{{Qian}, S.B.},
  \bibinfo{author}{{Zhang}, L.Y.}, \bibinfo{author}{{Dai}, H.F.},
  \bibinfo{author}{{Soonthornthum}, B.}, \bibinfo{year}{2013}c.
\newblock \bibinfo{title}{{Deep, Low Mass Ratio Overcontact Binary Systems.
  XIII. DZ Piscium with Intrinsic Light Variability}}.
\newblock \bibinfo{journal}{\aj} \bibinfo{volume}{146}, \bibinfo{pages}{35}.
\newblock \DOIprefix\doi{10.1088/0004-6256/146/2/35}.
\bibitem[{{Yang} et~al.(2005b){Yang}, {Qian}, {Zhu}, {He} and
  {Yuan}}]{Yang2005-PASJ}
\bibinfo{author}{{Yang}, Y.G.}, \bibinfo{author}{{Qian}, S.B.},
  \bibinfo{author}{{Zhu}, L.Y.}, \bibinfo{author}{{He}, J.J.},
  \bibinfo{author}{{Yuan}, J.Z.}, \bibinfo{year}{2005}b.
\newblock \bibinfo{title}{{Photometric Investigations of Three Short-Period
  Binary Systems: GSC 0763-0572, RR Centauri, and {$\epsilon$} Coronae
  Australis}}.
\newblock \bibinfo{journal}{\pasj} \bibinfo{volume}{57},
  \bibinfo{pages}{983--993}.
\newblock \DOIprefix\doi{10.1093/pasj/57.6.983}.
\bibitem[{{Yang} et~al.(2011){Yang}, {Shao}, {Pan} and {Yin}}]{Yang2011-PASP}
\bibinfo{author}{{Yang}, Y.G.}, \bibinfo{author}{{Shao}, Z.Y.},
  \bibinfo{author}{{Pan}, H.J.}, \bibinfo{author}{{Yin}, X.G.},
  \bibinfo{year}{2011}.
\newblock \bibinfo{title}{{Orbital-Period Variations and Photometric Analysis
  for the Neglected Contact Binary EH Cancri}}.
\newblock \bibinfo{journal}{\pasp} \bibinfo{volume}{123}, \bibinfo{pages}{895}.
\newblock \DOIprefix\doi{10.1086/661527}.
\bibitem[{{Yang} et~al.(2010){Yang}, {Wei}, {Kreiner} and {Li}}]{Yang2010-AJ}
\bibinfo{author}{{Yang}, Y.G.}, \bibinfo{author}{{Wei}, J.Y.},
  \bibinfo{author}{{Kreiner}, J.M.}, \bibinfo{author}{{Li}, H.L.},
  \bibinfo{year}{2010}.
\newblock \bibinfo{title}{{Orbital Period Changes and Their Evolutionary Status
  for the Weak-Contact Binaries. III. AO Camelopardalis and AH Tauri}}.
\newblock \bibinfo{journal}{\aj} \bibinfo{volume}{139},
  \bibinfo{pages}{195--204}.
\newblock \DOIprefix\doi{10.1088/0004-6256/139/1/195}.
\bibitem[{{Yildiz} and {Do{\u{g}}an}(2013)}]{Yildiz2013-MNRAS}
\bibinfo{author}{{Yildiz}, M.}, \bibinfo{author}{{Do{\u{g}}an}, T.},
  \bibinfo{year}{2013}.
\newblock \bibinfo{title}{{On the origin of W UMa type contact binaries - a new
  method for computation of initial masses}}.
\newblock \bibinfo{journal}{\mnras} \bibinfo{volume}{430},
  \bibinfo{pages}{2029--2038}.
\newblock \DOIprefix\doi{10.1093/mnras/stt028},
  \href{http://arxiv.org/abs/1301.6035}{{\tt arXiv:1301.6035}}.
\bibitem[{{Zhou} et~al.(2017){Zhou}, {Qian} and {Zhang}}]{Zhou2017-PASJ}
\bibinfo{author}{{Zhou}, X.}, \bibinfo{author}{{Qian}, S.},
  \bibinfo{author}{{Zhang}, B.}, \bibinfo{year}{2017}.
\newblock \bibinfo{title}{{Multi-color photometric investigation of the totally
  eclipsing binary NO Camelopardalis}}.
\newblock \bibinfo{journal}{\pasj} \bibinfo{volume}{69}, \bibinfo{pages}{37}.
\newblock \DOIprefix\doi{10.1093/pasj/psx010},
  \href{http://arxiv.org/abs/1705.05736}{{\tt arXiv:1705.05736}}.
\bibitem[{{Zhou} et~al.(2016){Zhou}, {Qian}, {Zhang}, {Li} and
  {Wang}}]{Zhou2016-AdAst}
\bibinfo{author}{{Zhou}, X.}, \bibinfo{author}{{Qian}, S.B.},
  \bibinfo{author}{{Zhang}, J.}, \bibinfo{author}{{Li}, L.J.},
  \bibinfo{author}{{Wang}, Q.S.}, \bibinfo{year}{2016}.
\newblock \bibinfo{title}{{The Photometric Investigation of V921 Her Using the
  Lunar-Based Ultraviolet Telescope of Chang'e-3 Mission}}.
\newblock \bibinfo{journal}{Advances in Astronomy} \bibinfo{volume}{2016},
  \bibinfo{pages}{746897}.
\newblock \DOIprefix\doi{10.1155/2016/7468976},
  \href{http://arxiv.org/abs/1608.00398}{{\tt arXiv:1608.00398}}.
\bibitem[{{Zhu} et~al.(2005){Zhu}, {Qian}, {Soonthornthum} and
  {Yang}}]{Zhu2005-AJ}
\bibinfo{author}{{Zhu}, L.Y.}, \bibinfo{author}{{Qian}, S.B.},
  \bibinfo{author}{{Soonthornthum}, B.}, \bibinfo{author}{{Yang}, Y.G.},
  \bibinfo{year}{2005}.
\newblock \bibinfo{title}{{Deep, Low Mass Ratio Overcontact Binaries. II. IK
  Persei}}.
\newblock \bibinfo{journal}{\aj} \bibinfo{volume}{129},
  \bibinfo{pages}{2806--2814}.
\newblock \DOIprefix\doi{10.1086/430187}.
\bibitem[{{Zhu} et~al.(2013){Zhu}, {Qian}, {Zhou}, {Li}, {Zhao}, {Liu} and
  {Liu}}]{Zhu2013-AJ}
\bibinfo{author}{{Zhu}, L.Y.}, \bibinfo{author}{{Qian}, S.B.},
  \bibinfo{author}{{Zhou}, X.}, \bibinfo{author}{{Li}, L.J.},
  \bibinfo{author}{{Zhao}, E.G.}, \bibinfo{author}{{Liu}, L.},
  \bibinfo{author}{{Liu}, N.P.}, \bibinfo{year}{2013}.
\newblock \bibinfo{title}{{Properties of the Close-in Tertiary in the Quadruple
  System V401 Cyg}}.
\newblock \bibinfo{journal}{\aj} \bibinfo{volume}{146}, \bibinfo{pages}{28}.
\newblock \DOIprefix\doi{10.1088/0004-6256/146/2/28}.

\end{thebibliography}

\end{document}